\documentclass[aps,pra,superscriptaddress,nofootinbib]{revtex4}
\usepackage[intlimits]{amsmath}
\usepackage{amsfonts}
\usepackage{psfrag}
\usepackage{array}
\usepackage{float}
\usepackage{tikz}

\usepackage{subfigure}
\usepackage{enumitem}
\usepackage{color}
\usepackage{subfigure}
\usepackage{enumitem}
\advance\voffset by -1cm
\advance\hoffset by -0.5cm
\newcommand\be            {\begin{equation}}
\newcommand\bea           {\begin{equation}\begin{array}l\displaystyle}
\newcommand\ee            {\end{equation}}
\newcommand\bes           {\begin{subequations}}
\newcommand\esu           {\end{subequations}}

\renewcommand{\(}{\left(}
\renewcommand{\)}{\right)}
\renewcommand{\[}{\left[}

\newcommand{\bigx}[1]{\bBigg@{#1}}

\newcommand\p            {\partial}

\newcommand\ket[1]{|#1\rangle}
\newcommand\bra[1]{\langle #1 |}
\newcommand\braket[3]{\langle #1 | #2 | #3 \rangle }
\newcommand\tr {\mathrm{Tr}}
\newcommand\med[1]{\langle #1 \rangle }

\renewcommand\vec[1]{{\vert{#1}\rangle}}

\newcommand\no[1]{{\,:\!#1\!:\,}}

\def\3pt#1#2#3{{\langle{#1}\vert{#2}\vert{#3}\rangle}}

\newcommand\doi[2]        {\href{http://dx.doi.org/#1}{#2}}

\newcommand{\EQ}{\begin{equation}}
\newcommand{\EN}{\end{equation}}
\usepackage{epsfig}
\usepackage{color}
\usepackage{psfrag}
\usepackage{amsmath}
\usepackage{graphicx}
\usepackage{amsfonts}
\usepackage{amssymb}
\begin{document}
\bibliographystyle{plainnat}

\title{{\Large {\bf The Coprime Quantum Chain}}}

\author{G. Mussardo}
\affiliation{SISSA and INFN, Sezione di Trieste, via Bonomea 265, I-34136, 
Trieste, Italy}
\author{G. Giudici}
\affiliation{SISSA and INFN, Sezione di Trieste, via Bonomea 265, I-34136, 
Trieste, Italy}
\author{J. Viti} 
\affiliation{ECT \& Instituto Internacional de Fisica, UFRN, Lagoa Nova 59078-970 Natal, Brazil}

\begin{abstract}
\noindent

In this paper we introduce and study the coprime quantum chain, i.e. a strongly correlated quantum system defined in terms of the integer eigenvalues $n_i$ of the occupation number operators at each site of a chain of length $M$. The $n_i$'s take value in
the interval $[2,q]$ and may be regarded as $S_z$ eigenvalues in the spin representation $j = (q-2)/2$. The distinctive interaction of the model is based on the coprimality matrix $\bf \Phi$: for the ferromagnetic case, this matrix assigns lower energy to configurations where occupation numbers $n_i$ and $n_{i+1}$ of neighbouring sites share a common divisor, while for the anti-ferromagnetic case it assigns lower energy to configurations where $n_i$ and $n_{i+1}$ are coprime. The coprime chain, both in the ferro and anti-ferromagnetic cases, may present an exponential number of ground states whose values can be exactly computed by means of graph theoretical tools. In the ferromagnetic case there are generally also frustration phenomena. A fine tuning of local operators may lift the exponential ground state degeneracy and, according to which operators are switched on, the system may be driven into different classes of universality, among which the Ising or Potts universality class. The paper also contains an appendix by Don Zagier on the exact eigenvalues and eigenvectors of the coprimality matrix in the limit $q \rightarrow \infty$.

\vspace{3mm}
\noindent
Pacs numbers: 11.10.St, 11.15.Kc, 11.30.Pb

\end{abstract}
\maketitle

\section{Introduction}
\label{sec:intro}
\noindent
The question of divisibility is arguably among the oldest problems of mathematics being, as it is, an aspect deeply related to the cycles of nature.
There are numbers, such as 360 for instance, which have always  had a special appeal since they are divisible by many smaller integers. At the other extreme there are numbers with no smaller divisors except 1 -- the prime numbers -- that are, undeniably, even more appealing: not only the primes are indivisible
but, by a fundamental theorem, they may also be regarded as the atoms of arithmetic, since any natural number can be factorised in an unique way in terms of them. In contrast with the finitely many chemical elements, the number of primes is however infinite, as already proved by Euclid in his Elements.
On primes numbers, divisibility and the like there is of course a huge series of books and articles that the reader may find interesting and even amusing as, for instance, those of references \cite{Rimenboim1,Rimenboim2,Schroeder,Zagier,BD,Young}. 

Number Theory -- the branch of pure mathematics which studies the discrete properties of numbers, such as arithmetic functions, distribution of prime numbers, congruences, quadratic residues and many other of those properties-- seems to be at any rate the farthest subject from physics. This impression also hinges upon the distinction which exists between discrete and continuous mathematics: while the latter employs the concept of limit, the former uses induction, and in the traditional view in which space and time are continuous and the laws of nature are described by differential equations, Number Theory seems indeed to play no fundamental role in our understanding of the physical world. 

However, this is a superficial conclusion. First of all, at a deeper level there is no dividing line between discrete and continuous mathematics, as shown for instance by the well-known article by  Bernhard Riemann on prime numbers \cite{Riemann}, where key progresses were made using sophisticated methods from analysis. Nowadays the so called Analytic Number Theory -- the area which uses methods borrowed from analysis to approach properties of numbers -- not only is a well developed subject (see, for instance \cite{Hardy,Baker,Apostol,Manin}) but still remains a remarkable source of famous open problems and conjectures, such as for instance the generalised Riemann hypothesis about the zeros of the $\zeta(s)$ function and other Dirichlet series \cite{Edwards,Tichmaesh,Conrey,Bombieri,Sarnak,Mazur}. Secondly and even more importantly, the advent in physics of quantum mechanics -- in particular the emphasis given to the {\em discrete} spectrum of certain physical operators, like the Hamiltonian -- has drastically changed the classical prospective, stimulating over the years a very fertile exchange of ideas between number theory and quantum mechanics. Following for instance the original suggestion by Polya and Hilbert in 1910, there have been later several attempts to solve the Riemann hypothesis in terms of quantum mechanical models (see for instance \cite{Berry,Connes,Sierra,Leclair,Schumayer} and references therein). Similarly, some years ago there was a proposal by one of the authors of this paper \cite{GMussardo} to solve the primality problem, namely to determine whether a given integer is a prime or not, using a quantum mechanical scattering experiment for a properly designed semi-classical potential that has the prime numbers as its only eigenvalues.   

While in the reference \cite{GMussardo} the primality problem was translated into a {\em one-particle} quantum mechanical setup, this paper instead puts forward a 
{\em many-body} quantum Hamiltonian which exploits the coprimality between integer numbers. We believe that, with proper insights, such a quantum system can be experimentally realised by cold atoms and moreover in two equivalent ways: either by means of spinless atoms and their on-site integer occupation numbers $n_i$ with a maximum value $q$, or employing instead atoms with higher spin, which live in the spin representation $j = (q-2)/2$. In both cases, using a proper optical laser design, we can firstly accommodate the atoms on a regular lattice and secondly let them  interact through a next-neighbouring interaction tailored in such a way to be sensitive to the relative coprimality of the integer numbers $n_i$ and $n_{i+1}$: here we simply recall that two integers $a$ and $b$ are coprime if their greatest common divisor is just $1$. Contrary to other more familiar quantum chains, such as XXZ or the like, we will show that the coprime quantum chain has the notable property of presenting an exponential degeneracy of its ground state. However, a proper tuning of additional local operators may break such a huge degeneracy and lead to a closure of the mass gap, therefore driving the original coprime quantum chain into criticality: the interesting thing is that, depending both on the maximum value $q$ of the occupation numbers and the type of operators switched on, one can reach different classes of universality as, for instance, the one of the Ising model or the $3-$state Potts model. As largely discussed later, such predictions can be accurately checked by exploiting entanglement entropy measures \cite{ent1,ent2,ent3,ccee}. It is also worth to underline that it is for the huge degeneracy of the ground state that the two-dimensional classical analogue of the coprime chain is always disordered and it has only  a high temperature phase. In short, the coprime quantum chain seems to give rise to a quite rich physical scenario: a remarkable situation, given that the dynamics of this model is based on a condition so simple as the coprimality between integer numbers.  

The paper is organised as follows. In Section \ref{sec:def} we introduce the definition of the coprime quantum chain, i.e. its Hilbert space and Hamiltonian. 
In Section \ref{coprimality} we discuss the main properties of the coprimality matrix, underlying both the \lq\lq random'' nature of this matrix and its periodicities, as vividly shown by its discrete Fourier transform. In Section \ref{coprimality} we also recall some basic facts of prime numbers and we introduce the prime-number vectors whose overlaps capture the interactions encoded in the Hamiltonian. As it will become soon clear, to understand the dynamics of such a quantum chain an important point is the analysis of the \lq \lq classical" ground states of the coprime quantum chain, i.e. the states of minimal energy in the absence of operators in the Hamiltonian which induce transitions among the various occupation numbers $n_i$. For this reason, in Section \ref{classicalgroundstates} we address the problem of counting the number of classical ground states in the case of ferromagnetic interaction. In the subsequent
Section \ref{expofer}, using results from graph theory, we discuss the exponential degeneracy  of the classical ground states, whose precise number depends of course on the boundary conditions. In Section \ref{classicalanti} we repeat the analysis for the anti-ferromagnetic case. In Section \ref{spectrumQH} we discuss the phase diagram of the  coprime quantum chain in the ferromagnetic case and we show that, with an appropriate tuning of some local operators, we can drive the system into different classes of universality, including those of Ising or $3-$state Potts model.
In Section \ref{classicalcoprimeprime}, mimic a Peierls argument, we will prove that the classical analogue of the  coprime  quantum chain is always in its disordered high temperature phase. Finally, our conclusions are gathered in Section \ref{conclusions}. The paper also contains several appendices: Appendix A collects the main results of graph theory needed in the text; Appendix B shows the explicit calculation of the maximum degree of the graph associated to the coprime model, in the limit in which $q \rightarrow \infty$; Appendix C, written by Don Zagier, is concerned with the detailed analysis of the eigenvalues and eigenvectors of the coprime matrix in the limit $q \rightarrow \infty$.

\section{Definition of the coprime quantum chain}
\label{sec:def}
\noindent
In this section we  introduce the coprime quantum chain and its general quantum Hamiltonian for the case of a one-dimensional lattice consisting of 
$M$ sites. 

\begin{figure}[t]
\begin{center}
\psfig{figure=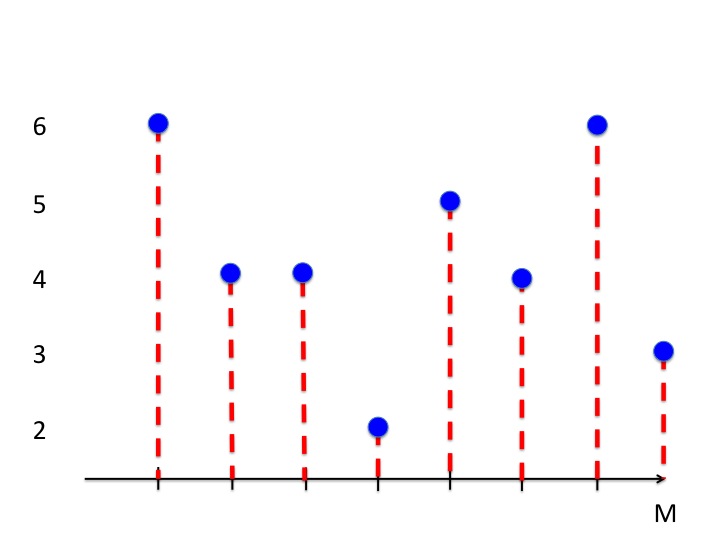,height=5cm,width=8cm}
\caption{{\em A configuration of the coprime model with $q=6$. In this example the various occupation numbers are $n_1=6,~n_2=4$ etc.}}
\label{coprimechainmodel}
\end{center}
\end{figure}

\vspace{3mm}
\noindent
{\bf Hilbert Space}. The fundamental degrees of freedom in the coprime chain are the occupation number operators $\hat n_i$ at each site $i$ of a one-dimensional lattice. These operators are characterised by their eigenvalues $n_i$, which take $(q-1)$ integer values
\be
n_i \,=\,2,3,\ldots, q \,\,\,.
\label{possiblevalues}
\ee 
For reasons that will become clear soon, we have shifted the more conventional interval of the  occupation numbers by $2$,
 so that the lowest possible value is $2$ while the maximum is $q$. We assume that the gas described by the occupation numbers~(\ref{possiblevalues}) obeys 
 a bosonic statistics, although the number of particles on a certain lattice site $i$ cannot exceed the value $q$ and be less than 2. In the limit $q\rightarrow\infty$ the system is a true one-dimensional Bose gas. As customary, we can also define at each site the annihilation and creation operators $c_i^-$ and $c_i^+ = (c_i^-)^\dagger$, with the properties  
\be 
c_i^-\,\mid 2 \rangle \,=\, 0 
\hspace{5mm}
,
\hspace{5mm}
c_i^+ \, \mid q \rangle \,=\, 0 
\,\,\,\,\,\,\,\,\,\,\,\,\,
,
\,\,\,\,\,\,\,\,\,\,\,\,\,
\forall i \,\,\,.
\ee
We can alternatively regard the $(q-1)$ possible occupation numbers~\eqref{possiblevalues} as the eigenvalues of the $S_z$ component of an ordinary spin in representation $j = (q-2)/2$. In order to match the eigenvalues $m$ of $S_z$ with the values (\ref{possiblevalues}), one needs the relation
\be 
m \,=\, n_i -\frac{q+2}{2} \,\,\,. 
\ee
Using this mapping of the occupation number operators onto a spin system, we can then define the action of $c_i^+$ and $c_i^-$ 
on each state as  
\be 
\begin{array}{l}
c_i^- \,\mid n_i \rangle \,=\,\sqrt{(n_i -2) \,(q - n_i +1)} \,\mid n_i -1\rangle\,\,\,, \\
\\
c_i^+ \,\mid n_i \rangle \,=\, \sqrt{(q - n_i) (n_i -1)} \,\,\,\,\mid n_i +1 \rangle\,\,\,. 
\end{array}
\ee
These operators satisfy the commutation relations 
\begin{eqnarray}
\left[\hat n_i , c_i^\pm \right] &\,=\,&  \pm \, c_i^\pm\,\,\,,  \label{commutationrelation}\\
\left[c_i^+ , c_i^-\right] & \,=\, & \hat n_i - \frac{(q-2)}{2}\,\,\,.   
\end{eqnarray}
Hence, on a chain of $M$ sites, the dimension of the Hilbert space is ${\rm dim}\, {\cal H} = (q-1)^M$ and its Fock space is spanned by the 
vectors 
\be 
\mid n_1, n_2, \ldots, n_M \rangle \,=\, \mid n_1 \rangle \, \otimes \,\mid n_2 \rangle \, \cdots \,\otimes \, \mid n_M \rangle \,\,\,
\label{FOCKSPACE}
\ee
associated to the occupation numbers at each site of the chain. A typical configuration of the coprime model is shown in Fig.~\ref{coprimechainmodel}. 
In the following we will consider various boundary conditions for the coprime chain, such as cyclic (periodic) or fixed boundary conditions, the former associated to the condition $n_{i+M} \,=\,n_i$, the latter to two fixed values of both $n_1$ and $n_M$. We will also consider free boundary conditions, where the values at the extreme of the chain are free to assume any
possible value in the interval $[2,q]$. 

\vspace{3mm}
\noindent
{\bf Local hermitian operators}. The generic form of a local hermitian operator acting on the vectors (\ref{FOCKSPACE}) is given by  
\be 
G_i \,=\, {\bf 1} \otimes {\bf 1} \cdots {\bf 1} \otimes \underset{\underset{i-site}{\uparrow}}{{\mathcal G}} \otimes {\bf 1} \otimes {\bf 1} \cdots \otimes {\bf 1}  
\label{notation}
\ee
where ${\mathcal G}$ is an hermitian matrix acting on the $(q-1)$ dimensional Hilbert space at the site $i$, while ${\bf 1}$ is the $(q-1)\times (q-1)$ identity matrix acting on each of the remaining sites. Let's remind that over the real numbers $\mathbb{R}$, the complex $(q-1) \times (q-1)$ hermitian matrices form a vector space of dimension $(q-1)^2$: denoting by $E^{ik}$ the $(q-1) \times (q-1)$ matrix with entry one in the position $(j,k)$  and zeros elsewhere, a canonical basis is given by 
\begin{equation}
\begin{array}{lllll}
{\mathcal D}^{(j)} \,\,= E^{jj} & &1 \leq j \leq q-1 & & \,(q-1 \,\,\,{\rm matrices}), \\
{\mathcal S}^{(jk)} \,=  (E^{jk} + E^{kj}) & & 1 \leq j <  k \leq q-1 & & \left( \frac{(q-1) (q-2)}{2} \,\,\, {\rm matrices} \right), \\
{\mathcal A}^{(jk)} = i (E^{jk} - E^{kj}) & & 1 \leq j <  k \leq q-1 & & \left( \frac{(q-1) (q-2)}{2} \,\,\, {\rm matrices} \right).
\end{array}
\label{magneticandallthat}
\end{equation}
Notice that the operators ${\mathcal D}^{(j)}$ play the role of magnetic fields: indeed, switching on one of them, say ${\mathcal D}^{(s)}$, the system tends to polarise the occupation numbers $n_i$ along the value $s$. The operators ${\mathcal S}^{(jk)}$ and ${\mathcal A}^{(jk)}$ play instead the same role of the Pauli matrices $\sigma_x$ and $\sigma_y$ for the spin 1/2 quantum spin chains, namely they mix the values of the occupation numbers at each site. To simplify the notation, in the following we will assume that the matrices given in eq.\,(\ref{magneticandallthat}) have been enumerated according to an index $\alpha =1,2,\ldots (q-1)^2$ and therefore generically denoted as ${\mathcal G}^{(\alpha)}$. Hence, with this new notation, a basis for the local hermitian operators is given by 
\be 
G_i^{(\alpha)} \,=\, {\bf 1} \otimes {\bf 1} \cdots {\bf 1} \otimes \underset{\underset{i-site} \uparrow}{{\mathcal G}}^{\hspace{-2mm}(\alpha)} \otimes {\bf 1} \otimes {\bf 1} \cdots \otimes {\bf 1}  
\hspace{6mm}
,
\hspace{6mm}
\alpha=1,2,\ldots, (q-1)^2\,\,\,.
\label{notationlocalbasis}
\ee

\vspace{3mm}
\noindent
{\bf Quantum Hamiltonian}. 
In order to introduce the quantum Hamiltonian of our model, it is convenient to consider initially the arithmetic function 
\be \label{coprimalityfunction}
\Phi(a,b) \,=\,
\left\{ 
\begin{array} {lll}
0 & & {\rm if} \, {\rm gcd}(a,b) =1 \\
1 & & {\rm if} \, {\rm gcd}(a,b) \neq 1 
\end{array}
\hspace{5mm}
,
\hspace{5mm}
a,b \in \mathbb{N}
\right. 
\ee
where ${\rm gcd}(a,b)$ stands for the greatest common divisor between the two natural numbers $a$ and $b$. In the following we will say that 
two integers $a$ and $b$ are coprime if their greatest common divisor is $1$. We call $\Phi(a,b)$ the {\em coprimality function} and 
its properties will be discussed in greater detail in Section \ref{coprimality}. 

The coprime quantum chain\footnote{In the following we will sometimes refer to the model as \lq\lq $q-$coprime chain", in particular if we want to emphasise the properties of the quantum chain with respect to parameter $q$.} is a local model whose  Hamiltonian is given, in the basis of the occupation numbers, by
\be 
H \,=\,-\sum_{i=1}^M \left[\Phi(n_i,n_{i+1})\, + 
\sum_{\alpha=1}^{(q-1)^2} \beta_{\alpha} G^{(\alpha)}_i   \right] \,\,\,. 
\label{quantumhamiltonian}
\ee
Let's stress that the fingerprint of this model is the omnipresence of the first term that is diagonal in the basis of the occupation numbers\footnote{
The coprimality function in the Hamiltonian is formally multiplied by a tensor product of the identity operators on next neighbouring sites.}. Notice that this kind of interaction makes the model qualitatively different from any other more familiar spin chain considered in the literature, such as XXZ, Heisenberg or Potts spin chain, etc. The parameters $\beta_{\alpha}$ are genuine coupling constants whose values determine the different phases of the model. It will be especially interesting to see later how, by defining a suitable combination of these couplings, we will be able to filter particular ground states of the quantum chain. 

Last comment: as it is written, the quantum Hamiltonian \eqref{quantumhamiltonian} refers to the {\em ferromagnetic case}, since it privileges equal or common divisible values
of the occupation numbers of neighbouring sites. The {\em antiferromagnetic case} can be easily obtained by changing 
in~\eqref{quantumhamiltonian} the diagonal interaction as 
\be 
\Phi(a,b) \rightarrow \overline\Phi(a,b) = 1 - \Phi(a,b)\,\,\,. 
\label{antiferromagnetic}
\ee
After this transformation the configurations which become more favourable are obviously those in which two nearby sites have numbers which share 
no common divisors.

\section{The Coprimality matrix}\label{coprimality}
\noindent

\vspace{3mm}
\noindent
{\bf Basic Arithmetic}. 
Before discussing in greater detail the coprimality function $\Phi(a,b)$, let us recall that a fundamental result in number theory is the unique decomposition of  a natural number $n$ into its prime factors $p_i$, counted with their relative multiplicities $\sigma_i$ 
\be 
n \, =\, p_1^{\sigma_1} \, p_2^{\sigma_2} \, \cdots p_l^{\sigma_l} 
\label{primedecomposition}
\,\,\,.
\ee
Simple as it is, this theorem will be the basis for what follows. Moreover, it is also useful to recall two other related properties of the prime numbers: the first, 
known as Bertrand's theorem \cite{Erdos}, states that, for any integer $n$, there is always a prime $p$ in the interval $(n, 2n)$, alias 
\be 
n < p < 2n \,\,\,. 
\label{doppio}
\ee 
The second property, somehow equivalent to the previous one, concerns a bound on the $(k+1)$-th prime number in terms of $p_k$
\be
p_{k+1} < 2 \,p_k \,\,\,.
\label{twiceprime}
\ee
Finally, let's remind that a pretty simple approximate expression for the $n$-th prime number is given by 
\be 
p_n \sim n \, \log n\,\,\,,
\label{approximateprime}
\ee
the above statement is equivalent to the celebrated prime number theorem (see \cite{BD} for an historical survey).

\vspace{3mm}
\noindent
{\bf Coprimality}. 
We now turn our attention to the coprimality function: once fixed the maximum eigenvalue $q$ of the number operators $\hat n_i$, 
we can define the $(q-1) \times (q-1)$ ferromagnetic
{\em coprimality matrix} $\bf\Phi$ whose matrix elements
are expressed by the coprimality function $\Phi(a,b)$
\be 
\left[{\bf \Phi}\right]_{ab} \,=\, \Phi(a,b) \,\,\,.
\label{coprimalitymatrix}
\ee
Notice that in our convention the indices of the coprimality matrix run from $2$ to $q$, for instance the top-left element is $\bf\Phi_{22}$. The matrix  
$\bf\Phi$ is a real and symmetric matrix made of $0$ and $1$, with some peculiar properties which can be unveiled using well known results in number theory. First of all, as it follows from its very definition, the function $\Phi(a,b)$ is testing whether or not the two integer numbers $a$ and $b$ have some common divisor greater than $1$: when such a number exists its output is $1$, otherwise it is $0$\footnote{For the peculiar role played by the integer number $1$, which acts as a \lq\lq neutral" divisor of all natural numbers, it seems wiser to exclude it from the list of possible values assumed by the occupation numbers and therefore to start their values from $2$, as we actually do. In this way, a-priori there is no privileged value among the entire set of occupation numbers.}. Hence, given two numbers $a$ and $b$, $\Phi(a,b)$ is checking a looser property of these numbers rather than their individual primality:  indeed it scrutinizes their common prime number content. So, if $a$ and $b$ were both primes, say $a = 3$ and $b=11$, obviously $\Phi(3,11) =0$ but an output equal to 0 could also result from two composite numbers that do not share any common divisor, as for example would happen choosing $a = 30 = 2 \times 3\times 5$ and $b = 77 = 7 \times 11$. In other words, the coprimality matrix $\bf\Phi$ is sensitive to the {\em multiplicative} structure of the natural numbers rather
than their {\em additive} structure. Notice that, with the definition \eqref{coprimalitymatrix} adopted for $\bf\Phi$, all the diagonal elements of this matrix are equal to $1$, so that ${\rm Tr}\, \,{\bf\Phi} = (q-1) $. 

We can also define the coprimality matrix $\bf{\overline \Phi}$ of the antiferromagnetic case as
\begin{equation}
\bf{\overline \Phi}\,=\, J - \bf\Phi \,\,\,,
\label{antifercoprim}
\end{equation}
where $J$ is the $(q-1)\times (q-1)$ matrix with all entries equal to one  
\begin{equation}
J \,=\, \left( 
\begin{array}{llllllll}
1 & 1 & 1 & 1  & . & .& 1 & 1\\ 
1 & 1 & 1 & 1 & . & .& 1 & 1\\ 1 & 1 & 1 & 1  & . & .& 1 & 1 \\ . & . & . & . & . & . & . & .\\ . & . & . & . & . & . &. &. \\
 1& 1 & . & . & 1  &  1& 1 & 1 \\ 
1 & 1 & . & . & 1 & 1  & 1 & 1 
\end{array}
\right).
\end{equation}
With respect to $\bf \Phi$, the matrix $\bf{\overline \Phi}$ have all $0$'s and $1$'s swapped and in this case,  ${\rm Tr}\, \,{\bf{\overline \Phi}} = 0$.

\vspace{3mm}
\noindent
{\bf Prime-Number Vectors}. Given the multiplicative nature of the function $\Phi(a,b)$,  it is useful to introduce an alternative representation 
for the $(q-1)$ numbers involved in the coprimality matrix $\bf\Phi$. The first step for doing so is to identify the set of the $l$ prime numbers less than $q$ 
which then are also among the allowed occupation numbers in (\ref{possiblevalues})
\be
{\mathcal P} \,=\,\{2,3,5,\ldots p_{l-1}, p_l\}  
\hspace{5mm}
,
\hspace{5mm} p_l \leq q \,\,\,. 
\ee
The total number $l$ of these primes -- as a function of $q$ -- is given by the prime-counting function $\pi(q)$ (see, for instance \cite{Schroeder,Zagier})
which, for our present purposes, can be approximated by the {\em logarithmic integral} ${\rm Li}(q)$
\be
\pi(q) \simeq \int_2^q \frac{dt}{\log(t)} \equiv {\rm Li}(q)\,\,\,.
\label{countingprime}
\ee
Since ${\rm Li}(x) \simeq x/\log(x)$, the number of primes present in the interval $[2,q]$ is thus roughly $l \simeq q/\log q$.  
This estimate tells us that there is always a fair number of primes in each interval $[2,q]$ of the possible values of the occupation numbers, although their number is (logarithmically) smaller than $q$ itself. 

Consider now a series of $l$-dimensional boolean vectors (which we called {\em prime-number vectors}) associated to $l$ boxes in 
correspondence to the $l$ primes in the interval $[2,q]$ as in the figure \ref{boxes} below. 
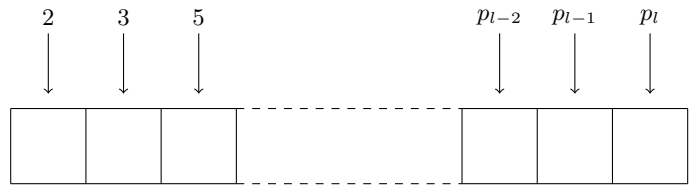
\begin{figure}[b]
\begin{center}
\begin{tikzpicture}
\node[above] at (0.5,1) {$2$};
\node[above] at (0.5+1,1) {$3$};
\node[above] at (0.5+2,1) {$5$};
\draw[->](0.5,1)--(0.5,0.2);
\draw[->](0.5+1,1)--(0.5+1,0.2);
\draw[->](0.5+2,1)--(0.5+2,0.2);
 \draw(0,0)--(3,0);
 \draw(0,-1)--(3,-1);
 \draw(0,0)--(0,-1);
 \draw(1,0)--(1,-1);
 \draw(2,0)--(2,-1);
 \draw(3,0)--(3,-1);
 \draw[dashed] (3,0)--(6,0);
 \draw[dashed] (3,-1)--(6,-1);
 \draw(6,0)--(9,0);
 \draw(6,-1)--(9,-1);
 \draw(6,0)--(6,-1);
 \draw(1+6,0)--(1+6,-1);
 \draw(2+6,0)--(2+6,-1);
 \draw(3+6,0)--(3+6,-1);
 \draw[->](0.5+6,1)--(0.5+6,0.2);
\draw[->](0.5+7,1)--(0.5+7,0.2);
\draw[->](0.5+8,1)--(0.5+8,0.2);
\node[above] at (6.5,1) {$p_{l-2}$};
\node[above] at (6.5+1,1) {$p_{l-1}$};
\node[above] at (6.5+2,1) {$p_{l}$};
\end{tikzpicture}
\caption{{\em The prime-number vector with $l$ entries associated to the integer $n$. The $k$-th entry is one if $n$ is divisible by the
prime $p_k$, it is zero otherwise.}}
\label{boxes}
\end{center}
\end{figure}
Using the prime decomposition (\ref{primedecomposition}), we can associate to each number $n$ in the interval $ [2,q]$ a prime-number vector: this vector is simply obtained by filling the $k$-th box with $1$ if the prime $p_k$ is present in the decomposition of $n$ (independently of its multiplicity), or filling the $k$-th box with $0$ otherwise. In other words, this assignment  flattens the various powers $\sigma_n$ of the prime number decomposition (\ref{primedecomposition}) of $n$;  
in this way we only keep track of the divisibility of $n$ by $p_k$. Consider for instance 
when $q = 37$: in this case the set ${\mathcal P}$ has cardinality $l = 12$ and consists of the prime numbers 
$$
{\mathcal P} = \{2, 3, 5, 7, 11, 13, 17, 19, 23, 29, 31, 37\} \,\,\,.
$$
We have then a $12$-dimensional prime-number vector space and with the rule given above the number $36$, say, will be represented by a prime-number vector as 
$$ 
n = 36 = 2^2 \,\times\, 3^2 \,\,\,\longrightarrow \,\,\,(1,1,0,0,0,0,0,0,0,0,0,0) \,\,\,.
$$
Since the dimension of the prime-number vector space is smaller\footnote{For large values of $q$ the dimension of this space is computed below.} than $q$, and moreover not all $l$-dimensional boolean vectors are present in the prime-number vector space\footnote{It is obvious, for instance, that the vector $y = (1,1,1,\ldots, 1,1,1)$ made of all $1$'s cannot be in the prime-number space, because it would correspond, at least, to the natural number $n = p_1 p_2 \ldots p_l$ (i.e. to the number given by the product of all the $l$ primes) which is much greater than the maximum number $q$ of the interval. Similar consideration may be applied to other boolean $l$-dimensional vectors.}, these two facts taken together imply that there will be a certain degree of degeneracy in this mapping, namely different integers will be associated to the
same prime-number vector. 

\begin{figure}[t]
\vspace{8mm}
\psfig{figure=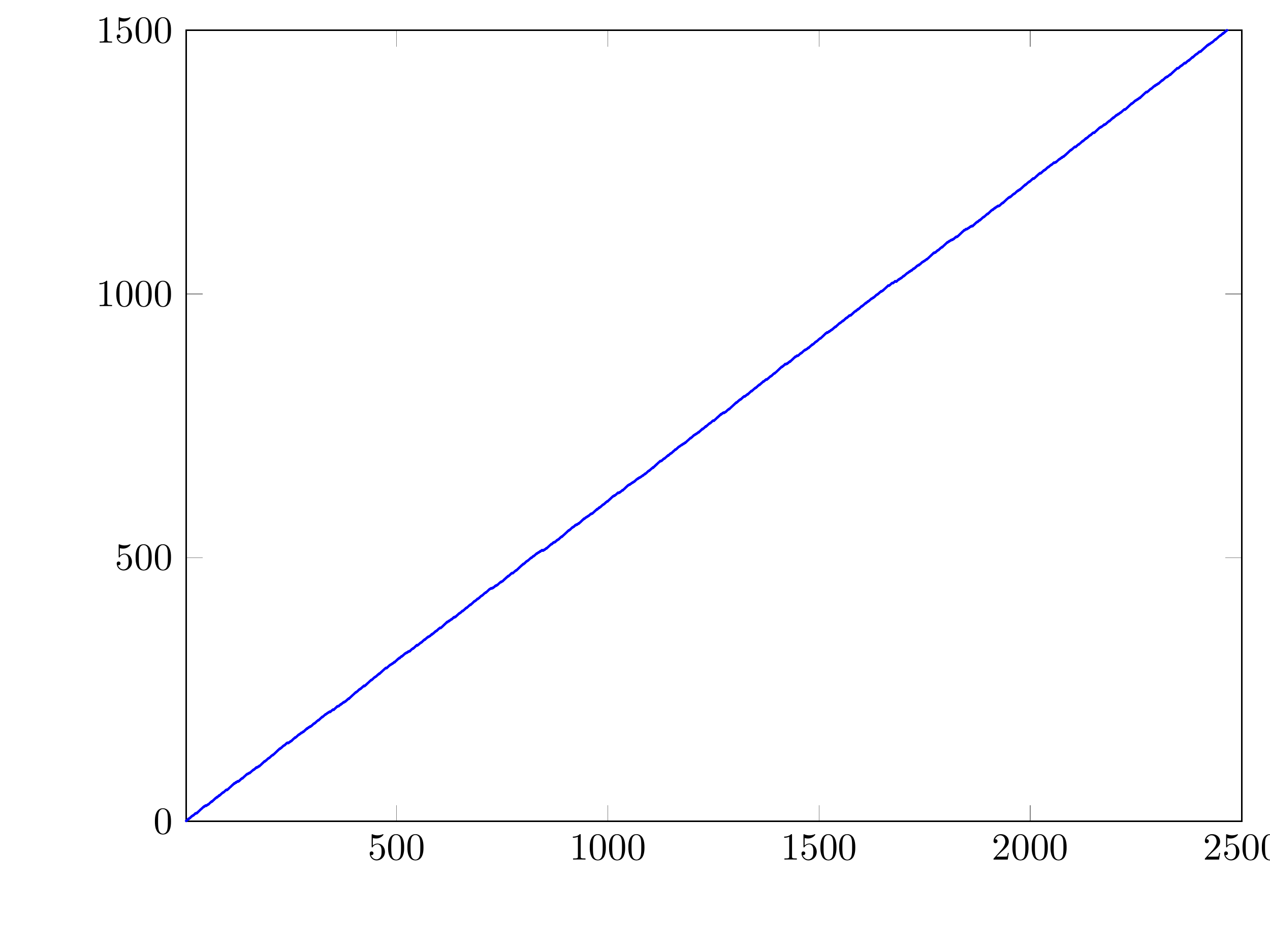, width=0.55\textwidth}
\caption{{\em Number of equivalence classes versus $q$ of the coprime chain. The dimension of the coprimality matrix $\bf\Phi$ is $(q-1)\times (q-1)$.}} 
\label{numberclasses}
\end{figure}

This means that all the integers in the interval $[2,q]$ fall into different {\em equivalence classes} which are identified by the their prime-number vectors. For instance, all numbers that are pure powers
of $2$ will belong to the same equivalence class associated to the same $l$-dimensional vector $v = (1,0,0,0,0,\ldots, 0)$, as well as all
the pure powers of $3$ pertain to another equivalence class associated to the $l$-dimensional vector $w = (0,1,0,0,0,\ldots,0)$, etc.   
In summary with this procedure, we can associate to each natural number $n$ its equivalence class and its  
prime number representative vector $v_n$ 
\be 
n = 2^{\sigma_2} \, \cdot\, 3^{\sigma_3} \ldots p_k^{\sigma_k} \cdot \, p_s^{\sigma_s} \,\,\,
\longrightarrow \,\,\, v_n \,=\, (1, 1, \ldots 
\underset{\underset{k}{\uparrow}}{1}, 0, \ldots, 0, \underset{\underset{s}{\uparrow}}{1}\ldots, 0) \,\,\,.
\ee
To make an explicit example, for $q=37$ we have the following $23$ equivalence classes  
\be
\begin{array}{llll}
(2,4,8,16,32) & & \rightarrow & (1,0,0,0,0,0,0,0,0,0,0,0) \\
(3,9,27) & &\rightarrow & (0,1,0,0,0,0,0,0,0,0,0,0) \\
(5,25) & &\rightarrow & (0,0,1,0,0,0,0,0,0,0,0,0) \\
(6,12,18,24,36) & & \rightarrow & (1,1,0,0,0,0,0,0,0,0,0,0) \\
(7) & & \rightarrow & (0,0,0,1,0,0,0,0,0,0,0,0) \\
(10,20) & & \rightarrow  & (1,0,1,0,0,0,0,0,0,0,0,0) \\
(11) & & \rightarrow &  (0,0,0,0,1,0,0,0,0,0,0,0) \\
(13) & & \rightarrow & (0,0,0,0,0,1,0,0,0,0,0,0) \\
(14,28) && \rightarrow & (1,0,0,1,0,0,0,0,0,0,0,0) \\
(15) & &\rightarrow &  (0,1,1,0,0,0,0,0,0,0,0,0) \\
(17) & &\rightarrow &  (0,0,0,0,0,0,1,0,0,0,0,0) \\
(19) & & \rightarrow &  (0,0,0,0,0,0,0,1,0,0,0,0) \\
(21) & & \rightarrow &  (0,1,0,1,0,0,0,0,0,0,0,0) \\
(22) & & \rightarrow &  (1,0,0,0,1,0,0,0,0,0,0,0) \\
(23) & &\rightarrow &  (0,0,0,0,0,0,0,0,1,0,0,0) \\
(26) & & \rightarrow &  (1,0,0,0,0,1,0,0,0,0,0,0) \\
(29) & & \rightarrow &  (0,0,0,0,0,0,0,0,0,1,0,0) \\
(30) & & \rightarrow &  (1,1,1,0,0,0,0,0,0,0,0,0) \\
(31) & &\rightarrow &  (0,0,0,0,0,0,0,0,0,0,1,0) \\
(33) & & \rightarrow &  (0,1,0,0,1,0,0,0,0,0,0,0) \\
(34) & & \rightarrow &  (1,0,0,0,0,0,1,0,0,0,0,0) \\
(35) & &\rightarrow &  (0,0,1,1,0,0,0,0,0,0,0,0) \\
(37) & & \rightarrow &  (0,0,0,0,0,0,0,0,0,0,0,1) 
\end{array}
\ee
It is easy to see that the number of classes, here denoted by ${\mathcal C}(q)$, coincides with the number of square-free integers\footnote{A square-free number is a number not divisible by a square. The function of number theory that identifies the square-free integers is the absolute value of the Moebius function $\mu(n)$, see \cite{Hardy}. Indeed $|\mu(n)|=1$ if and only if $n$ is a square-free number and zero otherwise.} less than $q$ and therefore, for large values
of $q$, it scales as \cite{Hardy}  
\be
{\mathcal C}(q) \simeq \frac{6}{\pi^2} \, q \,=\, 0.6079271019\cdots \,\, q. 
\label{lowerbound}
\ee
To show \eqref{lowerbound}, let us compute the probability that an integer $n$ randomly selected is square-free. The root of such a computation are
the loose correlations that exist among the primes, so that the probability that a given integer is divisible by the prime $p$ can be assumed to be $1/p$ 
(since in any sequence of natural numbers, one out of $p$ is divisible for $p$). Within this assumption, for an integer to be square-free,
it must not be divisible by the same prime $p$ more than once. Hence, either the number $n$ is not divisible by $p$ or, if it is, it is not divisible once again.
Therefore, denoting $P$ such a probability we have   
\begin{equation}
\label{square}
P \,=\, \left(1 - \frac{1}{p}\right) + \frac{1}{p} \,\left(1 - \frac{1}{p}\right) \,=\, 
1 - \frac{1}{p^2} \,\,\,.
\end{equation}
Recalling now the Euler infinite product representation of Riemann $\zeta(s)$ function
\begin{equation}
\zeta(s) \,\equiv \, \sum_{n=0}^\infty \frac{1}{n^s} \,=\, \prod_{p~\text{prime}} \frac{1}{1 - \frac{1}{p^s}} \,\,\,,
\label{Euler-Riemann}
\end{equation} 
and taking the product on all the possible primes in \eqref{square} (assuming independence of the divisibility by different primes), we end up with 
\be 
P_{tot} \simeq \prod_{p~\text{prime}} \left(1 - \frac{1}{p^2}\right) \,=\,\frac{1}{\zeta(2)} \,=\,\frac{6}{\pi^2} = 0.607927\cdots
\label{predictedslope}
\ee
Finally, since $P_{tot}$ in \eqref{predictedslope} is the fraction of square-free numbers, it coincides with $\lim_{q\rightarrow\infty}q^{-1}\mathcal{C}(q)$. An experimental
determination of the number of equivalence classes (obtained by really counting them) as a function of $q$ is shown in Fig.~\ref{numberclasses}.  One can obviously identify a linear behaviour in $q$, whose best fit produces a result quite close to the asymptotic exact formula (\ref{predictedslope}) 
\be 
{\mathcal C}(q) \simeq 
0.607 \, q  \,\,\, . 
\label{bestfitclasses}
\ee

From the point of view of the interaction dictated by the coprimality matrix~\eqref{coprimalitymatrix}, it is easy to realize that all vectors belonging to the same equivalence class are indistinguishable. Moreover, the coprimality matrix itself can be expressed in terms of the matrix of the overlaps of these prime-number vectors, i.e. their scalar products 
\be 
\phi(a,b) \,=\,\frac{\langle v_a \mid v_b \rangle}{d_{ab}} \,\,\,, 
\label{Gram}
\ee
where $d_{ab}$ is the total number of common divisors of the two numbers $a$ and $b$. Notice that the scalar product of coprime numbers 
simply vanishes. 

\vspace{3mm}
\noindent
{\bf Random Nature of the Coprimality Matrix}. 
The sensitivity to the multiplicative nature of the natural numbers awards to the matrix $\bf\Phi$ a certain degree of randomness. Indeed, assuming known 
the matrix element $\Phi(a,b)$, it would be impossible to predict just on the basis of this information the neighbouring matrix element $\Phi(a,b+1)$: passing from $b$ to $b+1$, we are in fact exploiting the {\em additive} nature of the natural numbers, while $\Phi(a,b)$ is  sensitive only to their {\em multiplicative} properties. So, it can easily happen that by adding $1$ to the number $b$ we can pass from a highly composite number to a prime number and vice-versa: take for instance the highly composite number $b = 2310 = 2 \times 3 \times 5 \times 7 \times 11$ and its consecutive number $b+1 = 2311$ which is instead prime. Therefore, spanning all the values along each row of the matrix, we will  essentially observe a random sequence of $0$'s and $1$'s, whose average however can be predicted with a reasonable accuracy by a simple argument. 

\begin{figure}[t]
\vspace{8mm}
\psfig{figure=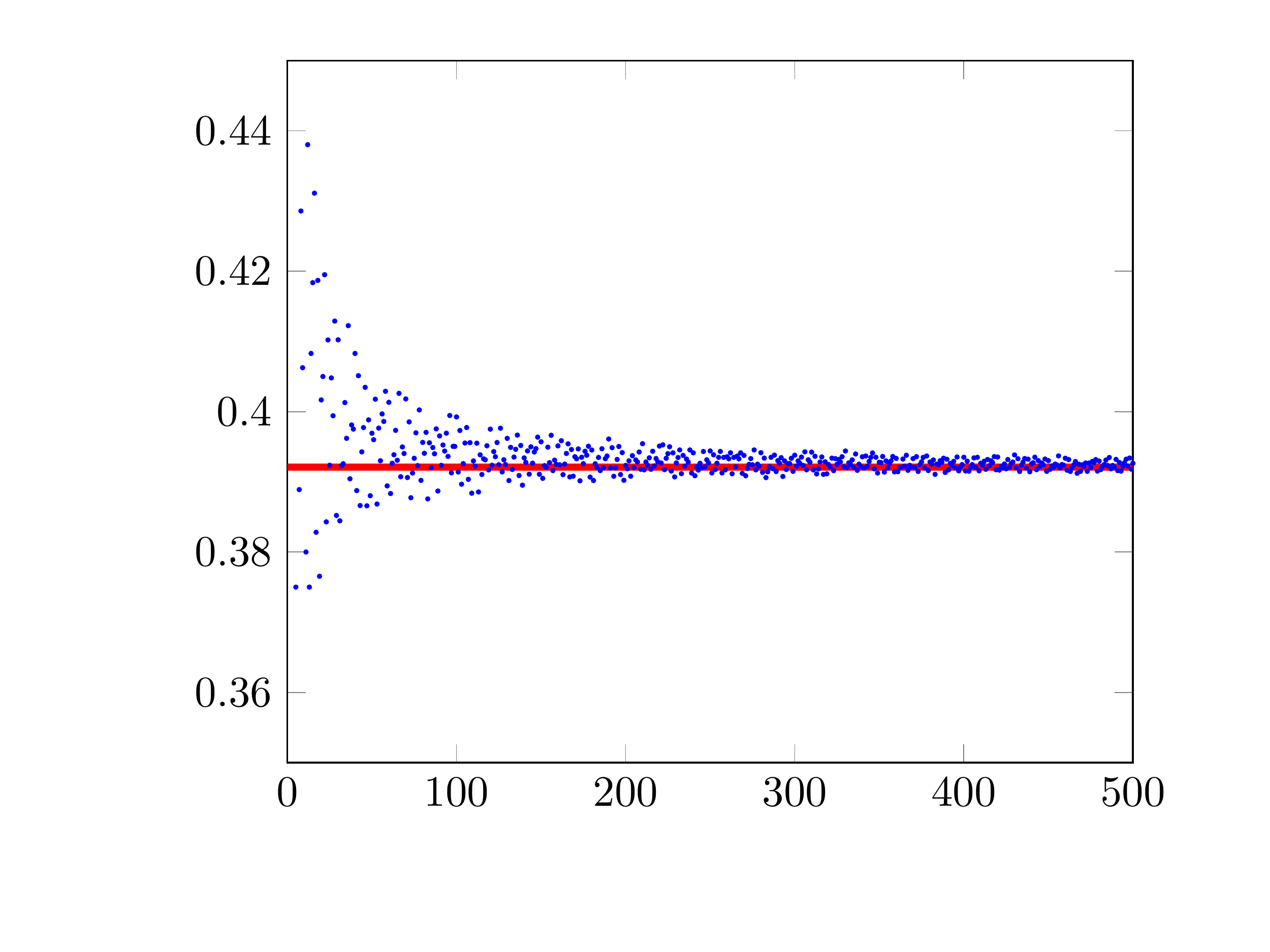, width=0.6\textwidth}
\caption{{\em Density $\rho_1$ of the $1$'s versus $q$ for the coprimality matrix $\bf\Phi$. The red line 
shows the theoretical value $\rho_1 \simeq 0.3920...$}}
\label{ratiozeros}
\end{figure}

Let us exploit once again the simple observation that the probability that a given integer $a$ is divisible by the prime $p$ is $1/p$. Therefore the joint probability that another number $b$ is also divisible by $p$ will be $1/p^2$, and the probability $P_{coprime}$ that both $a$ and $b$ are not divisible by the same set of primes\footnote{Assuming one again weak correlations among the primes.} $p$ is then 
\be 
\label{pcoprime}
P_{coprime} \simeq \prod_{p~\text{prime}} \left(1 - \frac{1}{p^2}\right) \,=\,\frac{1}{\zeta(2)} \,=\,\frac{6}{\pi^2} = 0.607927\cdots
\ee
Notice that eq.\,\eqref{pcoprime} involves the same value of the Riemann zeta function obtained earlier in \eqref{predictedslope}.
Given that $(q-1)^2$ is the total number of elements present in the matrix $\bf\Phi$, eq.~\eqref{pcoprime} leads to the following estimates of the densities $\rho_0$ and $\rho_1$ of $0$'s and $1$'s in the coprimality matrix 
\be
\label{density}
\rho_0 \,=\,\frac{N_0}{(q-1)^2} = P_{coprime} = 0.607927\cdots
\hspace{8mm}
, 
\hspace{8mm}
\rho_1 \,=\,\frac{N_1}{(q-1)^2} =1-P_{coprime} = 0.392073\cdots  \,\,\,, 
\ee
where $N_{0}$ and $N_1$ are the total numbers of $0$'s and $1$'s in $\bf\Phi$. These predictions can be easily tested by performing numerical experiments on the matrix $\bf\Phi$ by varying its dimensionality: some of the results that were obtained  with the aid of a computer are shown in the Table 1, while a more extensive analysis is reported in Fig.~\ref{ratiozeros}. As one can convince himself, the agreement between the probabilistic estimate based on the independence among the primes and the actual values of the densities  is reasonably good, of the order of few percent, particularly in light of the simple probabilistic argument used for this estimate.     

\begin{table}[b]\label{pifre}
\begin{center}
\begin{tabular}{|l|l|l|} \hline 
$q$ & Numbers of $1$'s \,\,\,& \,\,\,Estimate of $\rho_1$ \\
\hline \hline
$100 $ & \,\,\,\,\,\,\,\,\,\,\,$3913$ \,\,\,&  \,\,\,\,\,\,\,\,$ 0.399245$ \\
$ 150$ & \,\,\,\,\,\,\,\,\,\,\,$ 8785$ \,\,\,&  \,\,\,\,\,\,\,\,$ 0.395703$ \\
$ 200$ & \,\,\,\,\,\,\,\,\,\,\,$ 15537$ \,\,\,&  \,\,\,\,\,\,\,\,$0.392339 $ \\
$ 250$ & \,\,\,\,\,\,\,\,\,\,\,$24453 $ \,\,\,&  \,\,\,\,\,\,\,\,$0.394397$ \\
$ 300$ & \,\,\,\,\,\,\,\,\,\,\,$ 32205$ \,\,\,&  \,\,\,\,\,\,\,\,$0.393788 $ \\
$ 350$ & \,\,\,\,\,\,\,\,\,\,\,$ 47841$ \,\,\,&  \,\,\,\,\,\,\,\,$ 0.39279$ \\
$ 400$ & \,\,\,\,\,\,\,\,\,\,\,$ 62645$ \,\,\,&  \,\,\,\,\,\,\,\,$0.393496 $ \\
$ 450$ & \,\,\,\,\,\,\,\,\,\,\,$79233 $ \,\,\,&  \,\,\,\,\,\,\,\,$0.393019$ \\
$ 500$ & \,\,\,\,\,\,\,\,\,\,\,$ 97769$ \,\,\,&  \,\,\,\,\,\,\,\,$0.392645 $ \\
\hline 
\end{tabular} 
\end{center}
Table 1. Series of trials in order to test the goodness of the theoretical estimate of the total number of $1$'s present in the 
coprimality matrix $\bf\Phi$. 
\end{table}
\vspace{1mm}

\begin{figure}[t]
\psfig{figure=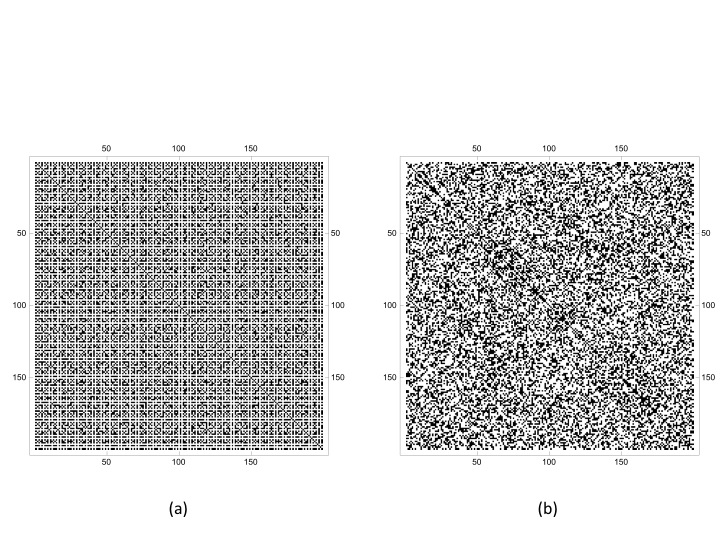,height=12.5cm,width=17cm}
\caption{{\em (a): Graphical representation of the coprimality matrix $\bf\Phi$ in comparison to (b) a similar representation of a random 
matrix with the same density of $0$'s.}}
\label{comparison}
\end{figure}

\vspace{3mm}
\noindent
{\bf Graphical Representation and Fourier Transform}. 
It is interesting to associate to the pair of natural numbers $(a,b)$ a point on the first quadrant of a  cartesian  plane. Notice that the two integers $a$ and $b$ are coprime if and only if the point with cartesian coordinates $(a, b)$ is \lq\lq visible" from the origin $(0,0)$, namely there is no point with integer coordinates lying on the segment that connects such a point to the origin. 
\begin{figure}[t]
\psfig{figure=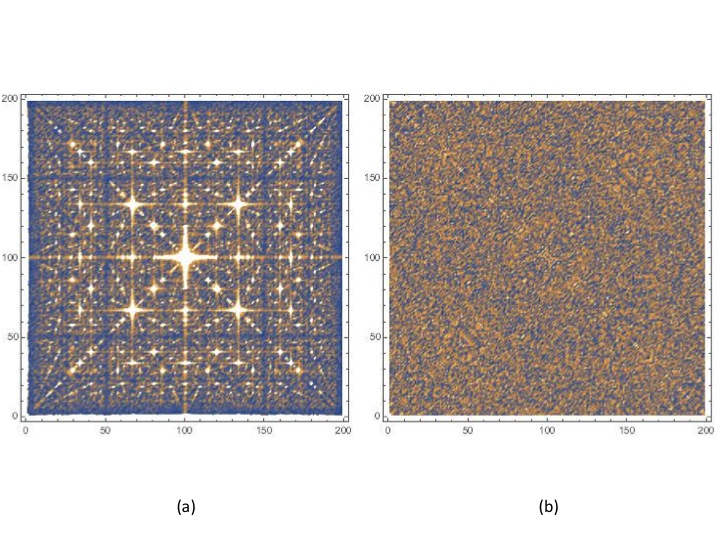,height=12.5cm,width=17cm}
\caption{{\em  Absolute value of the DFT of (a) the coprimality matrix $\bf\Phi$ compared with (b), the DFT of a random 
matrix with the same density of $0$'s. For both matrices, $q=200$.}}
\label{Fourier}
\end{figure}
This interpretation in the cartesian plane suggests a  graphical representation of the coprimality matrix, where all the entries equal to $1$ are coloured in black, while leaving white all the $0$'s. The result is shown in Fig.~\ref{comparison} compared with an analogous picture for a random matrix with entries $0$ and $1$ that has the same density $\rho_0$ of vanishing elements and all entries equal to one along the main diagonal. By looking at these two pictures, one can identify a certain degree of order in the coprimality matrix -- order that is on the contrary absent in the genuine random matrix with the same density of $0$'s. For spelling out in greater detail the texture of the coprimality matrix, let us first extend its linear dimension to arbitrarily large values of $q$: in this case it is easy to see that the matrix elements satisfy 
\be 
\label{prop_f}
\Phi(a, b) \,=\,\Phi(a^m, b^n) \,\,\,,
\ee
for any integer values $m$ and $n$. The property~\eqref{prop_f} appears as a sort of multiplicative periodicity of the coprimality matrix; however in this matrix there are more interesting additive periodicities, although approximate. Imagine to consider the matrix element $\Phi(a,b)$ where $a$ is one of the $l$ primes, say $p_f$, in
the interval $[2,q]$ while $b$ is coprime with $a$. If $b$ is itself another prime $p_k$, with $p_k \neq p_f$, it is obvious that we have the following additive periodicity properties 
\be 
\Phi(p_f, p_k) \,=\, \Phi(p_f + n p_f, p_k + m p_k)
\hspace{5mm}
,
\hspace{5mm}
n, m \in {\mathbb N} 
\,\,\,, 
\label{firstprimeperiodicity}
\ee
as far as $(n+1) \neq p_k$ and $(m+1) \neq p_f$. Consider now the case when $a$ is once again one of the $l$ primes, $p_f$, while $b$ is a generic composite number,~although coprime with $p_f$. In this case we have the property 
\be
\Phi(p_f,b)\,=\, \Phi(p_f + n p_f, b) 
\hspace{5mm}
,
\hspace{5mm} 
n \in {\mathbb N}
\label{secondperiodicity}
\ee
as far as $(n+1) \neq p_s$, where $p_s$ is one of the prime present in the decomposition of the number $b$. These two approximate periodicity conditions seem to be responsible for the pronounced peaks along the diagonals of the absolute value of the Discrete Fourier Transform (DFT)\footnote{The first paper where the DFT of the coprimality matrix was studied  is \cite{Schroeder2}.} of the coprimality matrix shown in Fig.~\ref{Fourier}. Notice that, by construction, the DFT $\widetilde \Phi(u,v)$, defined by 
\be 
\widetilde\Phi(u,v) \,=\,\sum_{a=2}^{q-1} \sum_{b=2}^{q-1} e^{2\pi i (u a + v b)/(q-1)} \,\,\,\Phi(a,b) 
\ee
shares the symmetries 
\be
\widetilde\Phi(u,v) \,=\,\widetilde\Phi(v,u) 
\hspace{5mm}
,
\hspace{5mm}
\overline{\widetilde\Phi(u,v)} \,=\,  \widetilde\Phi(-u, -v) \,\,\,.
\ee
Therefore, the absolute value of $\widetilde\Phi(u,v)$ is symmetric about the line $u = (q-1) - v$ as well. This means that the fundamental domain
of this function coincides with one of the four triangles identified by the two main diagonal, say the lowest one, the rest of the figure being simply a kaleidoscope effect. Understanding in detail the various peaks of the module of $\widetilde\Phi(u,v)$ is a task that goes beyond the present work. Here we would
like simply to underline that the series of the peaks (of decreasing amplitude) along the diagonal are placed at the frequency positions
$\left(\frac{(q-1)}{p_i},\frac{(q-1)}{p_i}\right)$ where $p_i$ are the consecutive prime numbers $p_i = 2, 3, 5, \ldots $.  

In Fig.~\ref{Fourier} we show, for comparison, the absolute value of the DFT of a random matrix that shares with the coprimality matrix the same density of $0$'s:
in this case, there is no sign of any particular frequency, i.e. the Fourier transform shows just white noise. 

\begin{figure}[t]
\center
\makebox[0pt][c]{
\hspace{-10mm}
\begin{minipage}{0.52\textwidth}
\includegraphics[scale=0.50]{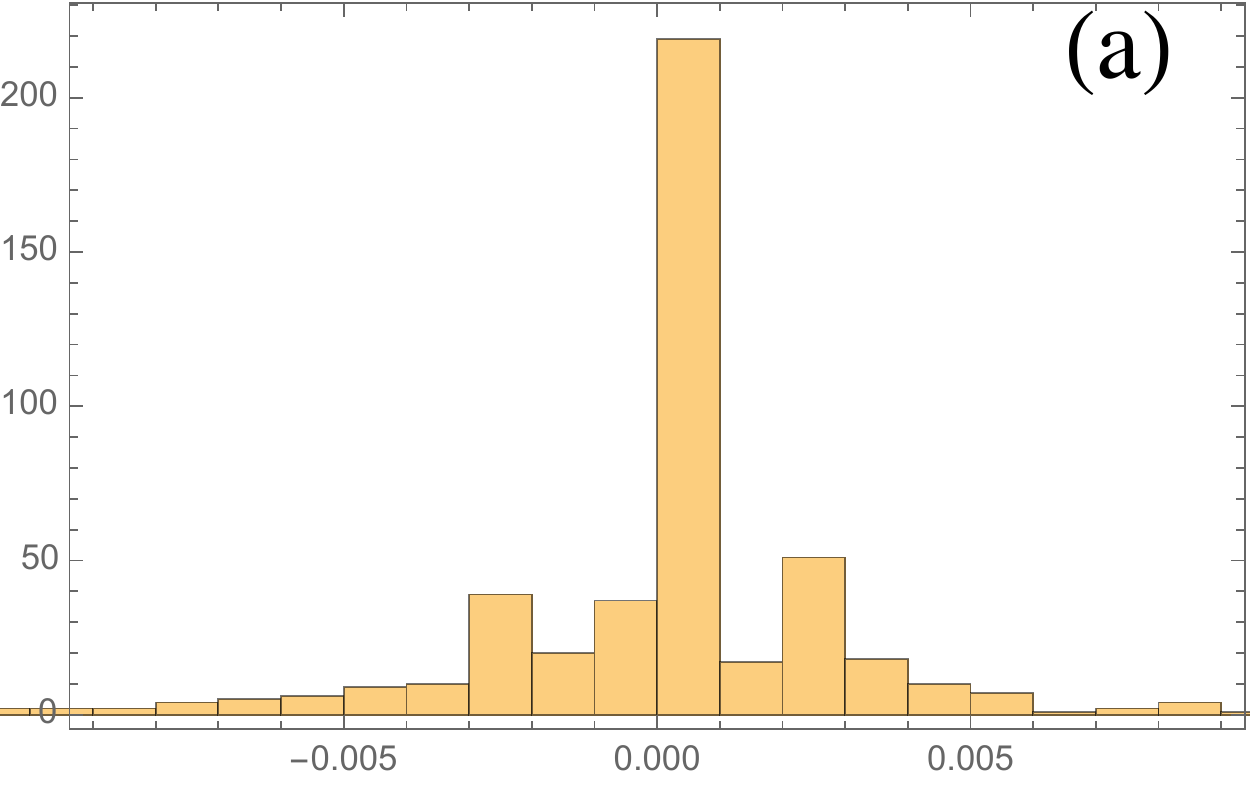}
\end{minipage}
\begin{minipage}{0.53\textwidth}
\includegraphics[scale=0.50]{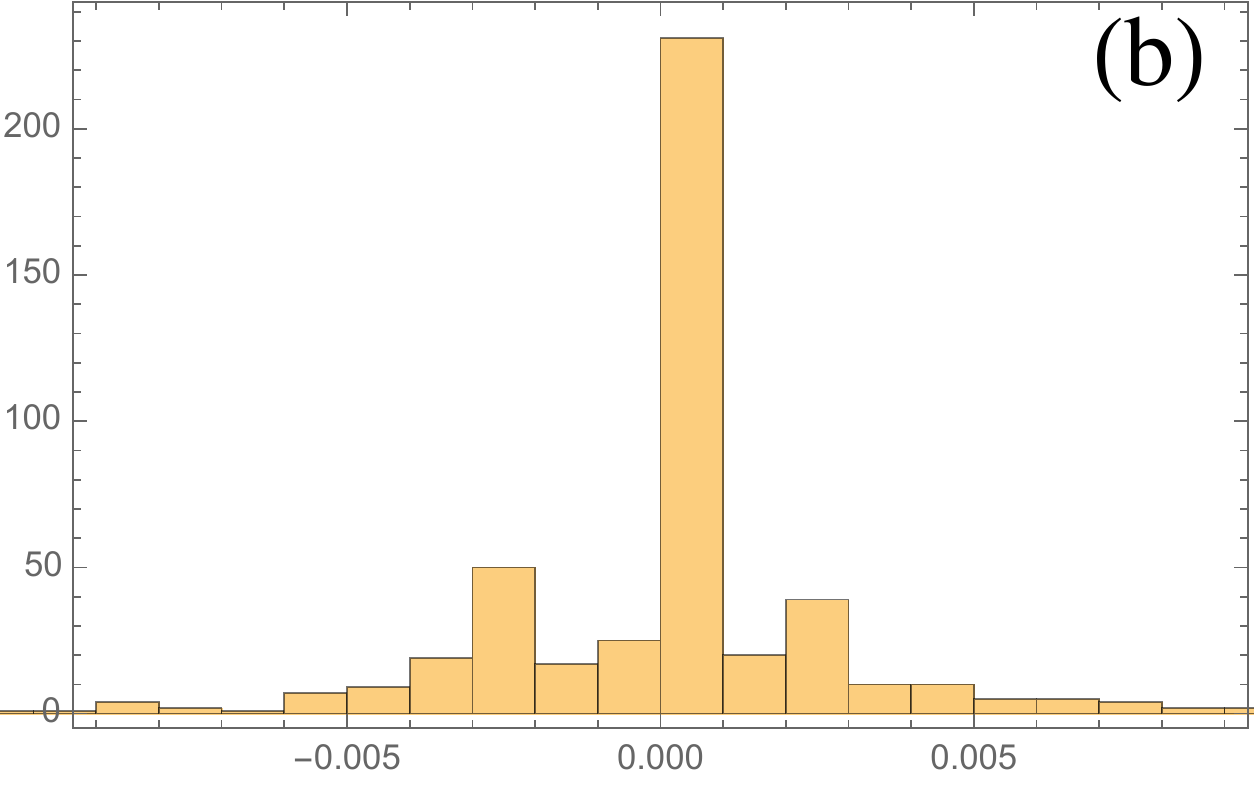}
\end{minipage}}
\caption{{\em Histogram of the normalised eigenvalues of the coprimality matrix of the ferromagnetic case (a) and anti-ferromagnetic case (b) for $q=500$.}}
\label{histogram}
\end{figure}

\vspace{3mm}
\noindent
{\bf Eigenvalues of the coprimality matrix}. There is a very interesting arithmetic pattern which emerges in the limit $q \rightarrow \infty$ for the coprimality matrix, its eigenvalues and eigenvectors, as discussed in great detail by Don Zagier in the appendix C of this paper. From the results of the Appendix C one can see that 
the lower and highest eigenvalues of the coprimality matrix, both in the ferromagnetic and anti-ferromagnetic case, scale with $q$. This permits to divide all 
eigenvalues by $q$: these new set of values (here called the {\em normalised eigenvalues}) live then on compact intervals which are  
\be
\begin{array}{lll}
{\cal I}_f &=&  (-0.00735 \,\,, \,\,0.5464) \\
{\cal I}_{af} &=& (-0.25937 \,\,, \,\,0.6787) 
\end{array}
\ee
for the ferromagnetic and anti-ferromagnetic cases respectively. In both cases, the spectrum is highly degenerate, with many zero eigenvalues. 
The histograms of the normalised eigenvalues of both cases (for $q=500$) are shown in Figure \ref{histogram}. Later we will use this information 
on the spectrum of the coprimality matrix to get various properties of the coprime quantum chain.  
 
\section{Classical Ground States of the Ferromagnetic Case}\label{classicalgroundstates}
Setting to zero all the couplings $\beta_{\alpha}$ relative to the operators $G_i^{(\alpha)}$ in the quantum Hamiltonian (\ref{quantumhamiltonian}),  
we essentially convert the original quantum chain to a one-dimensional classical model, with  Hamiltonian given by  
\be 
H_{cl} \,=\, -\sum_{i=1}^M \Phi(n_i,n_{i+1}) \,\,\,. 
\label{classical}
\ee
Studying the classical Hamiltonian~\eqref{classical}, we can identify the underlying structure of the vacuum states of the coprime quantum chain and, as we will see, this will turn out an interesting problem in itself. The ground states of the classical model are of course modified when we switch on the coupling constants $\beta_\alpha$ in~\eqref{quantumhamiltonian} although the conclusions contained in the next three sections can serve as a good starting point for characterising the  actual vacua of the quantum Hamiltonian~\eqref{quantumhamiltonian} as functions of the parameters $\beta_{\alpha}$. 

Notice that the coprime classical chain appears to be a generalisation of the $(q-1)$-state Potts model, with classical Hamiltonian 
\be
H_{Potts} \,=\, - \sum_{i=1}^M \delta(n_i,n_{i+1}) \,\,\,, 
\label{Potts}
\ee
with an important difference, though: while in the Potts model only equal occupation numbers at neighbouring sites have minimum energy, in the coprime chain instead minimal energy is assigned to all states with next neighbouring occupation numbers that share a common divisor. Even though this may appear only
a slight modification of the Potts model, yet it has profound consequences on the the vacuum structure, as discussed below. 

\vspace{3mm}
\noindent
{\bf A first look at the exponential degeneracy of the classical ground state energy}. The minimum of the classical Hamiltonian (\ref{classical}) is 
obtained by satisfying, for each pair of next-neighbouring sites, the condition 
\be 
\label{cond_loc}
\Phi(n_i,n_{i+1}) =1 \,\,\,.
\ee
The requirement \eqref{cond_loc} forces two next-neighbouring occupation numbers to have at least one common divisor.
Apart from the simplest coprime chains corresponding to $q=2$ and $q=3$, for $q > 3$ there
are several ways to satisfy \eqref{cond_loc} and this in general leads to an exponential proliferation of the ground states.
Consider, for instance, the case $q=5$: for this value of $q$, the local constraint $\Phi(n_i,n_{i+1}) =1$ is verified by the following pairs 
\be
(2,2)
\,\,\,\,\,
,
\,\,\,\,\,
(2,4) 
\,\,\,\,\,
,
\,\,\,\,\,
(4,2) 
\,\,\,\,\,
,
\,\,\,\,\,
(4,4) 
\,\,\,\,\,
,
\,\,\,\,\,
(3,3)
\,\,\,\,\,
,
\,\,\,\,\,
(5,5) \,\,\,. 
\ee
Once we have fixed the occupation number at the first site to be for instance $3$, i.e. $n_1 = 3$, to realise a ground
state compatible with this condition, the remaining occupation numbers on all the other sites must be $3$ as well.
Hence there is a unique possibility to construct a ground state 
with an occupation number equal to 3 and it is 
\be
3 \, 3\, 3\, 3\, \ldots 3\,\,3 \, 3\,\,\,\,.
\ee  
The same happens if we start with $5$ at the first site of the chain: in this case, we end up with a unique ground state given by a sequence of occupation numbers 
all equal to $5$ 
\be
5 \, 5\, 5\, 5\, \ldots 5\,\,5 \, 5\,\,\,\,.
\ee
However the situation changes for the other two values $\{2, 4\}$ of the occupation numbers: indeed, since they belong to the same equivalence class, they can be traded
one for the other on each site without altering the energy of the state. This hints at an exponential number of ground states which can be built by means of
arbitrary sequences of $2$'s and $4$'s, such as 
\be 
2 \, 2\, 4\, 2\, \ldots 4\,\,4 \, 2\,\,\,\,.
\ee
The multiplicity of the  ground states that contain only these two occupation numbers is easily computable: at each site we can have two possible choices (either $2$ or $4$) and therefore on a chain of $M$ sites their number is  $2^M$. Together with the other two ground states consisting of $3$'s and $5$'s, the total number ${\cal N}^f_M(q)$ of classical ground states of a coprime chain with $q=5$ and $M$ sites is then 
\be 
{\cal N}^f_M(q=5) \,=\, 2^M + 2 \,\,\,. 
\label{q=5}
\ee 
The reason of the superscript $f$ in \eqref{q=5} is that this calculation was tacitly performed assuming free boundary conditions at the ends of the chain.
Repeating the same analysis for a  $q=4$ coprime chain, one quickly realises that the number of ground states of this model grows as 
\be
{\cal N}^f_M (q=4) \,=\, 2^M +1 \,\,\,, 
\ee
simply because now the ground state made of $5$'s solely will be missing. For $q=2$ and $q=3$ we have of course only 
$2$ possible ground states for any number $M$ of the sites. 

The analysis of these two coprime chains, $q=4$ and $q=5$, turned out to be quite simple. However this simplicity is misleading, the calculation of the  ground state degeneracy for higher values of $q$ requires actually a more sophisticate set of mathematical techniques, especially those borrowed from graph theory.  

\begin{figure}[t]
\vspace{8mm}
\psfig{figure=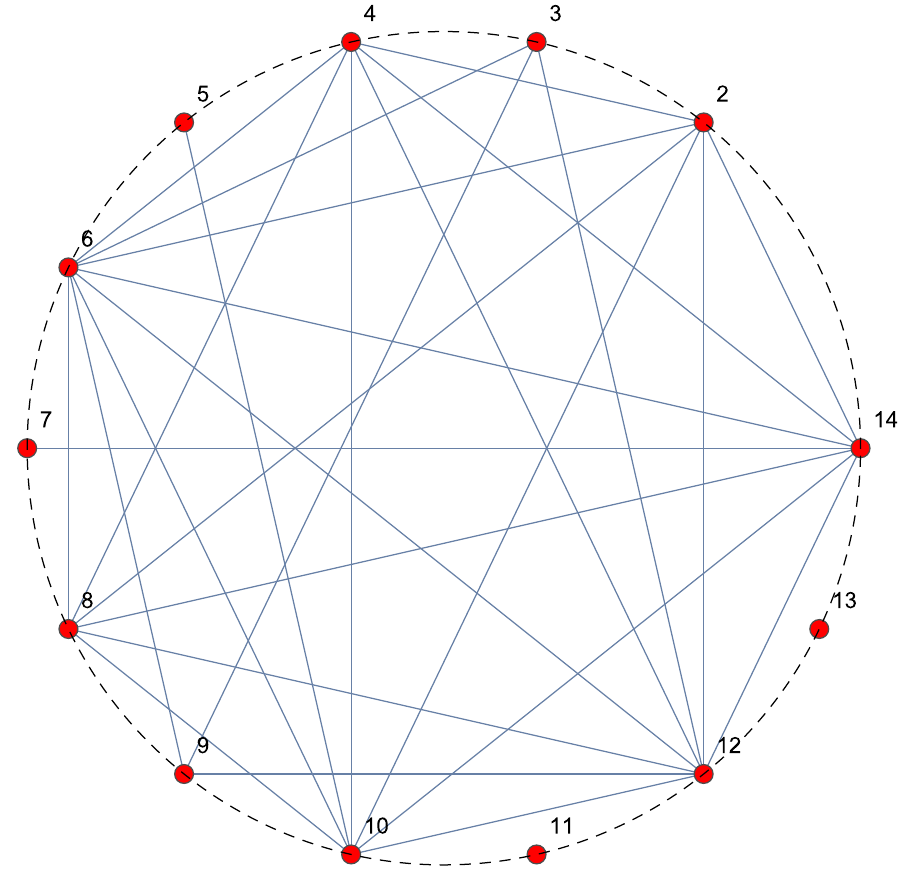,height=8cm,width=8cm}
\caption{{\em Incidence graph of the ferromagnetic coprime chain with $q=14$. Vertices are numbered from $2$ to $14$ and are connected if they satisfy the local constraint \eqref{cond_loc}.}}
\label{incidence}
\end{figure}

\vspace{3mm}
\noindent
{\bf Adjacency  matrix and graph theory}. In order to proceed further with the analysis, it is first convenient to extract the diagonal entries from the 
coprimality matrix and write it as 
\be 
\bf \Phi \,=\, {\bf 1} + {\bf A} \,\,\,.
\label{definitionadjacency}
\ee
The $(q-1) \times (q-1)$ symmetric matrix ${\bf A}$, whose only elements are $0$'s and $1$'s, is called the {\em adjacency matrix}
of the coprime chain. It is easy to realise  that the matrix ${\bf A}$ encodes the information about which pair of occupation numbers satisfy the constraint \eqref{cond_loc} and can therefore be neighbour in a ground state configuration.  We can then associate 
to each possible value of the $n_i=2,\dots,q$ the vertex of a graph, the so-called incidence graph,\footnote{ All the possible vertices can be conveniently represented as lying on circle and will be ordered as in Fig.~\ref{incidence}.} and connect by a line those vertices whose matrix element of ${\bf A}$ is equal to one. An example of this graphical construction with  $q = 14$ is shown in Fig.~\ref{incidence}; notice that the labels of the vertices are actually the occupation numbers. As we will see, we can use the incidence graph to infer some important features common to all the coprime chains for various values of $q$, features which will help us to carry on the general analysis of these models. For convenience, basic elements of graph theory  that will be useful in such a study are collected in Appendix  \ref{Appgraphtheory}. 
 
\vspace{3mm}
\noindent
{\bf Local, maximum and average degree}. Each vertex $a$ of a graph, see Fig.~\ref{incidence}, has its own {\em local degree} $d_a$ which
is the total number of lines coming out from it: in turn, the local degree is simply the sum of all elements of the adjacency matrix ${\bf A}$ along its $a$-th row ($2\leq a\leq q$)
\be
d_a \,=\,\sum_{b=2}^{q} {\bf A}_{a,b} \,\,\,. 
\label{localdegree}
\ee
Therefore in the example of Fig.~\ref{incidence}, the vertex $a=2$ has degree $d_2 = 6$, the vertex $3$ has degree $d_3 = 3$, etc. 

For any graph, we can also define two other useful quantities, the {\em maximum degree} $d_m$ -- which corresponds to the maximum among all the local degrees -- and the {\em average degree} ${\overline d}$, defined as the average of the local degrees  
\be 
d_m \,=\, {\rm Max} \, d_a \, 
\hspace{5mm}
\,, 
\hspace{5mm}
{\overline d}\,=\,\frac{1}{(q-1)} \, \sum_{a=2}^{q} d_a 
\,\,\,.
\label{degrees}
\ee
Therefore referring once again to the example of the graph in Fig.~\ref{incidence}, we have $d_m = 8$,
which corresponds to $a=6$, while $\overline d = 4.30769...$. 

\begin{figure}[t]
\vspace{8mm}
\psfig{figure=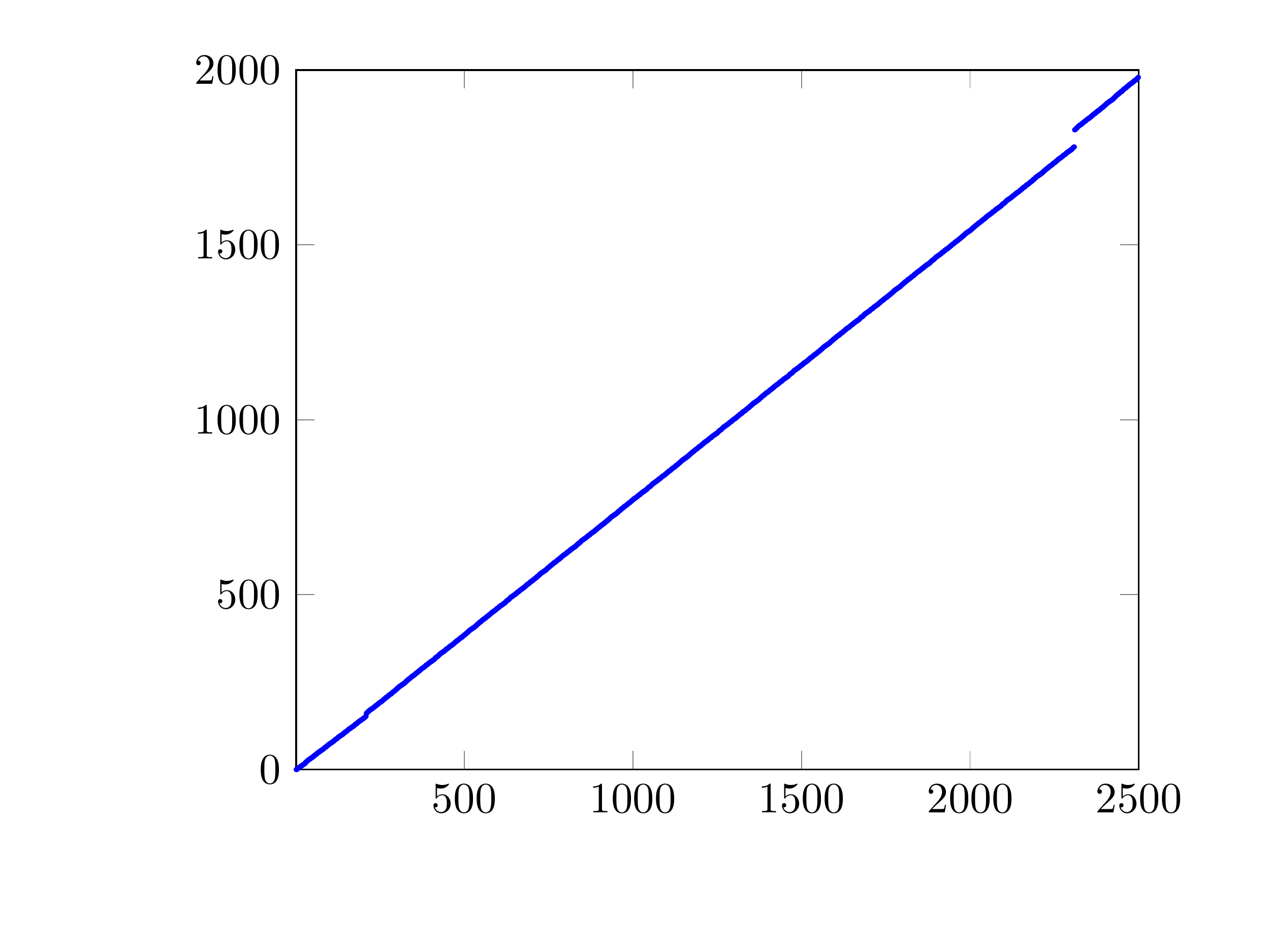,height=7.5cm,width=9.5cm}
\caption{{\em Maximum degree of the coprime chain versus $q$, up to $q=2500$. We can see a jump in the maximum degree at
 $q\simeq2300$, it corresponds to a value where the highly composite occupation number $2310=2\times3\times 5\times 7 \times 11$ becomes allowed.}}
\label{maxdegree}
\end{figure}

Recalling the approximate calculation of the density $\rho_1$ (see Sec.~\ref{sec:def}, eq.~(\ref{density}) in particular),
it is easy to argue that the average degree for the $q$-coprime chain shall scale with $q$ as 
\be
{\overline d}  \simeq \rho_1 (q-1) -1\simeq \overbrace{\left(1-\frac{6}{\pi^2}\right)}^{0.392073\dots}q-\overbrace{\left(2-\frac{6}{\pi^2}\right)}^{1.392073\dots} 
\,\,\,.
\label{average}
\ee
Concerning the maximum degree of the $q$-coprime chain, its explicit computation for several values of $q$ reveals that it also grows linearly 
with $q$: up to $q=2500$, the best fit of the slope extracted from  Fig.~\ref{maxdegree} is
\be
d_m\,\simeq \,  0.772312 \, q  \,\,\,.
\label{degreemax}
\ee 
However it is better to state straight away that the value $0.772312..$ given in \eqref{degreemax} is not the 
correct value of the slope since this quantity is strongly affected by finite size effects in the size of the adjacency matrix.
In particular, with a little bit of effort one can check that such a value tends to increase considering larger intervals $[2,q]$ and indeed,
as shown in Appendix \ref{Appmaxdegree}, for $q \rightarrow \infty$,  the slope is predicted to be {\em exactly} equal to $1$; namely for large enough $q$ we should expect 
\begin{equation}
d_m \simeq q \,\,\,.
\label{degreemaximumvero}
\end{equation}

\vspace{3mm}
\noindent
{\bf Eigenvalues and characteristic polynomials}. An important tool to evaluate the number of the classical ground states of the coprime chain is provided by the spectrum of the coprimality matrix $\bf \Phi$. Notice that, from the relation (\ref{definitionadjacency}), the eigenvalues $\lambda_i$ of $\bf\Phi$ differ from those $\eta_i$ of the adjacency matrix ${\bf A}$ simply by $1$ 
\begin{equation}
\lambda_i = \eta_i +1 \,\,\,.
\label{etalambda}
\end{equation}
In other words, the characteristic polynomials ${\cal C}_q(x)$ of the coprimality matrix $\bf\Phi$ are obtained from the characteristic polynomials ${\cal P}_q(x)$ of the 
adjacency matrix substituting $x \rightarrow x-1$, 
\begin{equation}
{\cal C}_q(x) \,=\, {\cal P}_q(x-1) \,\,\,.
\label{relationcharpolynomials}
\end{equation}
For any given incidence graph of the $q$-coprime chain, the characteristic polynomials of its adjacency matrix are special polynomials ${\cal P}_q(x)$ 
with integer coefficients whose first representatives are given by  
\be
\begin{array}{lllll}
q = 4 & & \,\,\,\longrightarrow \,\,\,& & {\cal P}_4(x) = x (x^2 -1)\\ 
q = 5 & & \,\,\,\longrightarrow \,\,\,& & {\cal P}_5(x) = x^2( x^2-1) \\
q =6 & & \,\,\,\longrightarrow \,\,\,& & {\cal P}_6(x) = x (x^4 - 4 x^2 - 2 x +1) 
\\
q =7 & & \,\,\,\longrightarrow \,\,\,& & {\cal P}_7(x) = x^2 ( x^4 - 4 x^2 - 2 x +1) \\
q = 8 & & \,\,\,\longrightarrow \,\,\, & & {\cal P}_8(x) = x^2 (x^5 - 7 x^3 - 8 x^2 +2) \\
q = 9 & & \,\,\,\longrightarrow \,\,\, & & {\cal P}_9(x)  = x^2 (x^6 - 9 x^4 - 10 x^3 + 9 x^2 + 18 x + 7) \\
q =10 & &\,\,\,\longrightarrow \,\,\,& &{\cal P}_{10}(x) = x (x^8 - 14 x^6 - 22 x^5 + 16 x^4 + 54 x^3 + 28 x^2 - 8 x - 7). 
\end{array}
\label{polynchar}
\ee
Notice that, from a purely algebraic point of view,  the eigenvalues of the adjacency and coprimality matrices have the amazing property to give rise to {\em integer} numbers ${\cal N}^{cyc}_M$ whenever we take the sum of {\em any} integer power $M$ of them as, for instance
\begin{equation}
{\cal N}^{cyc}_M = \sum_{i=1}^{q-1} \lambda_i^M \,\,\,.
\end{equation}
As shown below -- see the relation (\ref{totalnumbergroundstatetrace}) -- the integer nature of ${\cal N}^{cyc}_M$ simply comes from the observation that the total number of ground states of the coprime chain with periodic boundary conditions has to be a natural number for any length $M$ of the chain. However, this is a {\em physical} explanation: staring at this result from the bare point of view of the roots of a polynomial, it seems instead a pretty remarkable mathematical fact since such a property could be immediately spoiled, for instance, by just changing one coefficient of the polynomials listed above. 

Let's now focus the attention on the eigenvalues of the adjacency matrix ${\bf A}$ for the simple reason that the spectral theory of this kind of matrices is a quite well developed mathematical subject. In particular, there are interesting bounds on the largest eigenvalue $\eta_{max}$ given in terms of the maximum degree $d_m$ and the average degree ${\overline d}$ of the graph associated to the adjacency matrix \cite{Cvit}  
\be 
{\overline d} < \eta_{max} \leq d_{m} \,\,\,. 
\ee 
Since both ${\overline d}$ and $d_m$ scale with $q$, we see that also the maximum eigenvalue $\eta_{max}$ of our coprime chain must scale with $q$. Hence, 
for large $q$, we have $\lambda_{max} = \eta_{max} +1 \simeq \eta_{max}$ and therefore  
\be
\lambda_{max} \simeq \lambda_0 \, q  \,\,\,,
\ee
where, using both eqs. (\ref{average}) and (\ref{degreemaximumvero}), we arrive to the inequalities 
\be 
\left(1-\frac{6}{\pi^2}\right) < \lambda_0 < 1 \,\,\,.
\ee 
A direct numerical evaluation of the maximum eigenvalue gives, as the best values of the fit, the linear behaviour 
\be 
\lambda_{max} \,\simeq  \,  0.54636 \,q \,\,\,. 
\label{scalingmaxeig}
\ee
As one can learn reading the Appendix C, the exact value of the slope is actually $0.54637892502940\cdots$. 

\vspace{3mm}
\noindent
{\bf Inert vertices}. By looking at Fig.~\ref{incidence}, we see that the vertices associated to the occupation numbers $n_i = 11$ and $n_i=13$ are not connected to any other point: for any graph, vertices of this kind will be called {\em inert}. It is easy to identify them for a $q$-coprime chain. The  inert vertices are labelled by to those primes $p_i$ which satisfy the condition
\be 
\label{inert_con}
\frac{q}{2} < p_i \leq q \,\,\,,
\ee
since in the interval $[2,q]$ there are no integers that can share a common divisor with them. 
Indeed, the smallest composite number which contains them as factors is $\tilde n_i = 2 \times p_i$, but $\tilde n_i>q$ because of~\eqref{inert_con}. A rough estimation of the number of inert vertices present
in a $q$-coprime chain can be given in terms of the prime counting function $\pi(x)$: 
\be 
N_{inert}(q) \,=\, \pi(q) - \pi\left(\frac{q}{2}\right) \,\simeq \, \frac{q \,\log\frac{q}{4}}{2\,\log q\,\log\frac{q}{2}} 
\,\,\,.  
\label{numberinert}
\ee  
This formula predicts that the total number of inert vertices is larger than $2$ for $q > 17$ but one can directly check that this is already true for $q > 6$. While this 
result will be important later, for the time being notice that inert vertices  give rise to  vacuum configurations that are simply obtained repeating them. Using once again $q=14$ as an example, the two ground states produced by the sequences of inert vertices $11$ and $13$  are
\begin{eqnarray}
&& 11 \,\, 11\,\, 11\,\, 11\, \ldots\,  11\,\,11 \,\, 11\,\,\,\,  \\
&& 13 \,\, 13\,\, 13\,\, 13\, \ldots \, 13\,\,13 \,\, 13\,\,\,\,
\end{eqnarray}
For an algebraic characterisation of the inert vertices, notice that their values label the rows of the adjacency matrix 
${\bf A}$ with all entries equal zero, since they are disconnected from all the other vertices.  

\vspace{3mm}
\noindent
{\bf Vertices with the highest local degree}. 
In a generic $q$-coprime chain it is also easy to spot which vertex has the highest degree: it will be labelled by the number $h$ obtained as a product of the first consecutive $s$ primes 
\be 
h \,=\, 2 \times 3 \times \cdots \times p_s \leq q \,\,\,.
\label{highest}
\ee
\begin{figure}[t]
\psfig{figure=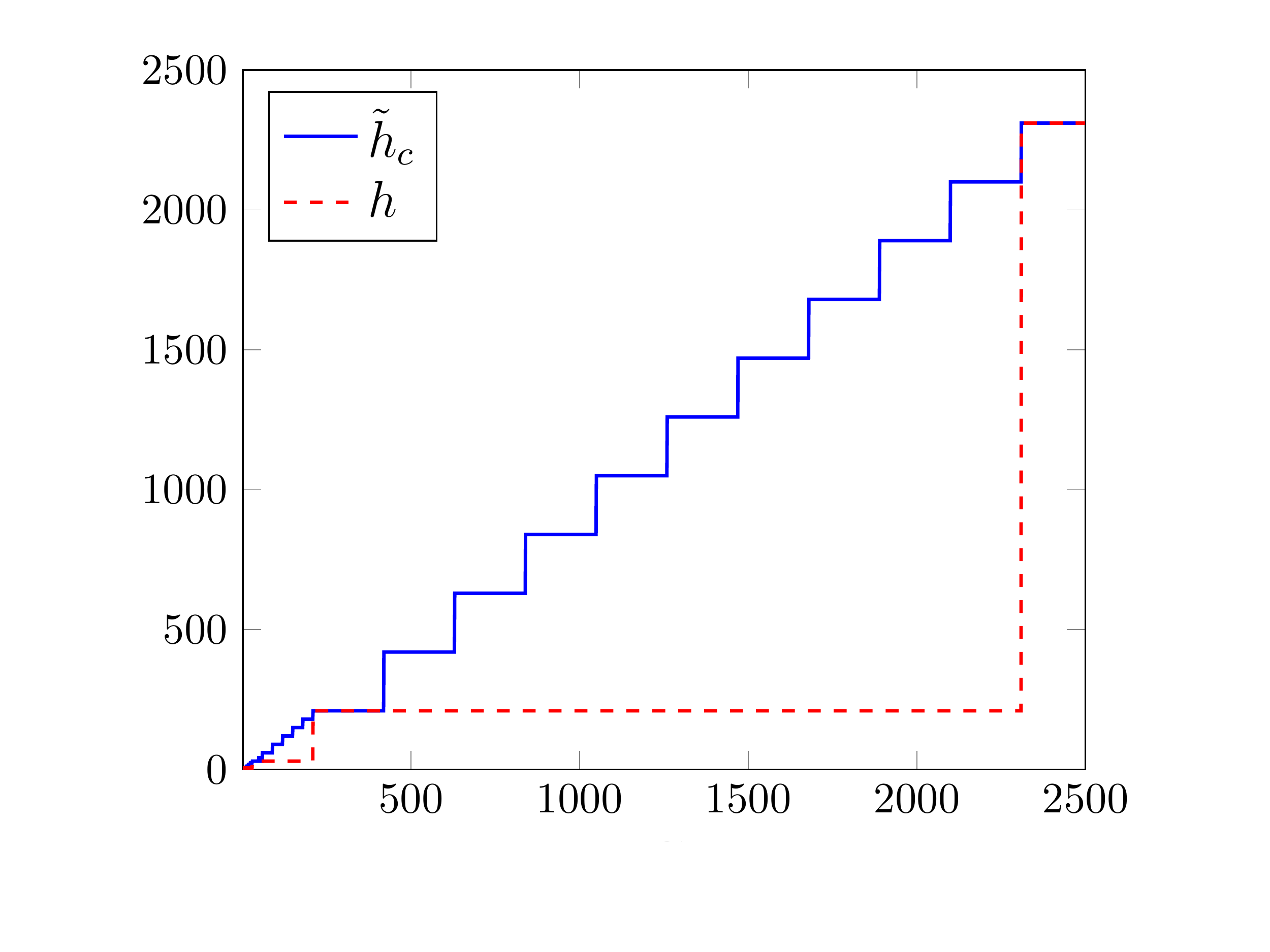,height=6.5cm,width=9cm}
\caption{{\em Minimum value $h$ (red dashed line) and maximum value $\tilde{h}_c$ (blue solid line) of the occupation number labelling a vertex with the highest  degree $d_m$, as discussed in eq.\eqref{minhigh}, versus $q$. }}
\label{MinMax}
\end{figure}

\noindent
The number $h$, indeed, has common divisors with all multiples of $2$, all multiples of $3$ etc., and therefore the vertex associated to it maximises the number of links with all the remaining vertices of the incidence graph. Equation \eqref{highest} in particular implies that there will be jumps in the value of $h$ each
time $q$ could be written as a product of consecutive primes, namely 
\be
\begin{array}{lll}
h = 2 \, &, &   \,\, 2 \leq q < 6 \\
h = 6 \, &, &   \,\, 6 \leq q < 30 \\
h =30 \, &, & \,\, 30 \leq q < 210 \\
h = 210 \, &, & \,\, 210 \leq q < 2310 \\
\,\,\,\cdots &  & \,\,\, \,\,\cdots 
\end{array}
\label{minhigh}
\ee
The analysis done so far, however, does not exclude that there may be other vertices with highest degree as well. Indeed, those are the vertices labelled by the values $\tilde h_c$ that have the same prime-number vector as the occupation numbers  $h$ in~(\ref{minhigh}).  It might also happen that many of such numbers will be present for a given $q$. Summarizing, the values of $h$ reported in eq. (\ref{minhigh}) correspond to the {\em minimum} label of the vertex with the highest possible degree, while at fixed $q$ we could have many other occupation numbers $\tilde h_c > h$, labelling vertices that also have degree $d_m$. In Fig.~\ref{MinMax} there are shown the minimum (red dashed curve) and the maximum (blue solid curve) values of the occupation numbers with maximum degree as functions of $q$. As argued above, Fig.~\ref{MinMax} confirms that in general more vertices share the same highest degree.   

\begin{figure}[ht]
\center
\includegraphics[scale=0.4]{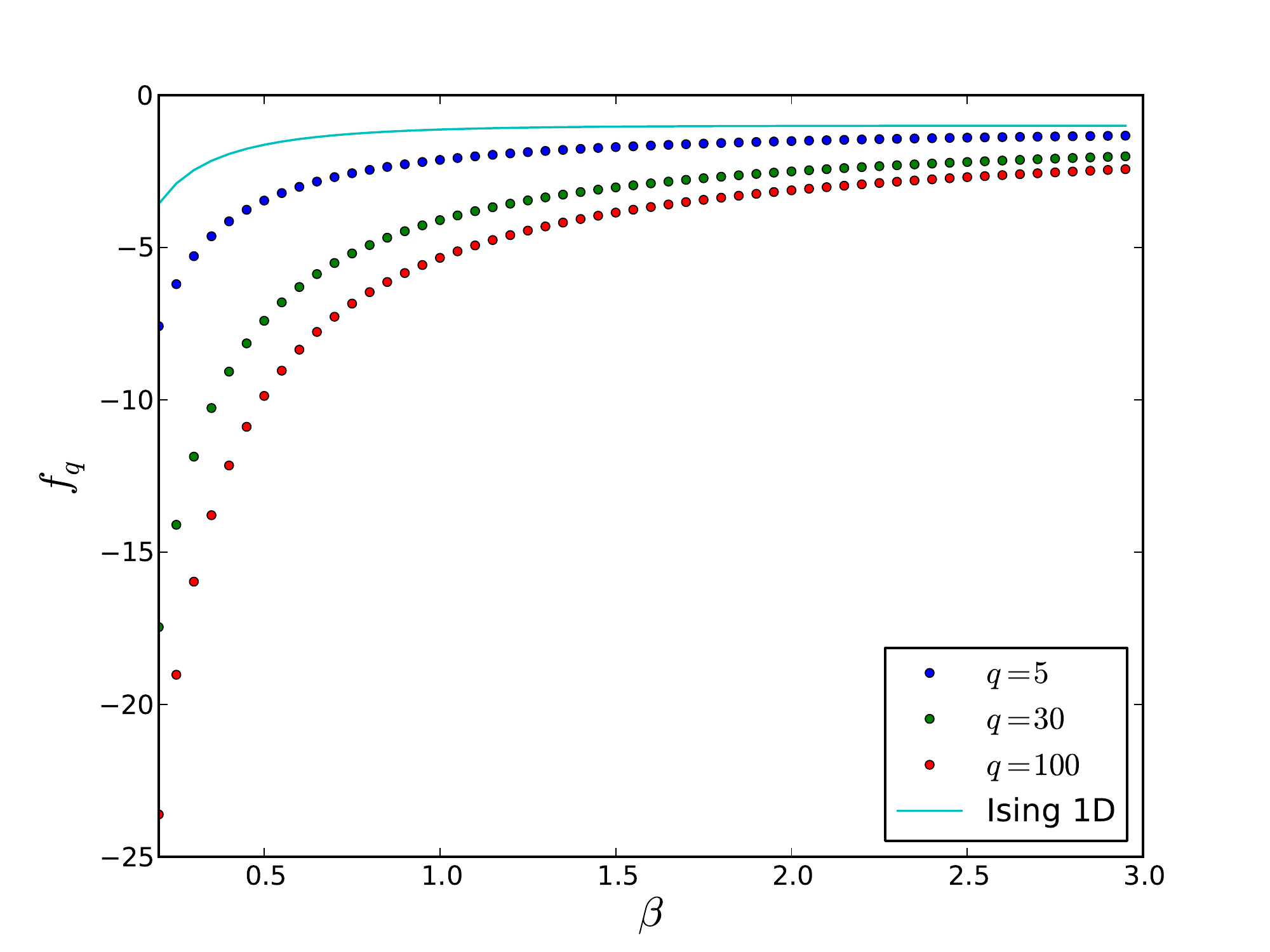}
\caption{{\em Free energy of the 1d classical coprime chain  versus the inverse temperature $\beta$. The points are the numerical data obtained from exact diagonalization
of \eqref{transfermatrix}. The solid curve represents the free energy of the 1d Ising model at zero external field.}}
\label{f1d}
\end{figure}

\vspace{3mm}
\noindent
{\bf Classical Free Energy}. 
The transfer matrix of the classical one-dimensional ferromagnetic coprime chain is given by 
\be \label{transfermatrix}
T_{a b} = \exp \( \beta \, \Phi(a,b) \)
\qquad , \qquad 
a,b = 2,3,\dots q \,\,\,. 
\ee
Hence, the partition function (with periodic boundary conditions) is expressed as 
\be
Z_M(\beta) \,=\, \mathrm{Tr}\, [T^M] \,=\, t_2^M(\beta) + t_3^M(\beta) + \cdots + t_{q}^M(\beta) \, \,\,,
\label{classss}
\ee
where $t_2 > t_3 \ldots > t_q$ are the $(q-1)$ eigenvalues of the matrix $T(a,b)$. Hence,  the free energy per unit site of the one-dimensional 
classical model reads 
\be 
\beta f_q = -  \log t_2(\beta) \,\,\,. 
\ee
As shown in Fig.~\ref{f1d} and as expected, the one-dimensional free energy exhibits no sign of non-analyticity, i.e. there is no phase transition for finite values of $\beta$.
Notice that taking the limit $\beta \rightarrow - \infty$, the only matrix elements of the matrix $T_{ab}$ which are different from zero (and equal to 1) are those relative to the numbers which are coprime. Hence, in this limit the transfer matrix $T_{ab}$ coincides with the coprimality matrix $\bf{\overline \Phi}$ of the antiferromagnetic case defined in eq.\,(\ref{antifercoprim}) and correspondingly, for $\beta \rightarrow -\infty$, the eigenvalues $t_i(\beta)$ go to the eigenvalues of the antiferromagnetic coprimality matrix $\bf{\overline \Phi}$. As discussed in the next Section, this means that in the limit $\beta \rightarrow - \infty$ the partition 
function (\ref{classss}) provides the number of ground states of the classical antiferromagnetic coprime chain of $M$ site with periodic boundary conditions. 

Vice-versa, if we start with the transfer matrix of the one-dimensional classical antiferromagnetic coprime chain 
\be \label{transfermatrixaf}
\overline T_{a b} = \exp \( \beta \, (1-\Phi(a,b)) \)
\qquad , \qquad
a,b = 2,3,\dots q \,\,\,,
\ee 
it is easy to see that in the limit $\beta \rightarrow -\infty$ this matrix reduces to the coprimality matrix $\bf \Phi$ of the ferromagnetic case and therefore 
in this limit the partition function simply counts the number of ground states of the classical ferromagnetic coprime chain of $M$ site with periodic boundary conditions.

\section{Classical ground states of the ferromagnetic case}\label{expofer}

In this Section we address the exponential degeneracy of the classical ground states in the ferromagnetic case postponing to the next section 
a similar analysis for the antiferromagnetic case. 

In the ferromagnetic case, all vertices that are not inert give rise to an exponential degeneracy of the classical ground states built out of them. The reason is that the interaction allows us to freely substitute at each site any possible value  $a$ of the occupation number with any other value $b$ provided $a$ and $b$ share at least a common divisor. The classical ground states of the chain can be conveniently associated to a path on a Brattelli diagram. The diagram contains on the horizontal axis the sites $i$ of the chain with $1\leq i\leq M$   and on the vertical axis the corresponding occupation number $n_i$, $2\leq n_i\leq q$. Starting from a given value $n_1$ on the initial site of the chain,  at each later step the path can either stay constant or jump to another value that is connected to the previous one by the adjacency matrix ${\bf A}$. As an example consider the adjacency matrix of the $q=6$ case  
\be
{\bf A} \,=\, 
\left(
\begin{array}{ccccc}
0 & 0 & 1 & 0 & 1 \\
0 & 0 & 0 & 0 & 1 \\
1 & 0 & 0 & 0 & 1 \\
0 & 0 & 0 & 0 & 0 \\
1 & 1 & 1 & 0 & 1 
\end{array}
\right) \,\,\,.
\label{incq6}
\ee  

\begin{figure}[h]
\vspace{8mm}
\psfig{figure=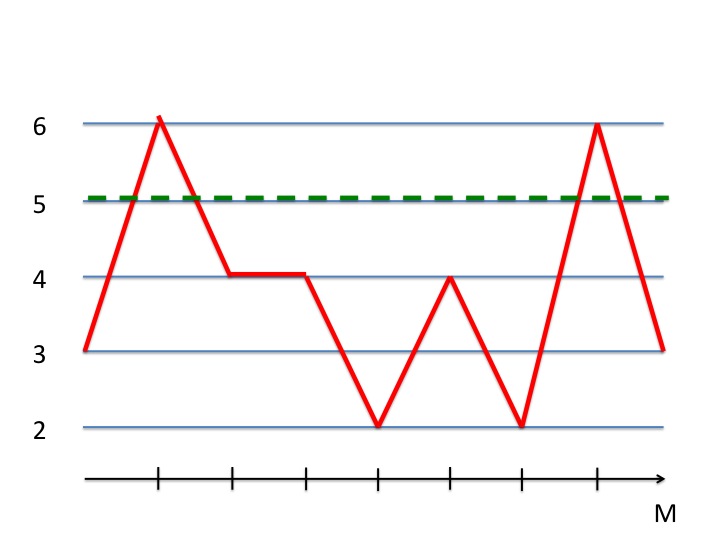,height=5cm,width=8cm}
\caption{{\em Brattelli diagram. Green Curve: path corresponding to the ground state of the inert vertex $5$. Red Curve: path corresponding to 
one of the exponentially numerous ground states generated by the other non-inert vertices. 
vertices. }}
\label{Brattelli}
\end{figure}

A possible  Brattelli diagram for a $q=6$ coprime chain is depicted in Fig.~\ref{Brattelli}. The green dashed line denotes the constant path 
associated to the ground state $555\dots$ whereas the red solid line corresponds to one of the exponentially numerous classical ground states, namely the sequence starting as $364\dots$. 

For an open chain of $M$ sites, the total number of classical ground states (including those coming from the inert vertices) corresponds to the 
total number of paths that can be drawn in the Brattelli diagram. The number of these paths can be easily computed with the aid of 
the coprimality matrix $\bf \Phi$. To this aim, let us denote by $N^{(a)}_t$ the total number of paths which have value $2\leq a\leq q$ at site $t$. In terms of these quantities  consider the $(q-1)$-dimensional vector 
\be
\vec {N_t} \,=\, 
\left(
\begin{array}{l}
N^{(2)}_t \\
N^{(3)}_t \\
N^{(4)}_t \\
\cdots \\
N^{(q-2)}_t \\
N^{(q-1)}_t \\
N^{(q)}_t 
\end{array}\right)
\,\,\,,
\label{tauvectors} 
\ee
with some initial boundary vector $\vec{N_1}$. 
The vector $\vec{N_t}$ evolves through multiplication by the matrix $\bf\Phi$ 
\be 
\vec{ N_{t+1}} \,=\, {\bf \Phi} \, \vec {N_t}\,\,\,. 
\label{recursive}
\ee
Indeed each of the new components  $N_{t +1}^{(a)}$ at site $(t+1)$ is obtained by summing over the paths $N^{(b)}_t$ at site $t$ whose final vertex $b$ is connected to $a$, i.e. those with $\Phi(a,b) = 1$. The total number of ground states for an open chain of $M$ sites (and $M-1$ links) is then 
\be 
N_M \,=\, \sum_{a=2}^q  N^{(a)}_{M-1} \,\,\,.
\ee
When $M=1$, the number of all classical ground states is simply equal to $(q-1)$, i.e. the number of all possible values of the occupation numbers. For a generic
$M$ the number of ground states can be easily extracted by noticing that the the matrix element $\left[{\bf \Phi}^k\right]_{ab}$ of the $k$-power of the matrix $\bf \Phi$ has the following interpretation 
\be
\left[{\bf \Phi}^k\right]_{a b} \,=\,\# \,{\rm \,of\, paths\, of\, length\, }\, k \, {\rm starting\, from \,the \,vertex~} \,a\, {\rm \,and \,ending\, at\, the\, vertex\,} \,b\,\,\,\,.
\label{totolab}
\ee 
It will be important though to  take into account the boundary conditions imposed at the ends of the chain. Let us discuss now some of them.  

\vspace{3mm}
\noindent 
{\bf Cyclic boundary conditions}. In this case, what matters are the diagonal matrix elements $\left[{\bf \Phi}^M\right]_{a a}$, corresponding to  the paths that start and end at the same value, and the sum thereof. Since there are $M$ links, the total number of ground states is given 
by  
\be 
{\cal N}_M^{cyc} \,=\,\sum_{a=2}^q \left[{\bf \Phi}^M\right]_{a a} \,=\, {\rm Tr} \, \left[(\bf\Phi)^M\right] \,\,\,. 
\label{totalnumbergroundstatetrace}
\ee
Some values of ${\cal N}_M^{cyc}$ varying the number of sites $M$ are collected  in Table 2. The number of ground state grows utterly fast and becomes soon exponentially large.  In fact, we can rewrite (\ref{totalnumbergroundstatetrace}) more explicitly as 
\be 
{\cal N}_M^{cyc} \,=\,{\rm Tr} \, \left[(\bf\Phi)^M\right] \,=\, 
\lambda_2^M + \lambda_2^M + \ldots + \lambda_{q}^N \,\,\,. 
\label{sumpow}
\ee
It is then obvious that for large values of $M$ the trace of $(\bf\Phi)^M$ is dominated by the largest eigenvalue $\lambda_2 \equiv \lambda_{max}$. Using the scaling law (\ref{scalingmaxeig}) established in Appendix C, we conclude that the number of ground states has for large $q$ asymptotically the exponential behaviour 
\be
{\cal N}_M^{cyc} \simeq (0.54636 \, q )^M \,\,\,. 
\label{estimateasym}
\ee

Finally let's notice that since the characteristic equation of the $q$-coprime chain is a polynomial of order $(q-1)$, it is enough to know the trace of the 
first $(q-2)$ powers of the matrix $\bf\Phi$ to know all its higher powers. Consider, for instance, the case $q=5$: from the characteristic polynomial 
of this model and its secular equation we have the relation 
\be
x^4 = 4 x^3 - 5 x^2 + 2x \,\,\,, 
\ee 
which is equivalent to the matrix identity for the matrix $\bf\Phi$
\be 
{\bf\Phi}^4 = 4 {\bf \Phi}^3 - 5 {\bf \Phi}^2 + 2 {\bf  \Phi }\,\,\,. 
\label{recrec}
\ee
Therefore the trace $t_4 = {\rm Tr}\,\left[\bf\Phi^4\right]$, is fully determined by the trace of the lower powers of $\bf \Phi$, i.e. 
$t_1 = {\rm Tr}\,\bf\Phi$, $t_2 = {\rm Tr} \, \left[{\bf \Phi}^2\right] $ and $t_3 = {\rm Tr} \, \left[{\bf \Phi}^3\right]$.  Hence, for this model it is enough 
to know these three integer numbers $t_1, t_2$ and $t_3$, in order to compute the trace of any other integer power of the matrix $\bf\Phi$. For instance, to get $t_5 = {\rm Tr}\, \left[{\bf \Phi}^5\right]$, it is sufficient to multiply the left and right terms of (\ref{recrec}) by $\bf\Phi$ and take the trace: in this way 
we get immediately the relation which links $t_5$ to the previous quantities $t_4, t_3$ and $t_2$.

\vspace{3mm}
\noindent
{\bf Fixed boundary conditions}. We now compute the number of classical ground states which start with $ n_1 = a$ and end with 
$n_M = b$. As shown in eq.\,(\ref{totolab}), the number of classical ground states in this case is given by 
\begin{equation}
{\cal N}_M^{a \rightarrow b} \,=\, \left[{\bf  \Phi}^{M-1}\right]_{a b} \,\,\,.
\label{cammini}
\end{equation}
We can further elaborate on \eqref{cammini} introducing the {\em boundary states} $| a \rangle $ and $| b \rangle$ that correspond to the 
two chosen boundary conditions: $|a\rangle$ and $|b\rangle$ are $(q-1)$ dimensional vectors with components $\langle j|a\rangle=\delta_{j,a-1}$ and $\langle j|b\rangle=\delta_{j,b-1}$, for $1\leq j\leq q-1$. In terms of these vectors, the number of classical ground states with fixed boundary conditions $a$ and $b$ at the 
two end-points can be written as 
\begin{equation}
{\cal N}_M^{a \rightarrow b} \,=\, \langle a |  {\bf \Phi}^{M-1} | b \rangle \,\,\,.
\label{sandwitch}
\end{equation}
Let $U$ be the unitary matrix that diagonalises the coprimality matrix $\widehat \Phi$
\begin{equation}
U^{\dagger} {\bf \Phi} \, \,U \,=\, {\cal D}  
\,=\, 
\left( 
\begin{array}{llllllll}
\lambda_2 &  &  &  &  &  &  \\
 & \lambda_3 &  &  &  &  &  \\
 &  & \lambda_4 &  &  &  &   \\
 &  &  & . &  &  &    \\
 &  &  &  & . &  &  &  \\
 &  &  &  &  & . &  &  \\
 &  &  &  &  &  & \lambda_{q} &  \\
\end{array}
\right)
 \,\,\,.
\end{equation}
Hence, we have (with standard labelling of the matrix elements of $U$) 
\begin{eqnarray}
{\cal N}_M^{a \rightarrow b} & \,=\, & \langle a |  {\bf \Phi}^{M-1} | b \rangle \,=\, 
\langle a | U U^{\dagger} {\bf \Phi}^{M-1} U U^{\dagger} | b \rangle \,\nonumber \\
& \,=\,&  \langle a | U \,{\cal D}^{M-1} \,U^{\dagger} | b \rangle \,=\, \sum_{j=2}^{q} 
U_{a-1,j-1} \, \lambda_j^{M-1} \, U^{\dagger}_{j-1,b-1} \,\,\,.
\end{eqnarray}
This formula can be further simplified in the limit $M \rightarrow \infty$, when the sum above is dominated by the largest eigenvalue $\lambda_{max} \equiv \lambda_2$  
\begin{equation}
{\cal N}_M^{a \rightarrow b} \simeq U_{a-1,1} U^{\dagger}_{1,b-1} \, \lambda_2^{M-1} = {\cal A}_{ab}\,
\, \lambda_2^M 
\,\,\,\,\,\,\,\,\,\,
,
\,\,\,\,\,\,\,\,\,\,
M \rightarrow \infty \,\,\,.
\end{equation}
where ${\cal A}_{ab} = (U_{a-1,1} U^{\dagger}_{1,b-1} \,\lambda_2^{-1})$. Therefore also in this case we have an exponential degeneracy of the number of classical ground states. Notice that 
\begin{equation}
\frac{{\cal N}_M^{a \rightarrow b}}{{\cal N}_M^{cyc}} = {\cal A}_{ab} \,\,\,, 
\end{equation}
it is an universal ratio, which depends however on the boundary conditions $a$ and $b$ chosen at the end of the chain.

\begin{table}[t]\label{groundstatenumbers}
\begin{center}
\begin{tabular}{||l|c|c|c|c|c|c|c|c|c|c||} \hline 
sites $M$ & 1 & 2 & 3 & 4 & 5 & 6 & 7 & 8 & 9 & 10\\
 \hline \hline 
 $\mathcal{N}_M^{cyc}(q=5)$ & \,\,\,\,\,\,4\,\,\,\,\,\, & \,\,\,\,\,\,6\,\,\,\,\,\, & \,\,\,\,\,10 \,\,\,\,\, & \,\,\,\,\,\,18\,\,\,\,\,\, & \,\,\,\,\, 34 \,\,\,\,\, & \,\,\,\,\, 66\,\,\,\,\,  &  \,\,\,\, 130 \,\,\,\, & 
 \,\,\, 258 \,\,\, & 514 &1026\\
 \hline
 $\mathcal{N}_M^{cyc}(q=6)$ & \,\,\,5\,\,\, & 13 & 35 & 105 & 325 & 1021 & 3225 & 10209 & 32345 & 102513 \\
 \hline 
 $\mathcal{N}_M^{cyc}(q=7)$ & 6 & 14 & 36 & 106 & 326 & 1022 & 3226 & 10210 & 32346 & 102514 \\
 \hline
 $\mathcal{N}_M^{cyc}(q=8)$ & 7 & 21 & 73 & 285 & 1147 & 4665 & 19033 & 77733 & 317575 & 1297581 \\
 \hline
 $\mathcal{N}_M^{cyc}(q=9)$ & 8 & 26 & 92 & 362 & 1478 & 6158 & 25922 & 109730 & 465914 & 1981586 \\
\hline
$\mathcal{N}_M^{cyc}(q=10)$ & 9 & 37 & 159 & 769 & 3859 & 19717 & 101537 & 524817 & 2717349 & 14081317\\
\hline
\end{tabular}
\end{center}
Table 2. Number of ground states with cyclic boundary conditions for various $q$-coprime chains by varying the length $M$ of the chain. 
\end{table}

\begin{figure}[b]
\center
\psfig{figure=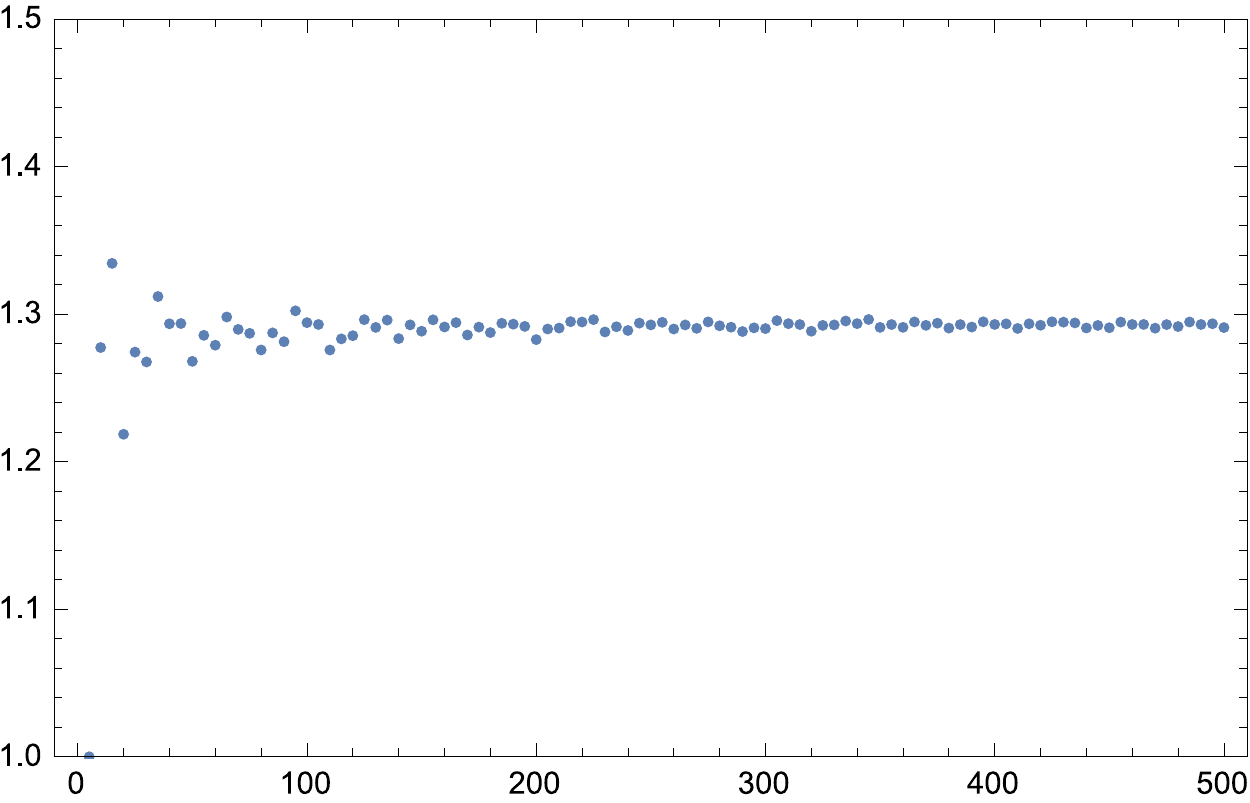,height=5cm,width=7cm}
\caption{{\em Universal ratio ${\cal R}$ defined in \eqref{URf} versus $q$.}}
\label{urf}
\end{figure}

\vspace{3mm}
\noindent
{\bf Free boundary conditions}. Choosing free boundary conditions at the ends of chain, the number of the classical ground states 
can be conveniently computed by means of the free boundary state
\begin{equation}
| f \rangle \,=\,\begin{pmatrix}1\\
                                     \vdots\\
                                     1
\end{pmatrix}\,\,\, .
\end{equation}
Indeed analogously to the case of fixed boundary conditions, we have  
\begin{eqnarray}
{\cal N}_M^f & \,=\, & \langle f |  {\bf \Phi}^{M-1} | f \rangle \,=\, 
\langle f | U U^{\dagger} {\bf \Phi}^{M-1} U U^{\dagger} | f \rangle \,\nonumber \\
& \,=\,&  \langle f | U \,{\cal D}^{M-1} \,U^{\dagger} | f \rangle \,=\, \sum_{k,j,l=2}^{q} 
U_{k-1,j-1} \, \lambda_j^{M-1} \, U^{\dagger}_{j-1,l-1} \,\,\,.
\end{eqnarray}
This formula simplifies when the chain is very large, since in the limit $M \rightarrow \infty$ we have 
\begin{equation}
{\cal N}_M^{f} \simeq 
\lambda_2^{-1}\, \left| \sum_{j=1}^{q-1} U_{1,j}\right|^2
\, \lambda_2^{M}\,\,\,. 
\end{equation}
Hence, the exponential growth of the number of classical ground states with free boundary conditions gives rise to the universal ratio 
\begin{equation}
{\cal R} = \lim_{M\rightarrow \infty} \frac{{\cal N}_M^{f}}{{\cal N}_M^{cyc}} = \lambda_2^{-1}\left| \sum_{j=1}^{q-1} U_{1,j}\right|^2 \,\,\,.
\label{URf}
\end{equation}
The plot of this quantity as a function of $q$ is given in Fig. \ref{urf}. The numerical extrapolation of the asymptotic value for $q \rightarrow \infty$ of these data, ${\cal R}_{\infty} \sim 1.294$, nicely matches with the theoretical value (\ref{univrat}) reported in the Appendix C.  
Let's note, \textit{en passant}, that in the graph theory jargon (see Appendix \ref{Appgraphtheory}) the quantity 
\begin{equation}
\beta_r=\frac{1}{\sqrt{q-1}}\sum_{j=1}^{q-1} U_{r,j}\,\,\,, 
\end{equation}
is also called the $r$-th angle of a graph.

\vspace{3mm}
\noindent
{\bf Frustration}. The ferromagnetic coprime chain can display the phenomenon of frustration, namely the impossibility to solve the conditions 
$\Phi(n_i,n_{i+1})=1$ for all the links, since there may be obstructions coming from the boundary conditions. This is particularly true in the case of fixed 
boundary conditions. Using what we learnt before on the relation between ground states and paths on Brattelli diagrams, it is easy to give an algebraic 
characterisation when a frustration is going to occur. Such a characterisation involves the coprimality matrix $\bf\Phi$:  
for fixed boundary conditions of type $a$ and $b$, and for an open chain of $M$ sites there will be frustration when 
\begin{equation}
\label{algebraicfrustrutation}
\left[\bf\Phi^{M -1}\right]_{ab}=0
\end{equation}
Geometrically the relation~\eqref{algebraicfrustrutation} expresses the absence of any path in the incidence graph starting from a vertex labelled by $a$ and ending to a vertex labelled by $b$ in exactly $M-1$ steps. It is easy to see that there will be frustration each time $a$ will label an inert vertex, while $b$ will be any other number $b\not=a$: for example if $q=14$, $a = 13$ and $b=6$, there is no path that can connect the corresponding vertices on the incidence graph. 

For the ferromagnetic chain we expect to have no frustration for both periodic and free boundary conditions. Namely, we expect that the equation
\begin{equation}
{\rm Tr}\, \left[{\bf \Phi}^{M}\right] \,=\, 0\,\,\,, 
\label{perfrus}
\end{equation}
relative to the periodic boundary conditions, as well as the equation
\begin{equation}
\langle f |  {\bf \Phi}^{M-1} | f \rangle \,=\,0 \,\,\,,
\label{perfrus2}
\end{equation}
relative to the free boundary conditions, will never have a solution. Indeed,  among the configurations that contribute to eqs.(\ref{perfrus}) and (\ref{perfrus2}) there are always the trivial ground states obtained repeating the same value of the occupation number on each lattice site: the existence of these paths makes both the expressions (\ref{perfrus}) and (\ref{perfrus2}) strictly positive.

\section{Classical Ground States of the Anti-Ferromagnetic Case}\label{classicalanti}
\begin{figure}[t]
\vspace{8mm}
\psfig{figure=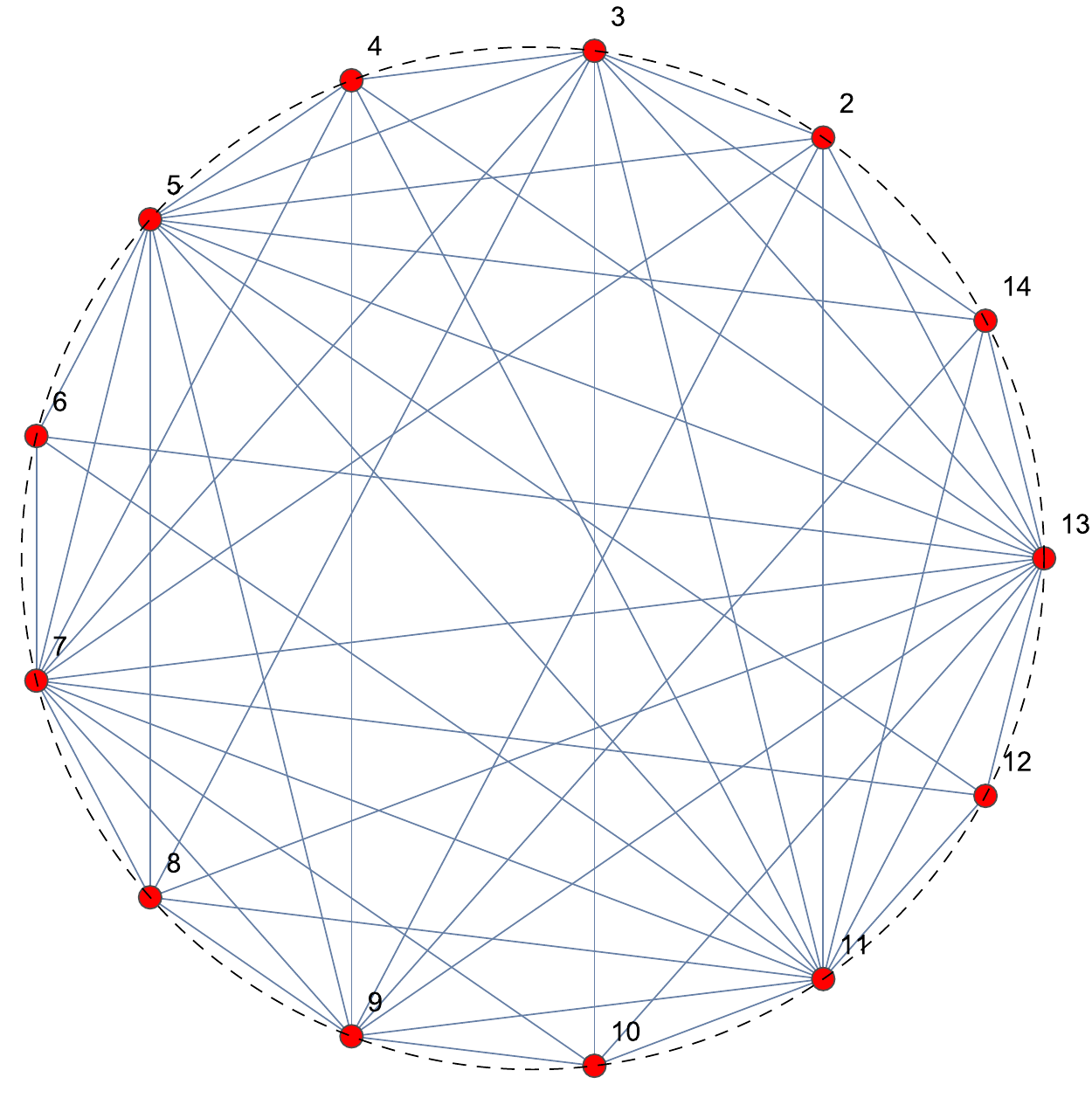,height=8cm,width=8cm}
\caption{{\em Incidence graph of the anti-ferromagnetic coprime chain with $q=14$.}}
\label{incidencecomplement}
\end{figure}

Let us now turn out attention to the classical anti-ferromagnetic case of the coprime chain. The classical Hamiltonian of the one-dimensional chain of $M$ sites is given by 
\begin{equation}
\label{hclass}
H^a_{cl}\,=\, -\sum_{i=1}^M \overline \Phi(n_i,n_{i+1}) \,=\, \sum_{i=1}^M (1- \Phi(n_i,n_{i+1}))\,\,\,.
\end{equation}
This time the Hamiltonian favours next-neighbouring occupation numbers that are coprime, i.e. $\Phi(n_i,n_{i+1}) = 0$. The incidence graph in the
anti-ferromagnetic chain is the complement graph of the  ferromagnetic chain (see Appendix A): namely a graph with the
same number of vertices  of the ferromagnetic graph  but with edges along the pairs $(i,j)$ which were originally missed, see Fig. \ref{incidencecomplement}
and compare with the previous Fig. \ref{incidence}. 

\begin{table}[b]\label{tgroundstatenumbersa}
\begin{center}
\begin{tabular}{||l|c|c|c|c|c|c|c|c|c||} \hline 
\text{sites} $M$ & 2 & 3 & 4 & 5 & 6 & 7 & 8 & 9 & 10  \\
 \hline \hline
$\mathcal{N}^{free}(q=5)$ & 10  & 26 & 66 & 170 & 434 & 1114 & 2850 & 7306 & 18706 \\
$\mathcal{N}^{cyc}(q=5)$ & 10 & 12 & 50 & 100 & 298 & 700 & 1890 & 4692 & 12250 \\
\hline
$\mathcal{N}^{free}(q=6)$ & 12  & 34 & 88 & 242 & 640 & 1736 & 4632 & 12492 & 33456 \\
$\mathcal{N}^{cyc}(q=6)$ & 12 & 12 & 64 & 120 & 408 & 952 & 2800 & 7104 & 19792 \\
\hline
$\mathcal{N}^{free}(q=7)$ & 22  & 88 & 338 & 1326 & 5146 & 20082 & 78146 & 304538 & 1185906 \\
$\mathcal{N}^{cyc}(q=7)$ & 22 & 48 & 250 & 860 &  3562 & 13468 & 53250 & 205860 & 804922 \\
\hline
\hline
\end{tabular}
\end{center}
Table 3: Number of ground states for $M = 1, 2, 3, \ldots 10$ for various
$q$-coprime chains with antiferromagnetic interactions for free (upper values of the columns)
and periodic boundary conditions (lowest values of the columns).  
\end{table} 

In contrast with the ferromagnetic one, the  anti-ferromagnetic incidence graph does not posses any inert vertex. Moreover, its vertices have, in general, higher
degree: indeed, as shown in Sec.~II, the probability that two random integers are coprime is $\frac{6}{\pi^2}>1/2$. Roughly speaking we should expect that the ground
state degeneracy  will be larger now than with ferromagnetic interactions. This is indeed the case, as shown by the values in Tab.~3 and further confirmed by the scaling law of the highest eigenvalue of the anti-ferromagnetic coprimality matrix, here denoted as $\xi_{max}$. In the limit $q \rightarrow \infty$ $\xi_{max}$ can be computed exactly in terms of an expression which is an infinite product over primes
\be 
\label{dz}
\lim_{q \to \infty} \frac{ \xi_\mathrm{max} }{ q } \, \equiv \lambda = \, \prod_{p ~ \mathrm{prime} } \( \frac{ p-1 + \sqrt{(p-1)(p+3)} }{2 p}  \) = 0.67846225243465570728 \cdots
\ee
We call the number $\lambda$ the {\em Zagier constant}. For a proof of eq. \eqref{dz} and other interesting related number theory results we defer to the Appendix C. 

\begin{figure}[b]
\center
\includegraphics[scale=0.6]{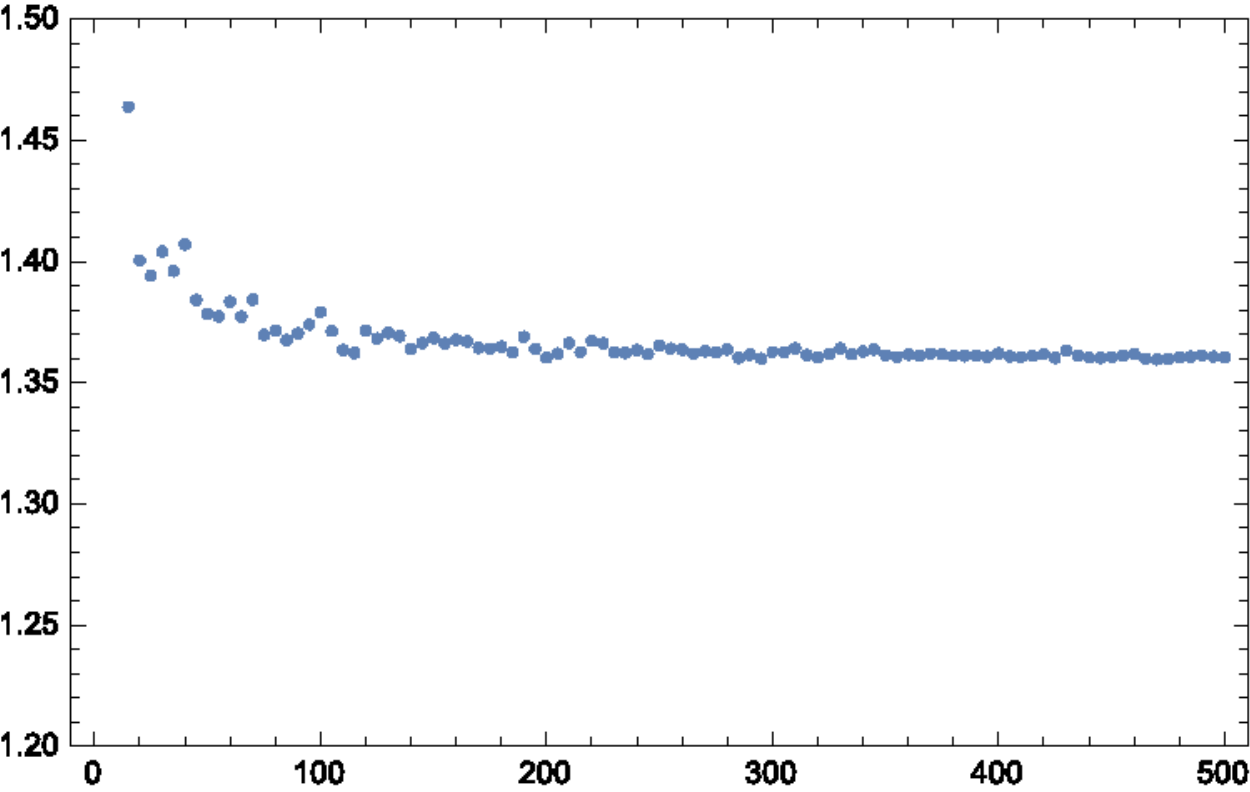}
\caption{{\em Universal ratio ${\cal R}^{(af)}$ defined in \eqref{URf2} versus $q$.}}
\label{urfa}
\end{figure}

All computations relative to the number of ground states with different boundary conditions
proceed in complete analogy with the ferromagnetic case with the only replacement  $\bf{\Phi}\rightarrow \bf{\overline \Phi}$ in the coprimality matrix. Also in this case there exists the universal ratio 
\begin{equation}
{\cal R}^{(af)} = \lim_{M\rightarrow \infty} \frac{{\cal N}_M^{f}}{{\cal N}_M^{cyc}} = \xi_{max}^{-1}\left| \sum_{j=1}^{q-1} \tilde U_{1,j}\right|^2 \,\,\,.
\label{URf2}
\end{equation}
where $\tilde U$ is the unitary matrix which diagonalises the antiferromagnetic coprimality matrix $\bf {\overline \Phi}$. The plot of this quantity as a function of $q$ is given in Fig. \ref{urfa}. The numerical extrapolation of the asymptotic value for $q\rightarrow \infty$ of these data, ${\cal R}^{(af)}_{\infty} \sim 1.3580$, nicely matches with the exact theoretical
value (\ref{UR}) derived in the Appendix C.  

\vspace{3mm}
\noindent
{\bf Frustration.} Contrary to the ferromagnetic case, the anti-ferromagnetic chain for $q > 6$ does not generally display frustration. 
The reason is basically the following: for $q > 6$, there are {\em always} at least {\em two} primes $p$ and $p'$ which fall in the  interval\footnote{A
slightly different viewpoint is to observe that the values at the end-points are not divisible by the largest prime $p_{max}$ that is certainly bigger
than $q/2$. On the other hand, these two numbers  cannot be divisible further by all the primes smaller than $p_{max}$ if  $q>6$.
It is not difficult to see that this circumstance leaves room for eliminating completely frustration in the antiferromagnetic case.}
${\cal I} =(\frac{q}{2}, q)$: these two primes  cannot enter as divisor of all numbers belonging to the range $[2,q]$. In other words, each of these two primes can be followed by {\em any} other number
in the range $[2,q]$ keeping the condition of minimal energy of the antiferromagnetic interaction intact. In particular, $p$ can be also followed by $p'$ and vice-versa. It is easy to show that these conditions automatically ensure that there could be no frustration for any choice of fixed boundary conditions selected for a system of length $M$ (and, a fortiori for periodic and free boundary conditions). But, how do we know that there are always at least two primes in the interval ${\cal I} = (\frac{q}{2},q)$ for $q > 6$? Because there is a theorem, due to Nagura \cite{Nagura}, which along the line of the Bertrand's theorem, ensures that for $q \geq 25$ there are at least {\em three primes} in the interval ${\cal I}$. For all the finitely many cases
with $q < 25$ not covered by the Nagura's theorem, one can make an explicit analysis and check that indeed for $q > 6$ there are always at least two primes in the  interval ${\cal I}$. 

We now discuss separately the lowest cases $q=3$, $q=4$, $q=5$ and $q=6$. 

\vspace{3mm}
\noindent 
{\bf ${\bf q=3}$ case.} For $q=3$, the anti-ferromagnetic coprimality matrix $\widehat \Phi^{(af)}$ is given by 
\begin{equation}
\bf{\overline \Phi}
=\left(
\begin{array}{ll}
0 & 1 \\
1 & 0 
\end{array}\right)  \,\,\,, 
\end{equation}
and therefore $\left[\bf{\overline\Phi}\right]^{2 N} = {\bf 1}$ while $\left[\bf{\overline \Phi}\right]^{2 N+1} = \bf{\overline \Phi}$.  In both cases there are 
matrix elements which are $0$ and therefore, according either to eq.\,(\ref{algebraicfrustrutation}) or eq.\,(\ref{perfrus}), one can have frustration. In particular, 
if the chain has $M=2n+1$ sites (and therefore $2 n$ links), choosing as boundary conditions $n_1 =2$ and $n_{M}=3$, we will have frustration. Vice-versa, if the chain has $M = 2n$ sites (and therefore $2n +1$ links), there will be frustration if we choose $n_1 = 2$ and $n_M = 2$ or $n_1 = 3$ and $n_M=3$.   

\vspace{3mm}
\noindent
{\bf ${\bf q=4}$ case.}
For $q=4$ and an open chain with $M=2k$ sites, all the classical ground states with free boundary conditions must necessarily have an alternating pattern of the type
\begin{equation}
 n_1\, 3\, n_3\, 3\, n_5\, 3\, n_7\, 3 \dots,
\end{equation}
where each number $n_i$ can be either 2 or 4. There is of course an additional symmetry under the exchange of the two numbers, namely the sequence
$3\,n_2\,3\,n_4\,3\,n_6\,3\dots$ is also a possible ground state. Hence, overall we have
\begin{equation}
 \mathcal{N}^{f}_M(q=4)=2^{k+1},
\end{equation}
possible ground states. For $M=2k-1$, the number of possible classical ground state is $3\times 2^{k-1}$. These considerations imply that, in the presence 
of certain fixed boundary conditions, there will be frustration: for instance, this will be the case if $M = 2k$ and if we choose as initial and
final values $n_1 = 2$ and 
as $n_M$ either $2$ or $4$. With periodic boundary conditions, the $q=4$ chain displays the same degeneracy of the free boundary conditions when $M$ is an 
even number, while it will be frustrated for $M$ being an odd number.

\vspace{3mm}
\noindent
{\bf ${\bf q=5}$ case.} In this chain there are always two primes, $3$ and $5$, that do not divide the other numbers of the chain. Therefore, as the general case discussed above, the $q=5$ antiferromagnetic case can never be frustrated.

\vspace{3mm}
\noindent
{\bf ${\bf q=6}$ case.} This is an interesting exceptional case: when $q=6$ there is only one prime in the interval ${\cal I}=(3,6)$, namely 
$p=5$. Notice that in order to avoid frustration the number $6$ can {\em only} be followed by $5$. Therefore,
if we enforce fixed boundary conditions that cannot meet this requirement, we will have frustration. By inspection, one can see that this
can  happen only for small chains. If $M =2$ we can exhibit many examples, for instance $n_1 = 2$ and $n_2 = 6$ is one of those. More in general it is
sufficient to spot the vanishing elements of the square of the antiferromagnetic coprimality matrix given by 
\begin{equation}
{\widehat \Phi^{(af)}} \,=\, 
\left(
\begin{array}{ccccc}
0 & 1 & 0 & 1 & 0 \\
1 & 0 & 1 & 1 & 0 \\
0 & 1 & 0 & 1 & 0 \\
1 & 1 & 1 & 0 & 1 \\
0 & 0 & 0 & 1 & 0 
\end{array}
\right). 
\label{incq62}
\ee  
For $M=3$, the only frustrated configuration is the one with fixed boundary conditions $n_1 = 5$ and $n_3=6$, because whatever the value 
of $n_2$ will be, it would be impossible to minimize the energy of all the two links. When $M =4$, the only frustrated configuration starts with
$n_1 = 6$ and ends with $n_4 =6$ and finally for $M > 4$ there will be no longer frustration. The simplest way to prove the last statement is to observe that the matrix elements of $(\bf{\overline \Phi})^k$, for $k > 3$ are all positive integers.


\section{Reaching criticality in the coprime quantum chain}\label{spectrumQH}

Switching on the operators $G^{\alpha}_i$ in the quantum Hamiltonian \eqref{quantumhamiltonian}, the structure of the classical ground states previously determined changes quite drastically, in particular their exponential degeneracy generally disappears. However peculiar situations might arise when performing a fine-tuning of the couplings of the operators $G^{\alpha}_i$. Rather than embarking on an exhaustive analysis of the  coprime quantum chain,
here we will focus only on those cases where it will be possible to reach various types of familiar criticalities: notably those of Ising or Potts quantum chains!  
  
In the following we will mainly consider the ferromagnetic coprime chain with $q=5$, for several reasons: firstly, because it is the simplest case where
the coprimality interaction gives rise to non-trivial effects, secondly because it is a case still manageable from a numerical point of view. Indeed, the exponential 
growth of the Hilbert space of the coprime quantum chain with the number of sites $M$, $d_H = (q-1)^M$, makes prohibitive any exact diagonalization procedure 
for large value of $q$ even for small $M$. In this respect, the dimension $d_H=4^M$ of the $q=5$ coprime chain permits to push the numerical analysis to sufficiently large $M$ and to extrapolate reliable properties in the thermodynamic limit through finite size tecnhiques. With this in mind, we also chose to work always with periodic boundary conditions.\footnote{A potential critical behavior cannot not be affected by the boundary condition employed.
Periodic boundary conditions are simply a way to make the finite size scaling as fast as possible.}

The simplest class of universality which can be realised in terms of the  coprime quantum chain is the one of the quantum Ising chain. In order to appreciate this point, let briefly remind its essential properties.
  
 \vspace{3mm}
\noindent 
{\bf Ising chain universality class}. In a nutshell, the class of universality of the quantum Ising chain consists of two phases, separated by a critical point in between: the low-temperature phase, characterised by two degenerate ground states; the high temperature phase characterised instead by only one ground state.
Such a scenario can be explicitly realised in terms of the Hamiltonian 
\be 
H \,=\,-\sum_{i=1}^N \left[ \hat\sigma^z_i \hat\sigma^z_{i+1} + \Delta \hat\sigma^x_i \right]  \,\,\,,
\label{quantumisingmodel}
\ee
which involves
the Pauli's matrices\footnote{Each operator $\hat\sigma^a_i$ has to be meant as in eq.\,(\ref{notation}), namely 
$\hat\sigma_i^{a} \,=\, {\bf 1} \otimes {\bf 1} \cdots {\bf 1} \otimes \underset{\underset{i-site}{\uparrow}}{\sigma^{a}} \otimes {\bf 1} \otimes {\bf 1} \cdots \otimes {\bf 1}  
$.}. 
For $\Delta < 1$ this Hamiltonian has two degenerate ground states which in the limit $\Delta \rightarrow 0$ can be written as 
\begin{equation}
| \Uparrow \rangle \,=\, \otimes_{i} | \uparrow \rangle_i 
\,\,\,\,\,\,\,\,
,
\,\,\,\,\,\,\,\,
| \Downarrow \rangle \,=\, \otimes_{i} | \downarrow \rangle_i 
\,\,\,, 
\end{equation}
where $|\uparrow \rangle_i$ and $| \downarrow \rangle_i$ are the two eigenvectors of the $\hat{\sigma}^z$ operator at the site $i$.
For $\Delta > 1$ the model is instead in its high temperature phase with only one ground state: when $\Delta \rightarrow \infty$ this ground state can be explicitly written and it is given by 
\begin{equation}
| \Rightarrow \rangle \,=\, \otimes_i | \rightarrow \rangle_i 
\,\,\,\,\,\,\,\,
,
\,\,\,\,\,\,\,\,
| \rightarrow \rangle_i = \frac{1}{\sqrt{2}} (| \uparrow \rangle_i + | \downarrow \rangle_i) \,\,\,.
\end{equation}
 Approaching the value $\Delta=1$, this model undergoes a quantum phase transition which is signalled by the closure of the gap in the energy spectrum.
 The critical point of the Ising model is well known to be described by the simplest minimal model of conformal field theory whose central charge $c$ is $1/2$ \cite{bpz}. Since the lattice model can be solved exactly \cite{kogut,sachdev}, the central charge at its critical point can be inferred  in many different ways, as  for instance finite size scaling of the ground state energy\cite{bcn,aff}. However, in order to compare later with the central charge characterizing criticality in the coprime chain, we found convenient to estimate $c$ numerically through the ground state entanglement entropy.

\vspace{3mm}
\noindent
{\bf Central charge and entanglement entropy}. As shown in \cite{ent1,ent2,ent3} and in particular in \cite{ccee},
for a critical one-dimensional spin chain of $M$ sites with periodic boundary conditions and bipartite 
in two subchains $A$ and $B$ whose length is $m$ and $M-m$, the entanglement entropy of the ground state reads
\be \label{entent}
S_A (m) \,=\, - \tr \, \hat{\rho}_A \log \hat{\rho}_A \,=\, \frac{c}{3} \log \( \pi M \sin \frac{ \pi m}{M} \) + \mathrm{const}
\ee
where $c$ is the central charge and  $\hat{\rho}_A$ the ground state reduced density matrix of the subsystem $A$
\be
\hat{\rho}_A = \tr_B \( \ket{ GS } \bra{ GS } \) \,\,\,. 
\ee
This formula can be used to fit numerical data for fixed number of sites $M$ or, for fixed size of the subsystem, choosing $m=M/2$ and varying $M$.

\vspace{3mm}
\noindent 
{\bf Ising critical point of the coprime chain}.
\begin{figure}[t]
\center
\makebox[0pt][c]{
\hspace{-10mm}
\begin{minipage}{0.52\textwidth}
\includegraphics[scale=0.45]{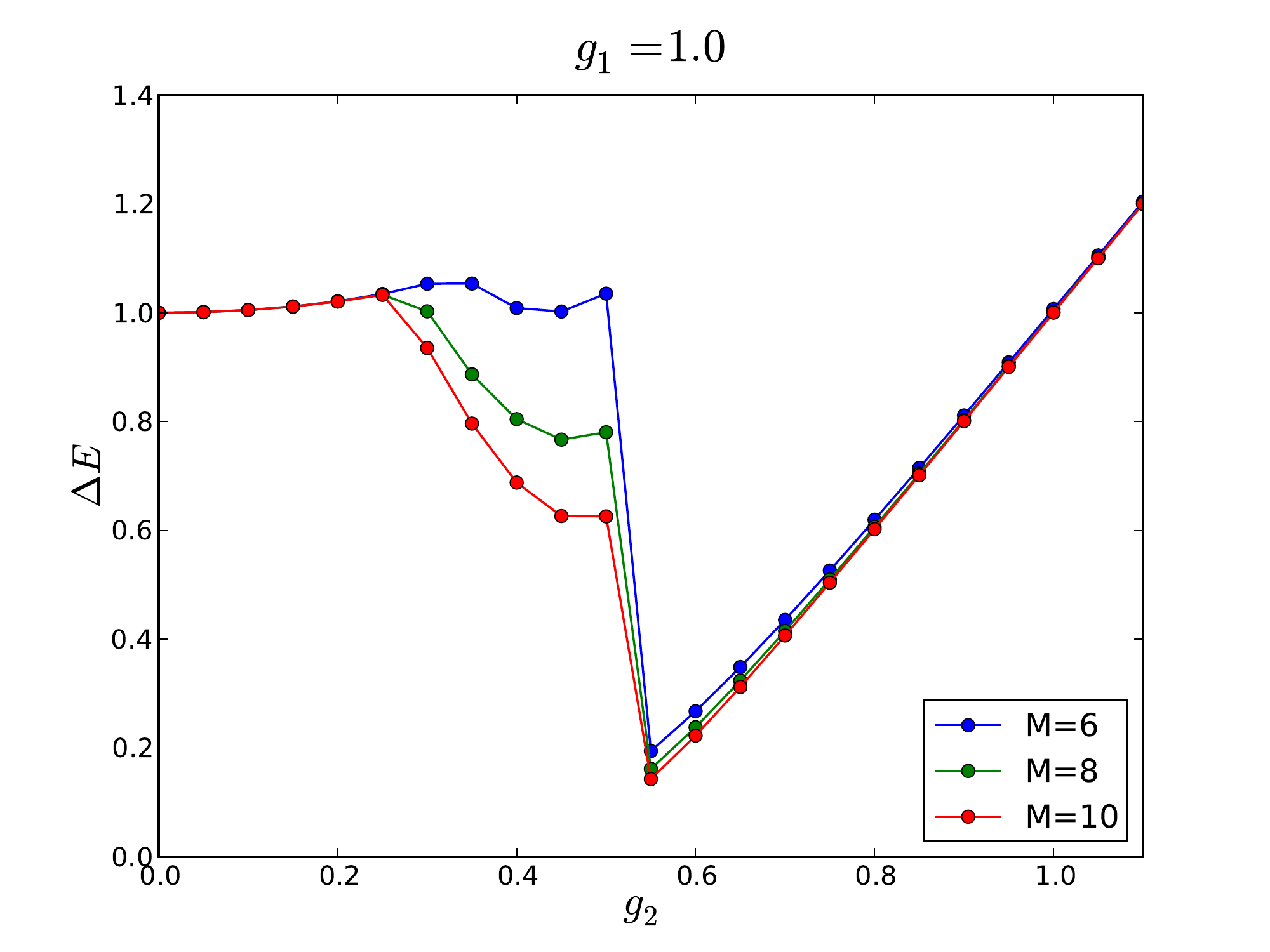}
\end{minipage}
\begin{minipage}{0.53\textwidth}
\includegraphics[scale=0.45]{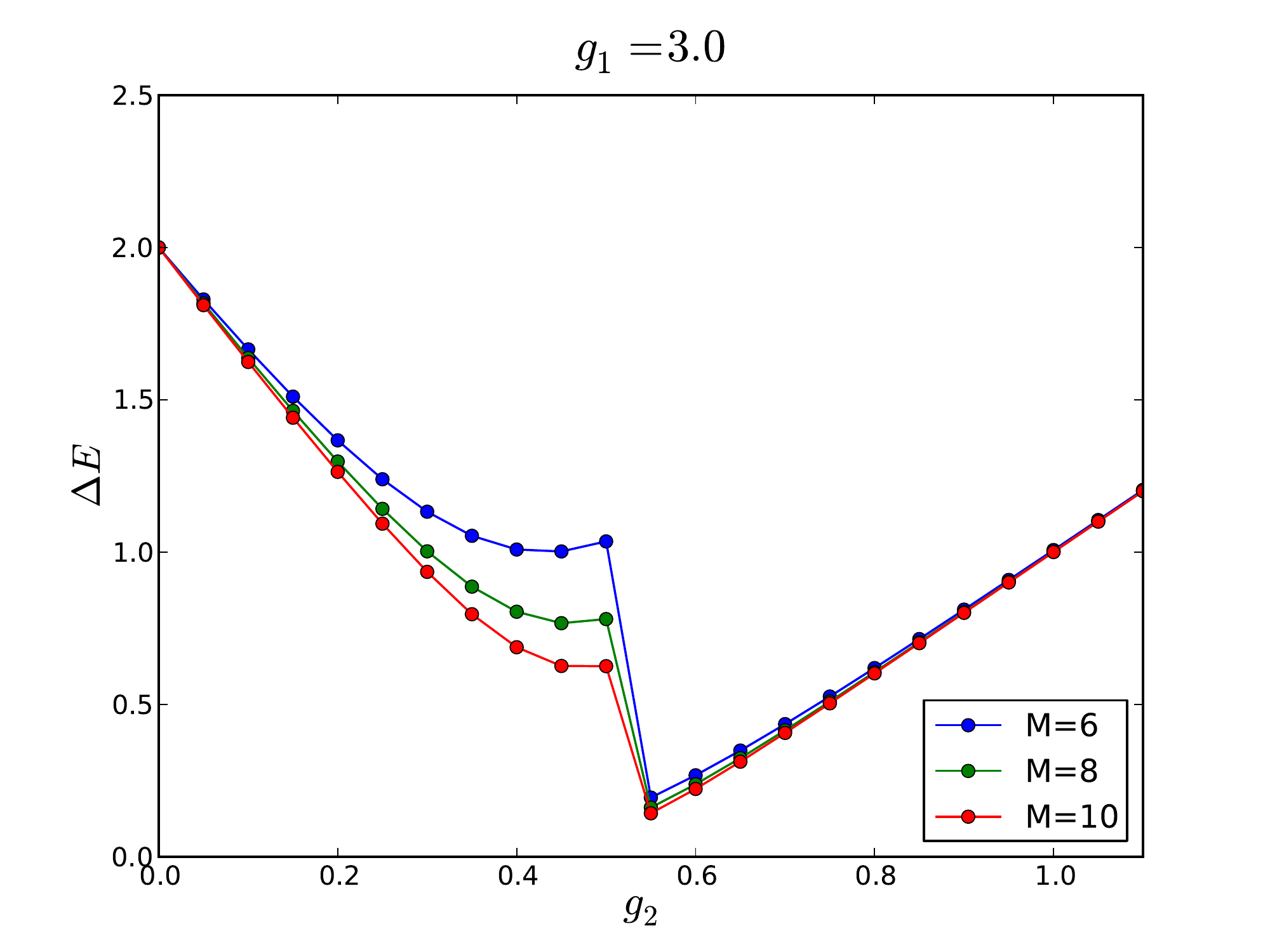}
\end{minipage}}
\caption{{\em Gap of the quantum Hamiltonian (\ref{qh1}) of the coprime chain $q=5$, for two different fixed values of $g_1=1.0$ and $g_2=3.0$ 
and varying $g_2$. The finite size scaling shows the closure of the gap when $g_2 \simeq 1/2$, independently of the value of $g_1$.}}
\label{gap2}
\end{figure}
In the $q=5$ coprime chain let us switch on the local operators $B^{(1)}_i$ and $B^{(2)}_i$, with the associated matrices given by (see the notations of eq.\,(\ref{notation}) and eq.\,(\ref{magneticandallthat})) 
\be \label{bope2}
{\cal B}^{(1)} = 
\begin{pmatrix}
1 & 0 & 0 & 0 \\
0 & 1 & 0 & 0 \\
0 & 0 & 0 & 0 \\
0 & 0 & 0 & 0 
\end{pmatrix} \equiv {\mathcal D}^{(1)} + {\mathcal D}^{(2)}   
\qquad
,
\qquad
{\cal B}^{(2)} = 
\begin{pmatrix}
0 & 1 & 0 & 0 \\
1 & 0 & 0 & 0 \\
0 & 0 & 0 & 0 \\
0 & 0 & 0 & 0 
\end{pmatrix} \equiv  {\mathcal S}^{(12)} 
\,\,\,, 
\ee
so that the Hamiltonian of such a  coprime quantum chain can be written as\footnote{Here and after, the $g_i$'s are obviously linear combinations of the 
previous coupling constants $\beta_{\alpha}$ introduced in eq.\,(\ref{quantumhamiltonian}).}
\be \label{qh1}
H \,=\,-\sum_{i=1}^M \left[\Phi(n_i,n_{i+1})  +  g_1 \,B^{(1)}_i   + g_2 \,B^{(2)}_i \right].
\ee
Moreover, we assume from now on all the couplings to be non-negative. We firstly consider the case in which $g_2 =0$: since the operator ${\cal B}^{(1)}$ consists of the two magnetic fields ${\mathcal D}^{(1)}$ and ${\mathcal D}^{(2)}$ which have the effect to lower the single-site energy of the two states $\ket{2}$ and $\ket{3}$, globally this leads to a reduction of the exponentially large number of the classical ground states to just two degenerate ground states, namely
\be 
|\tilde 2 \rangle = \ket{2\,2\,2\,2\,\dots 2}
\qquad {\rm and} 
\qquad
|\tilde 3\rangle = \ket{3\,3\,3\,3\,\dots 3}\,\,\,. 
\ee 
It is natural to think that these two degenerate states may play the same role of the two degenerate ground states $| \Uparrow \rangle$ and $|\Downarrow\rangle$ 
of the Ising chain in its low temperature phase.
The energy of the ground states $ | \tilde 2\rangle$ and $|\tilde 3\rangle$ depends on $\lambda_1$,  being $E_{GS} = - (1+g_1)M$ but their existence
does not rely on the actual value of $g_1$ as far as $g_1>0$. The value of $g_1$ also enters the first excited level: indeed the natural candidates for
the first excited states are the $N$-fold degenerate states
\be \label{n-deg}
\ket{ 2 \, 2 \, \dots 2 \, 4 \, 2 \, \dots 2 }
\ee
whose energy is $E_1 = - M - (M-1) g_1$, and the $M(M-1)/2$-fold degenerate states with two domain walls 
\be \label{2p-deg}
\ket{ 2 \, 2 \, \dots 2 \, 3 \, \dots 3 \, 2 }
\ee
whose energy is $E_2 = -(M-2) - M g_1$. Then if $g_1 < 2$ one has $E_1 < E_2$ and the first excited states are \eqref{n-deg},
while if $g_1 > 2$ the states in \eqref{2p-deg} have smaller energy. Thus, the gap of the coprime Hamiltonian (\ref{qh1}) when $g_2 = 0$ is given by 
\be
\Delta E(g_1,0) \,= \, E_{1st} - E_{GS} \,= \, \begin{cases} g_1, & \mbox{if } g_1 < 2 \\ \,\,2, & \mbox{if } g_1 > 2 \end{cases}\,\,\,. 
\ee
Let now us switch on the operator $B^{(2)}_i$: notice that, at each site of the lattice, the corresponding operator ${\cal B}^{(2)}_i$ mixes locally two states
(here associated to the vectors $| 2 \rangle$ and $| 3 \rangle$), as it also does the operator $\sigma^x_i$ 
in the Hamiltonian of the quantum Ising chain (\ref{quantumisingmodel}).
Therefore, one could expect that by varying the coupling constant  $g_2$  in \eqref{qh1} one could come across a quantum phase transition
in the Ising universality class. This is indeed the case and by exact diagonalization it is possible to show that
the ground state degeneracy persists (up to terms exponentially small in $M$) until $g_2$ reaches the critical value $g_2^{*} = 1/2$,
irrespectively of the value of $g_1$. For $g_2=g_2^*$ the ground state is no more degenerate and the gap of the Hamiltonian
\eqref{qh1} closes,
namely $\Delta E (g_1,1/2) = 0$ for any $g_1$. When $g_2 > 1/2$ there
is an unique ground state, as in the paramagnetic phase of the Ising chain \eqref{quantumisingmodel}.
Part of the numerical analysis is reported in Fig.~\ref{gap2}, where the gap is plotted as a function of $g_2$ for fixed $g_1$.

\noindent

Once the critical point has been located, we can proceed to identify its universality class by calculating the ground state entanglement entropy. 
As shown in Fig.~\ref{ee2} and Fig.~\ref{ee22}, the quantum critical point corresponds
to a second order phase transition, since the entanglement entropy diverges logarithmically with $M$, and the central charge that is extracted from \eqref{entent}
is $c = 1/2$, i.e. the one of  Ising  universality class.  

\begin{figure}[t]
\center
\makebox[0pt][c]{
\hspace{-15mm}
\begin{minipage}{0.33\textwidth}
\includegraphics[scale=0.32]{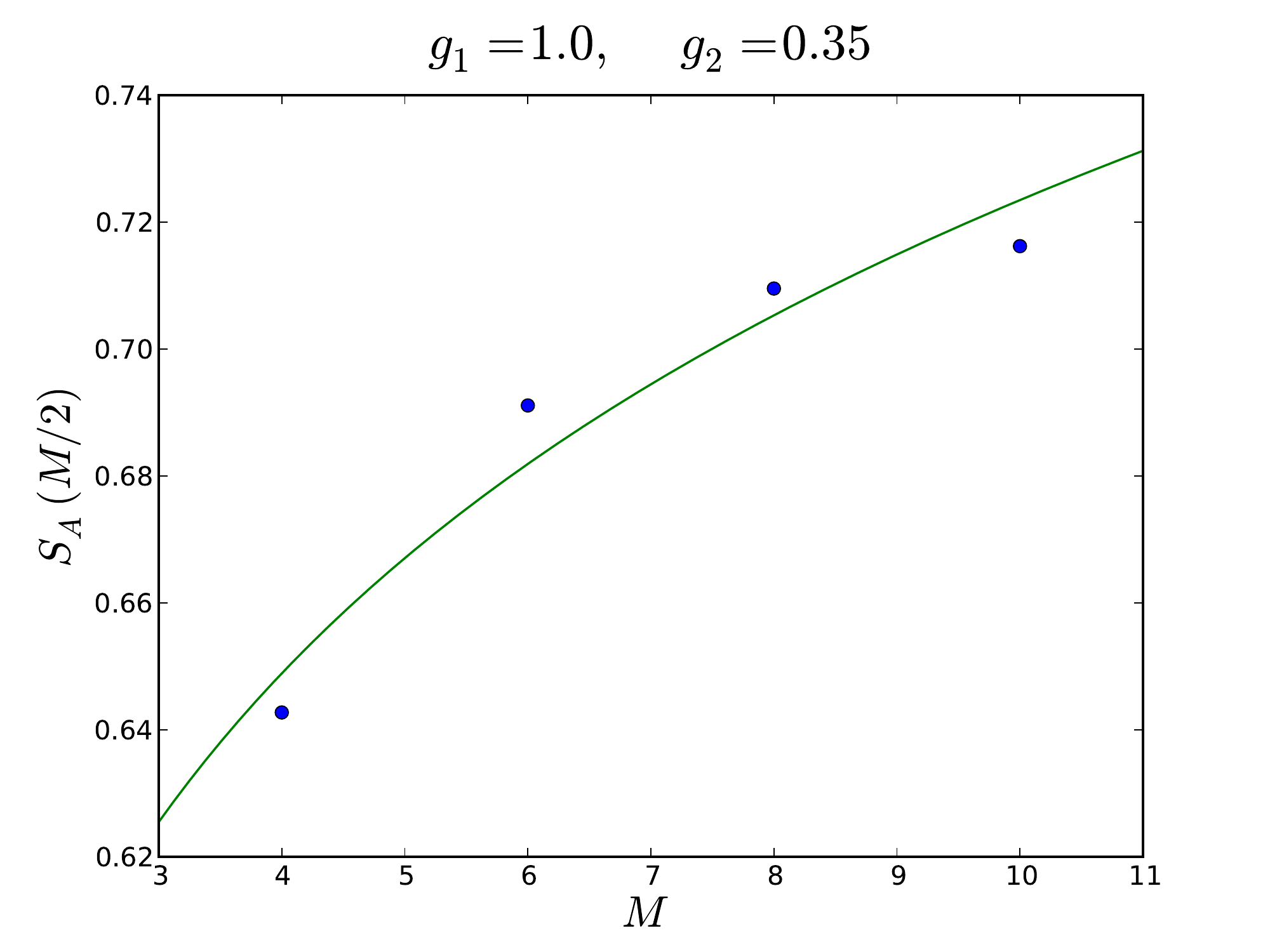}
\end{minipage}
\hspace{4mm}
\begin{minipage}{0.33\textwidth}
\includegraphics[scale=0.32]{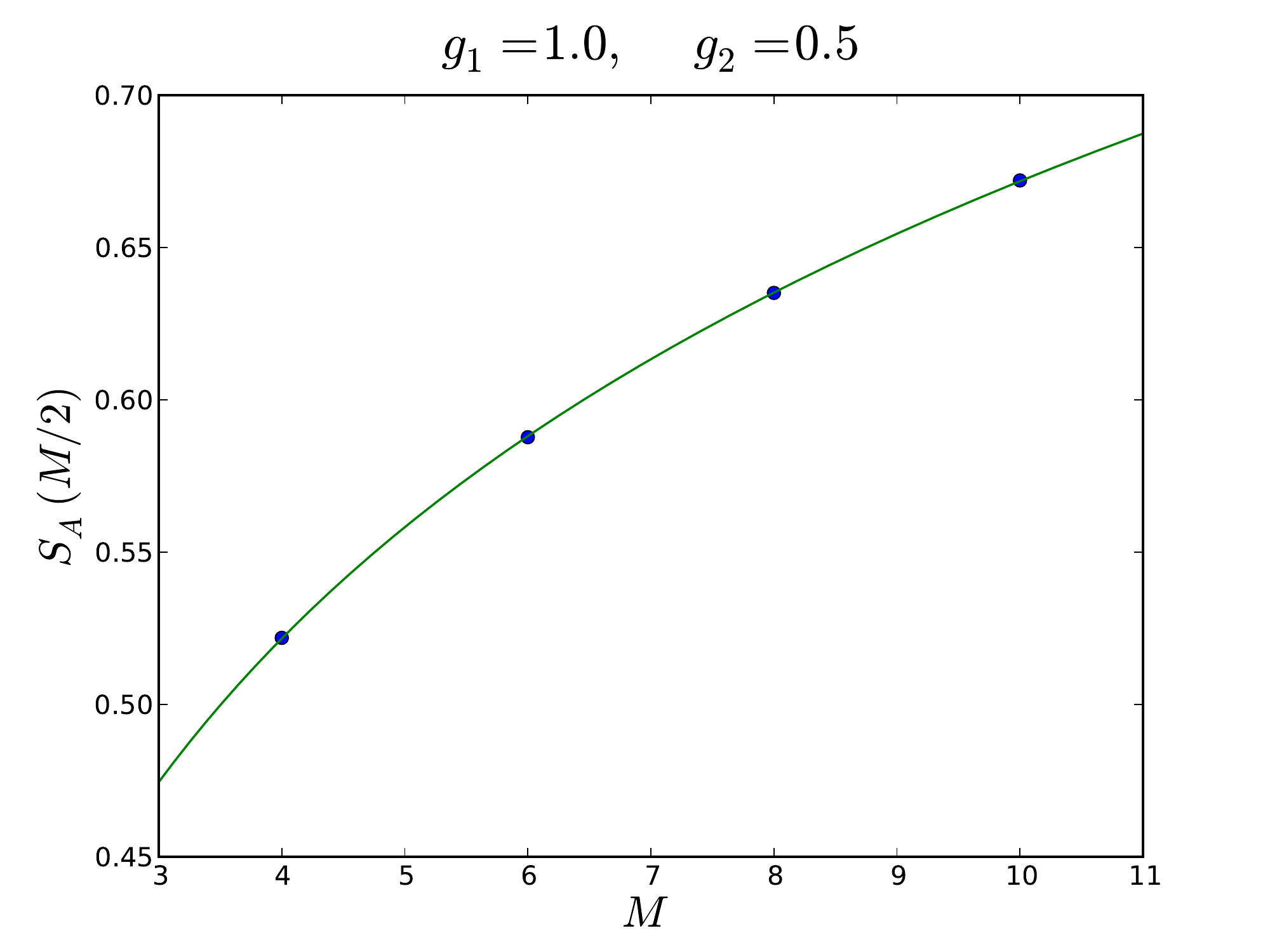}
\end{minipage}
\hspace{4mm}
\begin{minipage}{0.33\textwidth}
\includegraphics[scale=0.32]{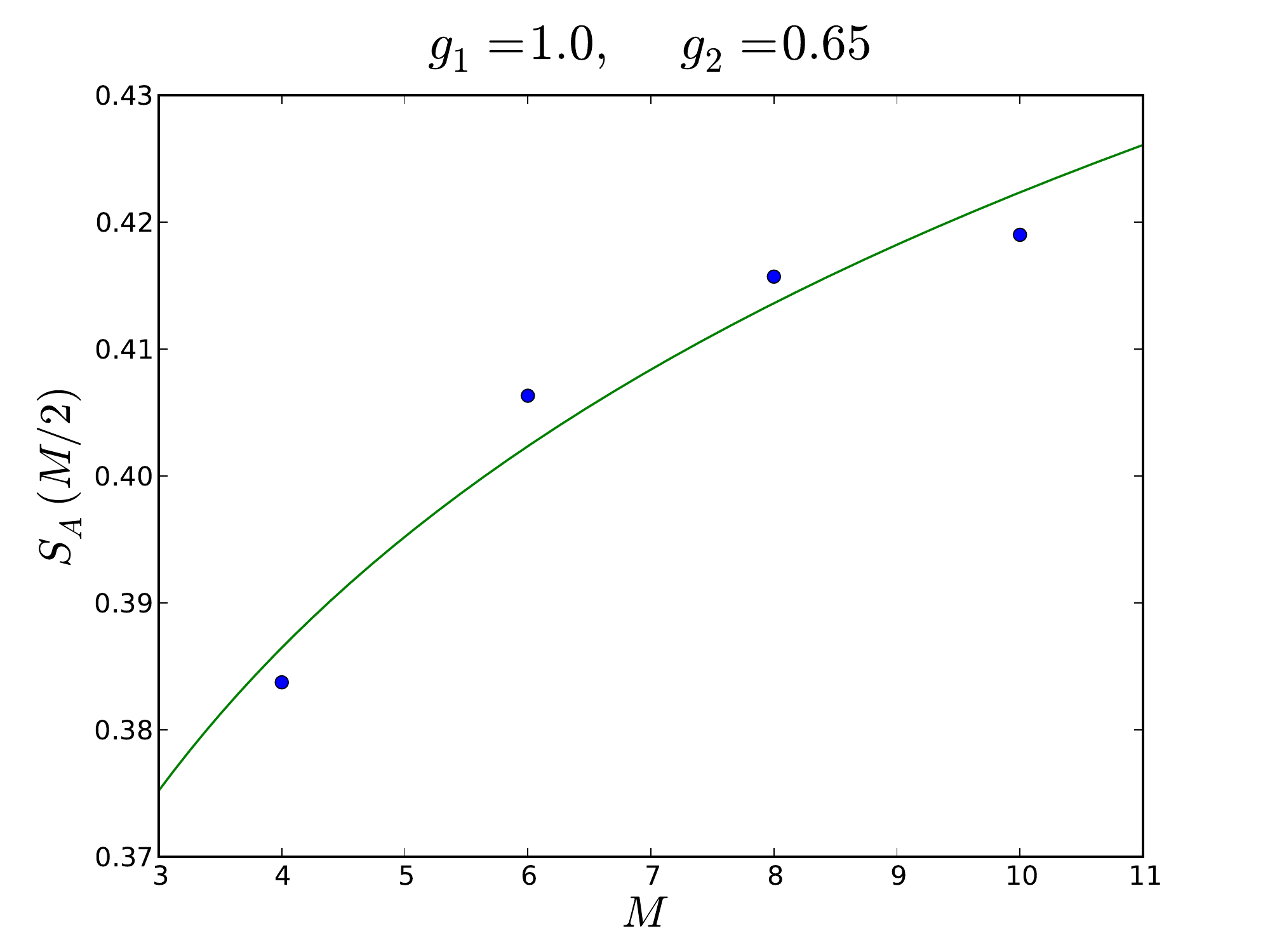}
\end{minipage}}\\
\makebox[0pt][c]{
\hspace{-15mm}
\begin{minipage}{0.33\textwidth}
\includegraphics[scale=0.32]{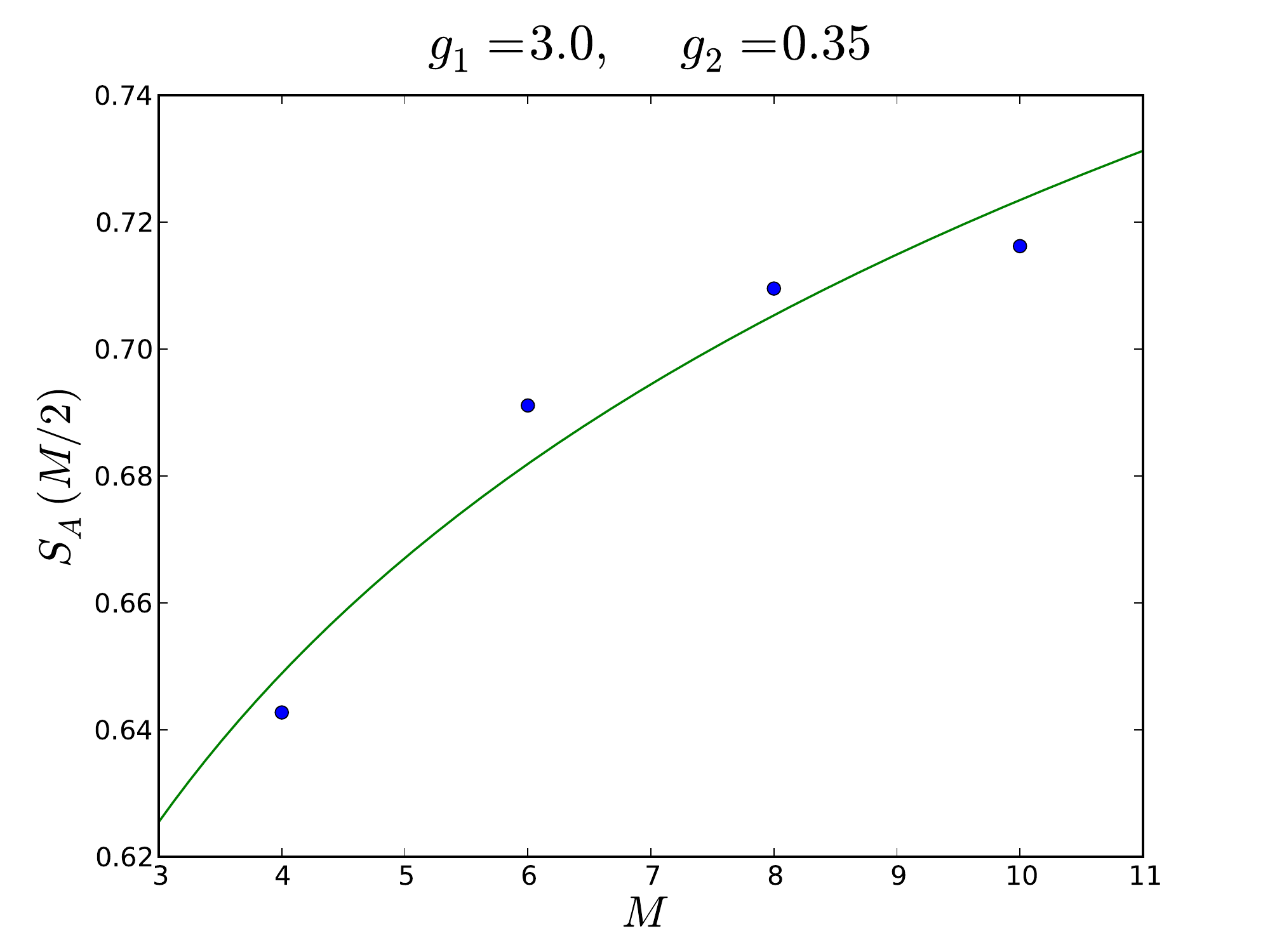}
\end{minipage}
\hspace{4mm}
\begin{minipage}{0.33\textwidth}
\includegraphics[scale=0.32]{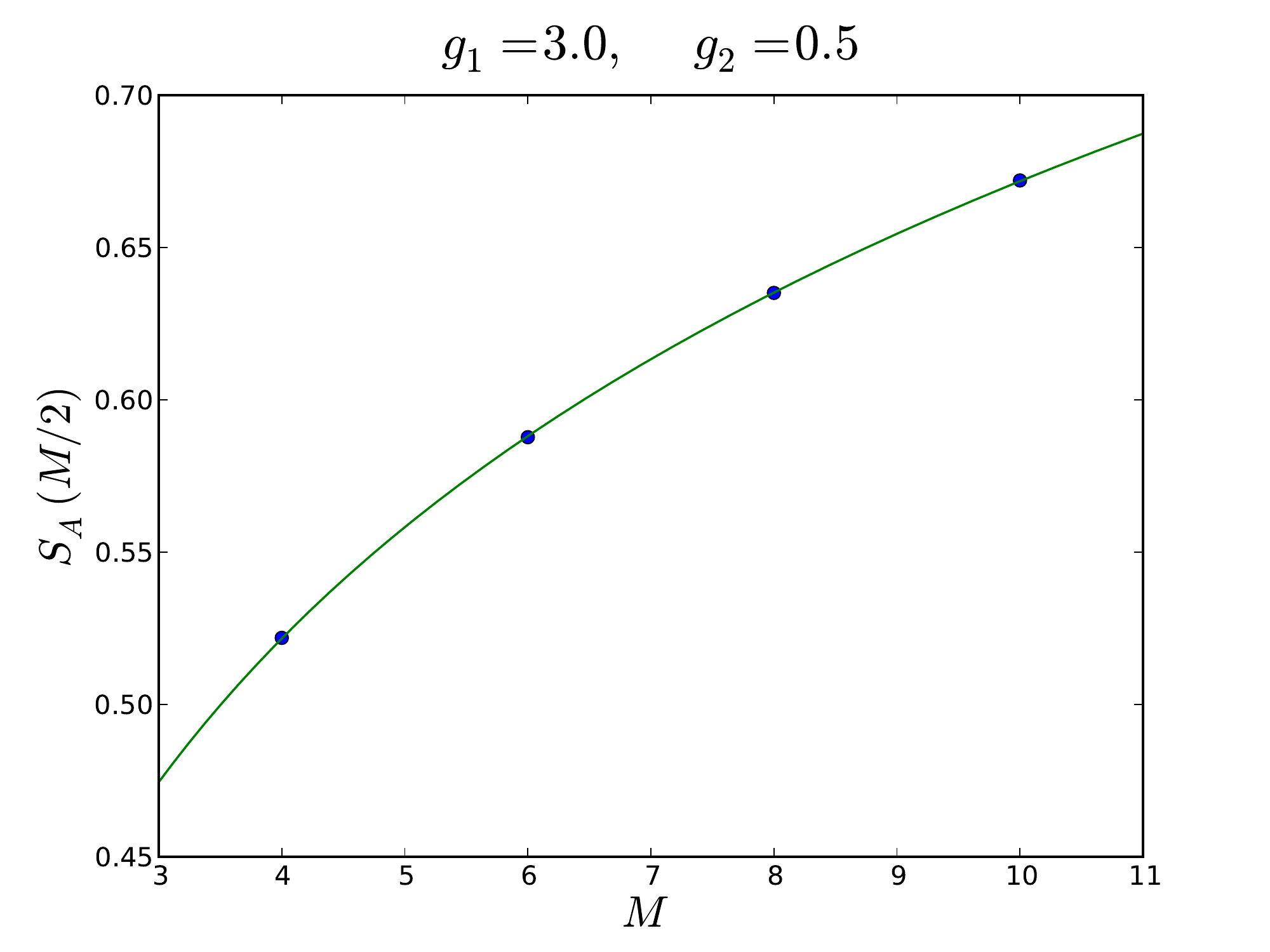}
\end{minipage}
\hspace{4mm}
\begin{minipage}{0.33\textwidth}
\includegraphics[scale=0.32]{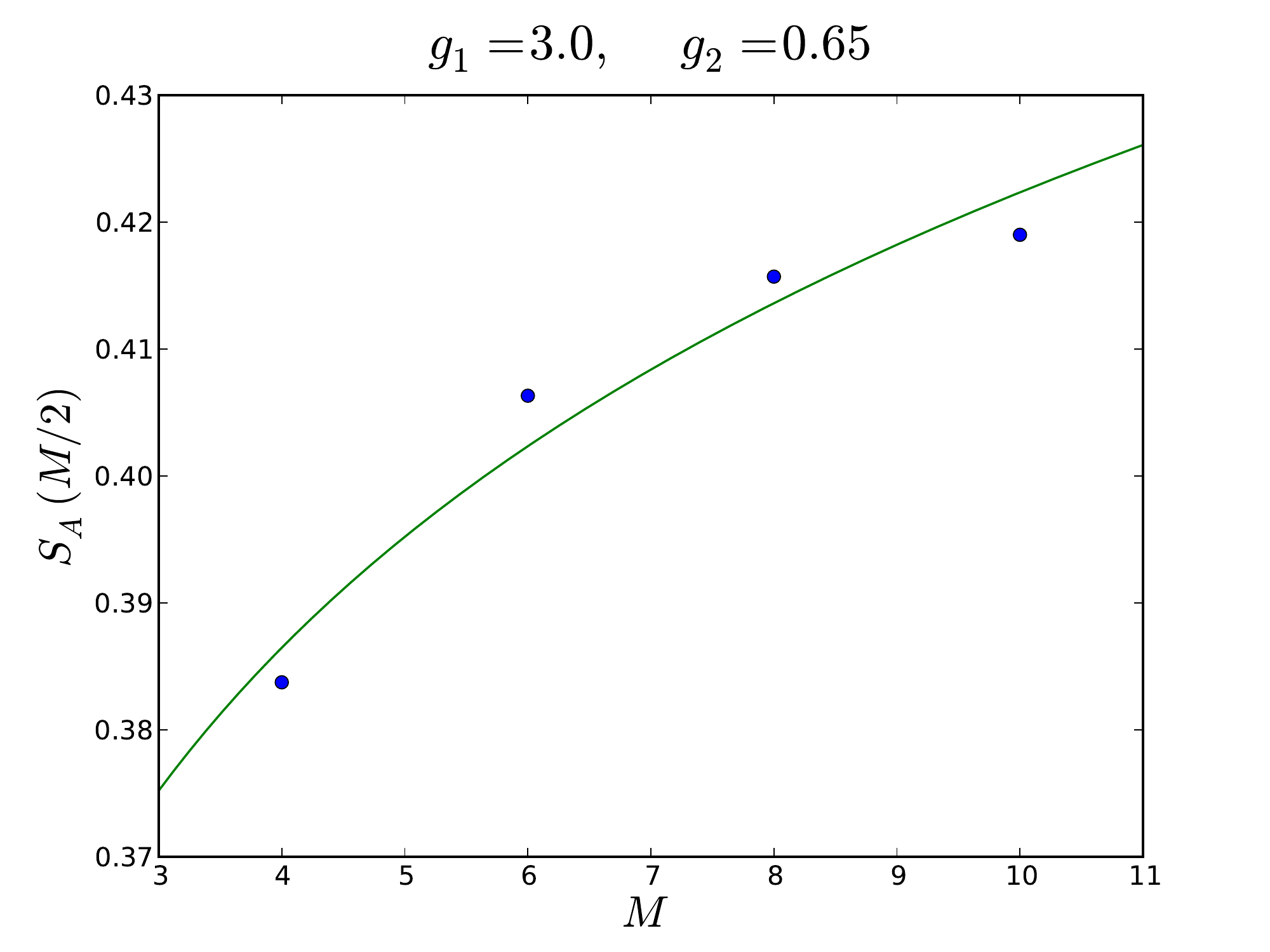}
\end{minipage}}
\caption{{\em Finite size scaling of the entanglement entropy of the coprime
chain for $q=5$ with Hamiltonian (\ref{qh1}) near the transition point $g_2 = 1/2$.
The scaling of the data in the plot in the middle 
fits the entanglement entropy formula \eqref{entent} with $c=0.49..$, when either $g_1=1.0$ or $g_1=3.0$. As soon as $g_2$ detaches from the critical value the entanglement entropy saturates very rapidly to a value proportional to the logarithm of the correlation length.}}
\label{ee2}
\end{figure}
\begin{figure}[t]
\center
\makebox[0pt][c]{
\hspace{-10mm}
\begin{minipage}{0.52\textwidth}
\includegraphics[scale=0.45]{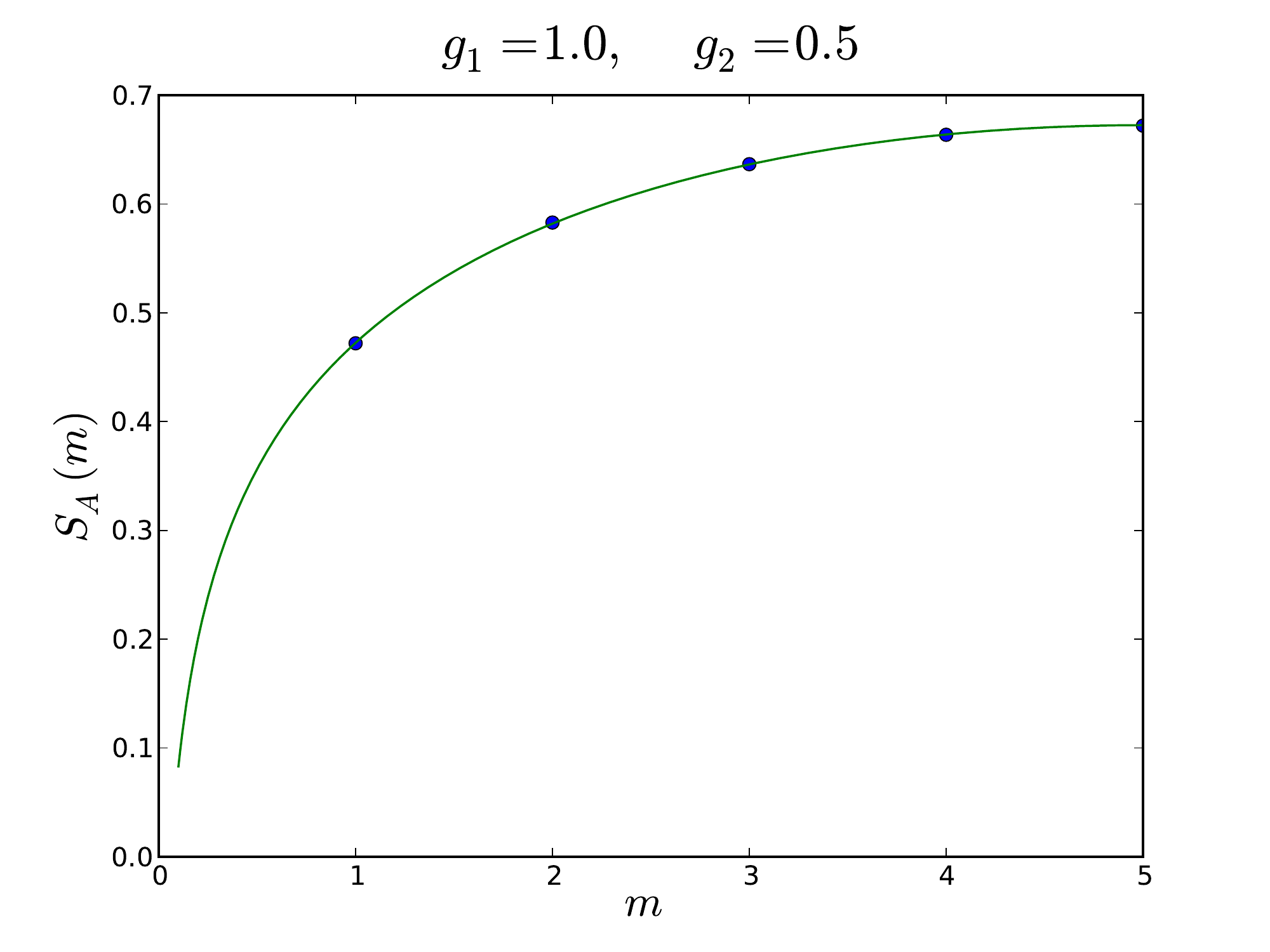}
\end{minipage}
\begin{minipage}{0.53\textwidth}
\includegraphics[scale=0.45]{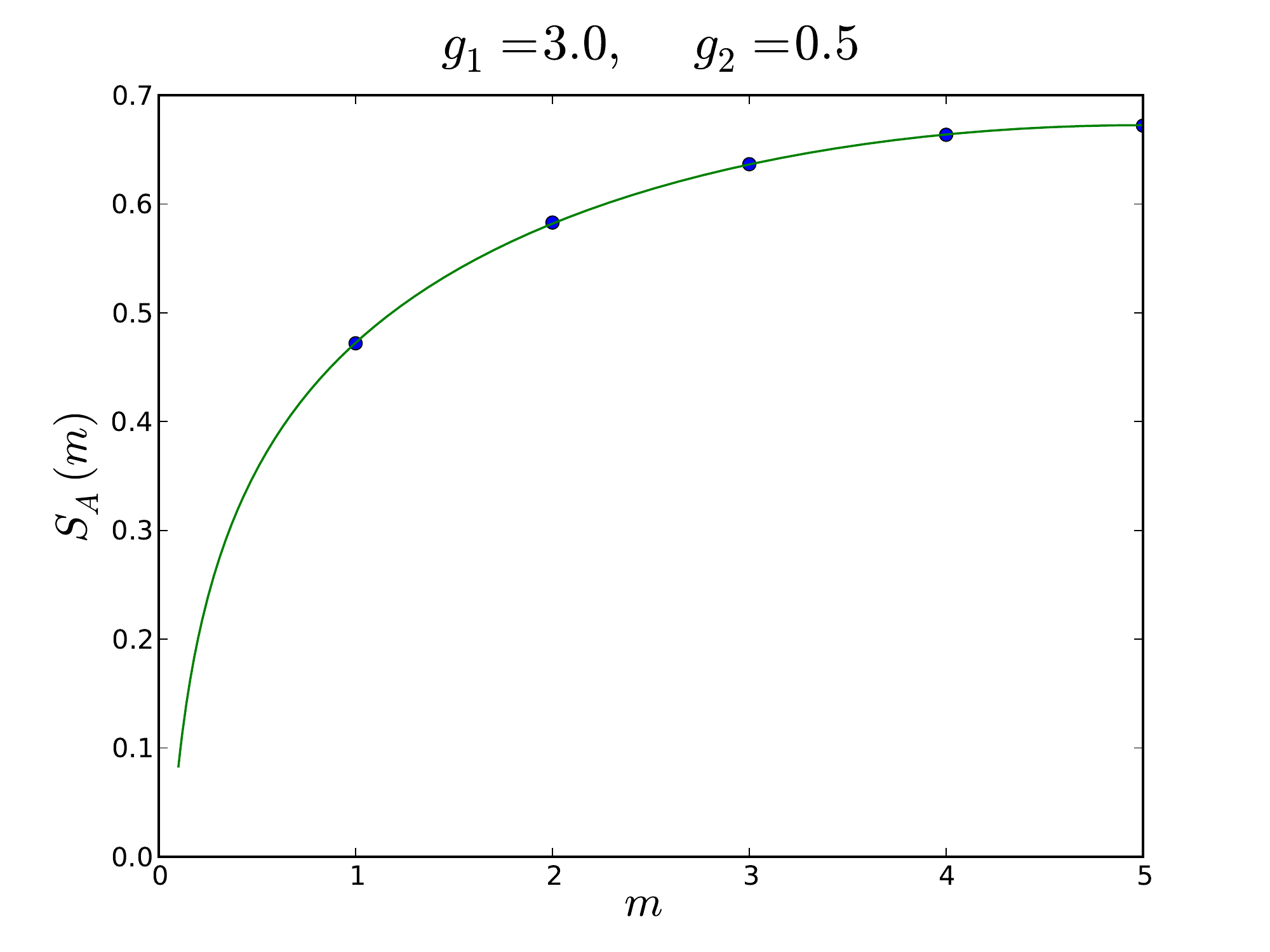}
\end{minipage}}
\caption{{\em Entanglement entropy for the coprime chain $q=5$ with Hamiltonian (\ref{qh1}) and $M=10$ sites, by varying the size $m$ of the subsystem $A$ at the transition 
point $g_2 = 1/2$. The fit with \eqref{entent} produces the value of the central charge $c=0.510$ both when $g_1=1.0$ and when $g_2=3.0$. }}
\label{ee22}
\end{figure}
In summary: starting from the highly degenerate set of ground states of the classical coprime chain with $q=5$, by means of the operators $B^{(1)}_i$ we can firstly 
remove the original degeneracy and remain with only two ground states. Switching on after the other operators $B^{(2)}_i$ and increasing the value $g_2$ 
of their coupling, we can reach a critical point $g_2^{*}$ where the mass gap of the system closes while for $g_2 > g_2^{*}$ there is 
only one ground state. The features just described are the
same of the quantum Ising chain and indeed the numerical determination of the entanglement entropy confirms
that the critical points of \eqref{qh1} and \eqref{quantumisingmodel} are in the same universality class.

\vspace{3mm}
\noindent 
{\bf Universality class of the 3-state Potts chain}.
\begin{figure}[t]
\center
\makebox[0pt][c]{
\hspace{-10mm}
\begin{minipage}{0.52\textwidth}
\includegraphics[scale=0.4]{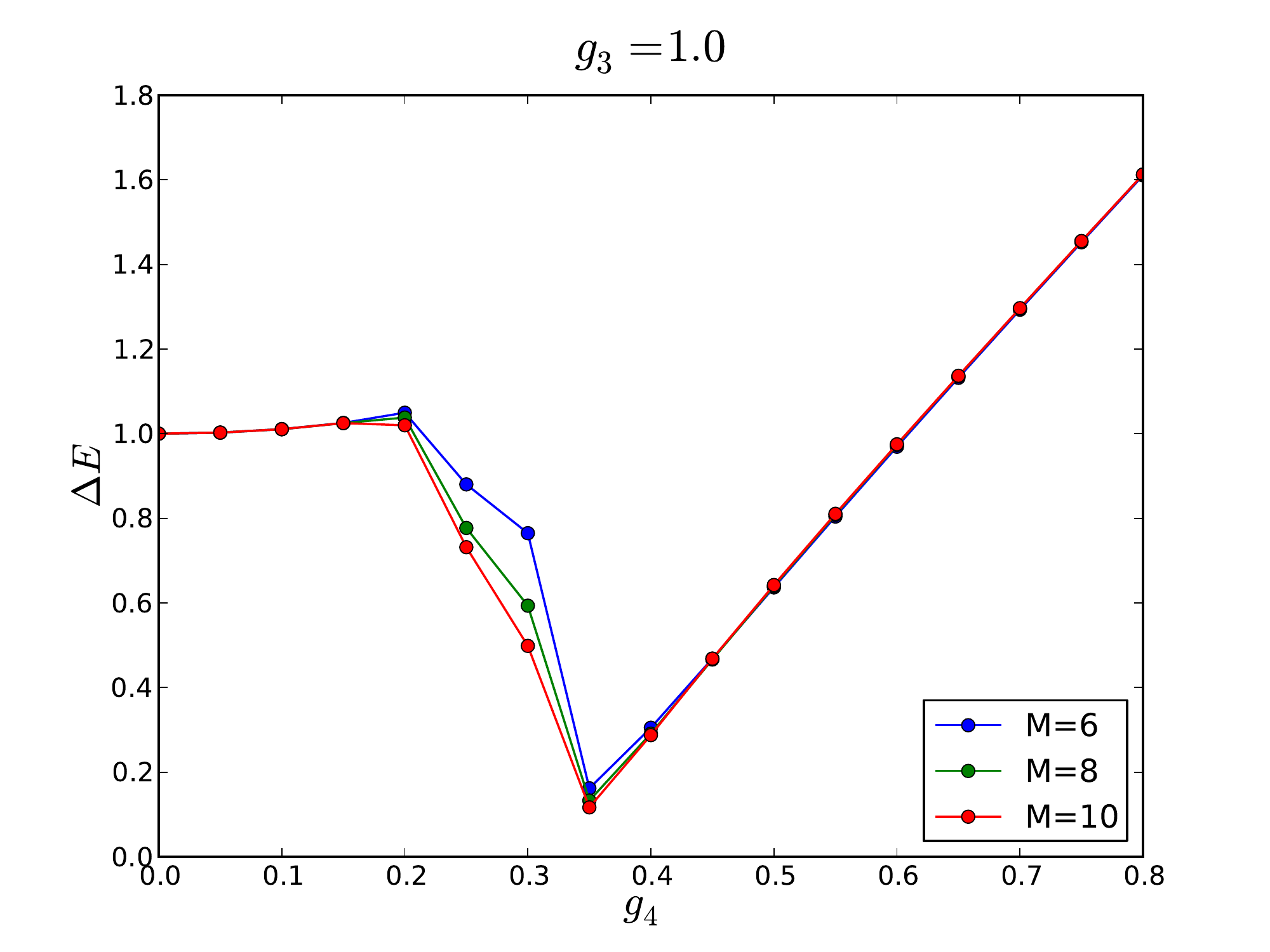}
\end{minipage}
\begin{minipage}{0.53\textwidth}
\includegraphics[scale=0.4]{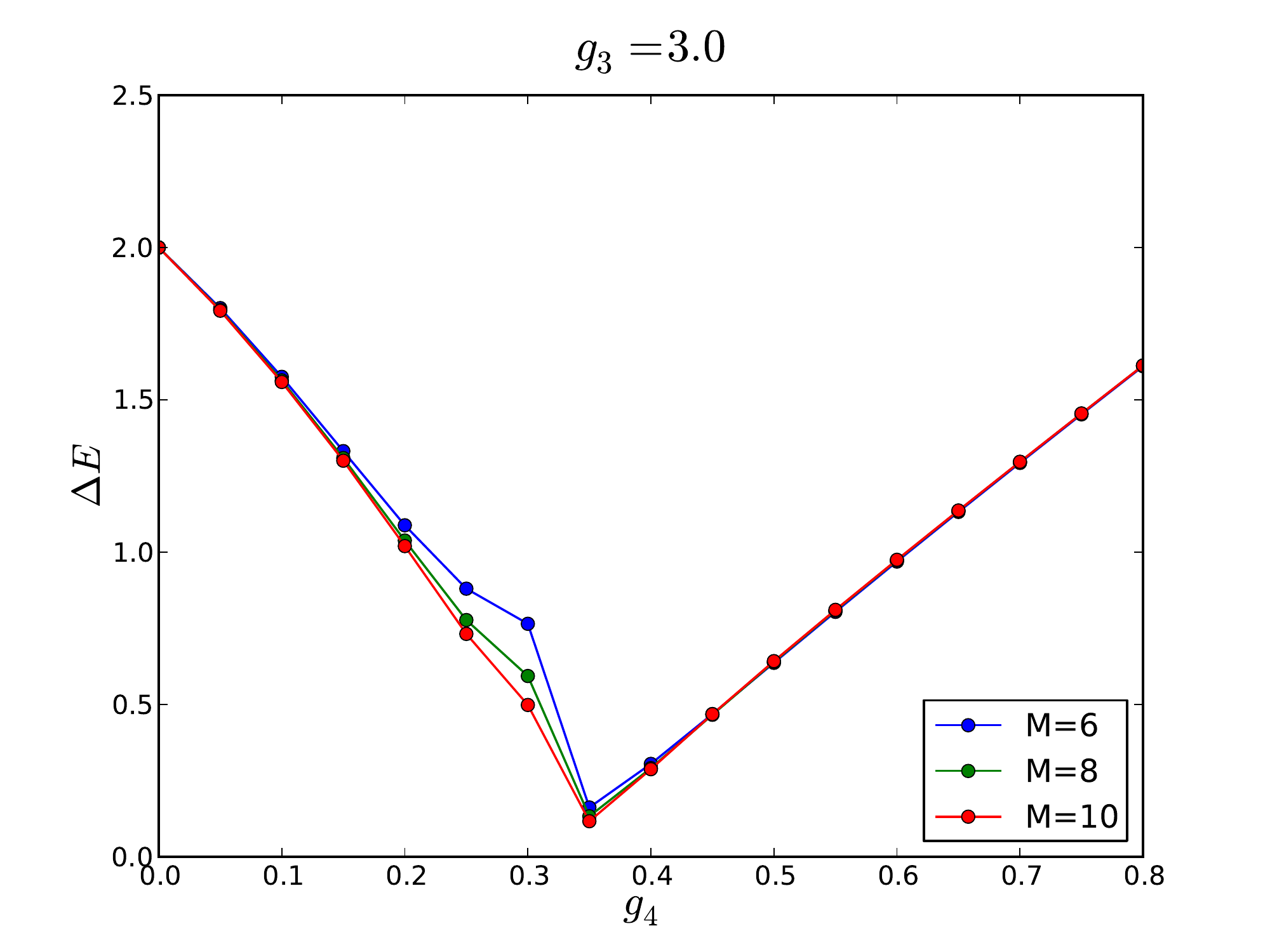}
\end{minipage}}
\caption{{\em Gap of the  quantum coprime Hamiltonian $q=5$ with $B^{(3)}$ and $B^{(4)}$ switched on. The data are for two fixed values of $g_3=1.0$, 
$g_3=3.0$ and varying $g_4$. The finite size scaling shows the closure of the gap when $g_4\simeq 1/3$, independently of the value of $g_3$. }}
\label{gap3}
\end{figure}
Let us now show that it is possible to use different operators in the $q=5$ coprime quantum
chain to reach another critical point, this time associated to the class of universality of the $3$-states Potts model. We briefly remind \cite{Wu} that the 
class of universality of this model consists of two phases: a low temperature phase where there are three equivalent ground states, here
denoted as $| \tilde R\rangle$, $| \tilde G \rangle$ and $| \tilde Y\rangle$ (for Red, Green and Yellow), and an high temperature phase where there is
an unique ground state, here denoted by $| \tilde W \rangle$ (for White). The two phases are separated by a critical point where the mass gap closes. 
Such a scenario is encoded into the quantum Hamiltonian
symmetric under the permutation group $S_3= \mathbb{Z}_3\times\mathbb{Z}_2$ \cite{HamquantumPotts}  
\begin{equation}
H_{3Potts} \,=\, - \sum_i\left[\hat \tau_i \hat\tau^\dagger_{i+1} +  \hat \tau_i^\dagger \hat\tau_{i+1} + g \hat \eta_i\right]\,\,\,,
\label{3PottsH}
\end{equation}
where the operators $\hat\tau_i$ and $\hat \eta_i$ have the general form of eq\,(\ref{notation}) and are expressed in terms of the matrices 
\begin{equation}
\tau = \left(
\begin{array}{ccc}
1 & 0 & 0 \\
0 & e^{2 \pi i/3} & 0 \\
0 & 0 & e^{4 \pi i/3} 
\end{array}
\right) 
\,\,\,\,\,\,\,\,
, 
\,\,\,\,\,\,\,\,
\eta = \left(\begin{array}{lll}
0 & 1 & 1 \\
1 & 0 & 1 \\
1 & 1 & 0 
\end{array}
\right) 
\,\,\,.
\end{equation}
For $g \rightarrow 0$, the so-called low-temperature phase, there are three degenerate ground states of the Hamiltonian (\ref{3PottsH}) 
expressed in terms of the eigenvectors $| R\rangle$, $| G \rangle$ and $| Y\rangle$ of the $\tau$ matrix   
\begin{equation}
|\tilde R \rangle \,=\, \otimes_i | R \rangle_i 
\,\,\,\,\,\,\,
,
\,\,\,\,\,\,\,
|\tilde G \rangle \,=\, \otimes_i | G \rangle_i 
\,\,\,\,\,\,\,
,
\,\,\,\,\,\,\,
|\tilde Y \rangle \,=\, \otimes_i | Y \rangle_i \,\,\,.
\end{equation}
For $g \rightarrow \infty$, the so-called high-temperature phase, there is instead an unique ground state fully symmetric under the $S_3$ group 
\begin{equation}
| \tilde W\rangle \,=\,\otimes_i | W \rangle_i 
\,\,\,\,\,\,\,
,
\,\,\,\,\,\,\,
| W\rangle_i \,=\,\frac{1}{\sqrt 3}(| R \rangle_i + | G \rangle_i + | Y \rangle_i) 
\,\,\,. 
\end{equation}
Between the low and high temperature phase there is a phase transition which occurs for the critical value $g = 1$. At the critical point the model
is described by a conformal field theory with central charge  $c=4/5$ \cite{dots}. It is worth to underline that, contrary to the Ising chain,
the 3-state Potts chain with  Hamiltonian 
(\ref{3PottsH}) cannot be solved exactly.

Let us now see how we can realise such class of universality in terms of the $q=5$  coprime quantum chain. First of all,
we can add to the classical Hamiltonian of the model \eqref{classham} the operators $B^{(3)}_i$ made by the one-site matrix ${\cal B}^{(3)}$ 
\be 
{\cal B}^{(3)} = 
\begin{pmatrix}
1 & 0 & 0 & 0 \\
0 & 1 & 0 & 0 \\
0 & 0 & 0 & 0 \\
0 & 0 & 0 & 1 
\end{pmatrix} \equiv {\mathcal D}^{(1)} + {\mathcal D}^{(2)} + {\mathcal D}^{(4)}\,\,\,.  
\ee
The presence of the magnetic fields ${\mathcal D}^{(1)}$, ${\mathcal D}^{(2)}$ and ${\mathcal D}^{(4)}$ into the quantum coprime Hamiltonian 
\be 
H \,=\,-\sum_{i=1}^M \left[\Phi(n_i,n_{i+1})\, + 
g_{3} B^{(3)}_i   \right] \,\,\,
\label{quantumhamiltonianB3}
\ee
immediately reduces the exponentially large degeneracy of its classical ground states to just three states, given by  
\be \label{inertgs}
| \tilde 2 \rangle = \ket{2\,2\,2\,2\,\dots 2}
\qquad
,
 \qquad
| \tilde 3 \rangle = \ket{3\,3\,3\,3\,\dots 3}
\qquad 
,
\qquad
| \tilde 5 \rangle = \ket{5\,5\,5\,5\, \dots 5} \,\,\,.
\ee
These states can be put in correspondence with the three degenerate ground states $| \tilde R \rangle$, $| \tilde G \rangle$ and $| \tilde Y\rangle$ of the 3-state Potts model. 

Next, we can switch on the additional operators $B^{(4)}_i$ whose associated one-site matrix is the linear combination   
\be
{\cal B}^{(4)} = 
\begin{pmatrix}
0 & 1 & 0 & 1 \\
1 & 0 & 0 & 1 \\
0 & 0 & 0 & 0 \\
1 & 1 & 0 & 0 
\end{pmatrix}\equiv {\mathcal S}^{(12)} + {\mathcal S}^{(14)} + {\mathcal S}^{(24)} \,\,\,.
\ee
These operators mix symmetrically on each site the occupation number $\ket{2},\ket{3}$ and $\ket{5}$ and therefore we expect that increasing the value of their coupling constant $g_4$
in the quantum Hamiltonian 
\be 
H \,=\,-\sum_{i=1}^M \left[\Phi(n_i,n_{i+1})\, + 
g_{3} B^{(3)}_i  + g_4 B^{(4)}_i \right] \,\,\, 
\label{quantumhamiltonian34}
\ee
we shall meet a quantum phase transition.
This is indeed the case and numerically we estimated that the model is critical for $g_4^{*} = 1/3$, irrespectively of the value of  $g_3$.
As in the Ising case, the mass gap of the chain closes for such a value of the coupling and the central charge extracted
at this critical point from the entanglement entropy is perfectly compatible with the value $c=4/5$ of the 3-state Potts model.
The numerical results are reported in Fig.~\ref{gap3} and Fig.~\ref{ee3}. For $g_4 > g_4^{*}$, the original three ground states disappear and the system presents only
one ground state, exactly as the physical scenario of the 3-state Potts model.


\begin{figure}[t]
\center
\makebox[0pt][c]{
\hspace{-10mm}
\begin{minipage}{0.52\textwidth}
\includegraphics[scale=0.45]{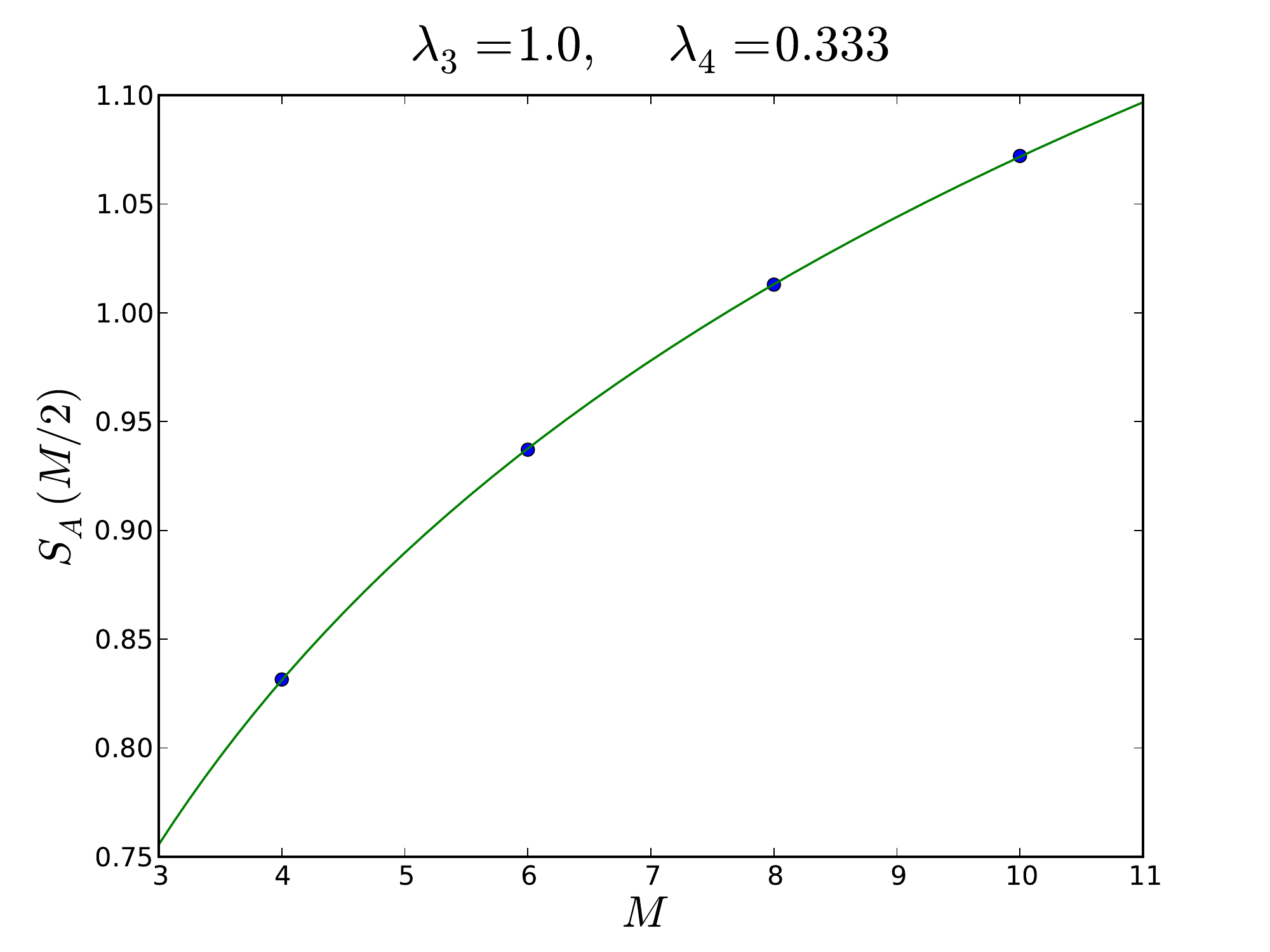}
\end{minipage}
\begin{minipage}{0.53\textwidth}
\includegraphics[scale=0.45]{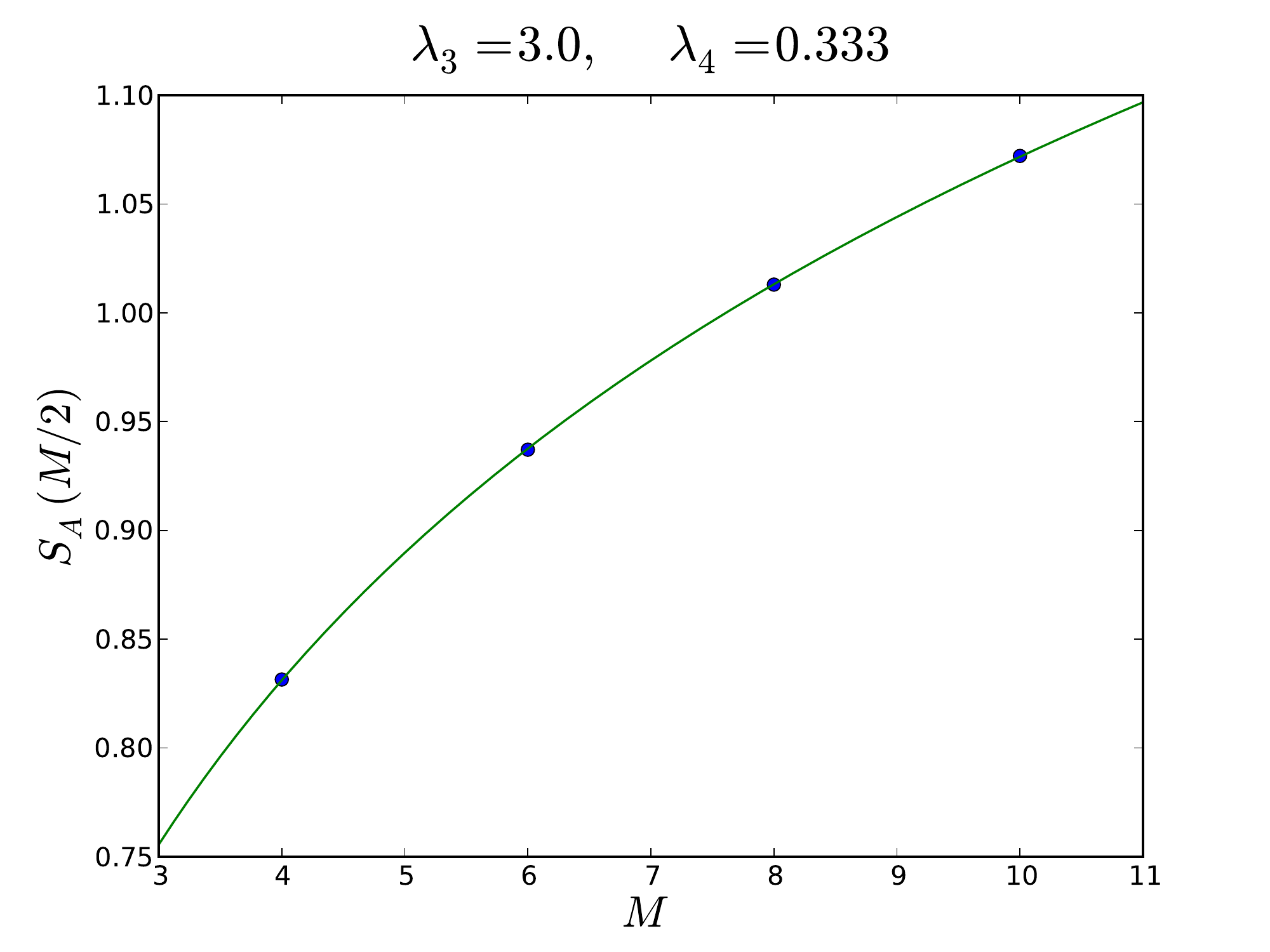}
\end{minipage}}
\caption{{\em Finite size scaling of the entanglement entropy of the $q=5$ coprime chain at the transition point $g_4 = 1/3$. The fit with \eqref{entent} gives a central charge $c=0.7876,,$ independently of $g_3$. }}
\label{ee3}
\end{figure}

\begin{figure}[H]
\center
\makebox[0pt][c]{
\hspace{-10mm}
\begin{minipage}{0.52\textwidth}
\includegraphics[scale=0.45]{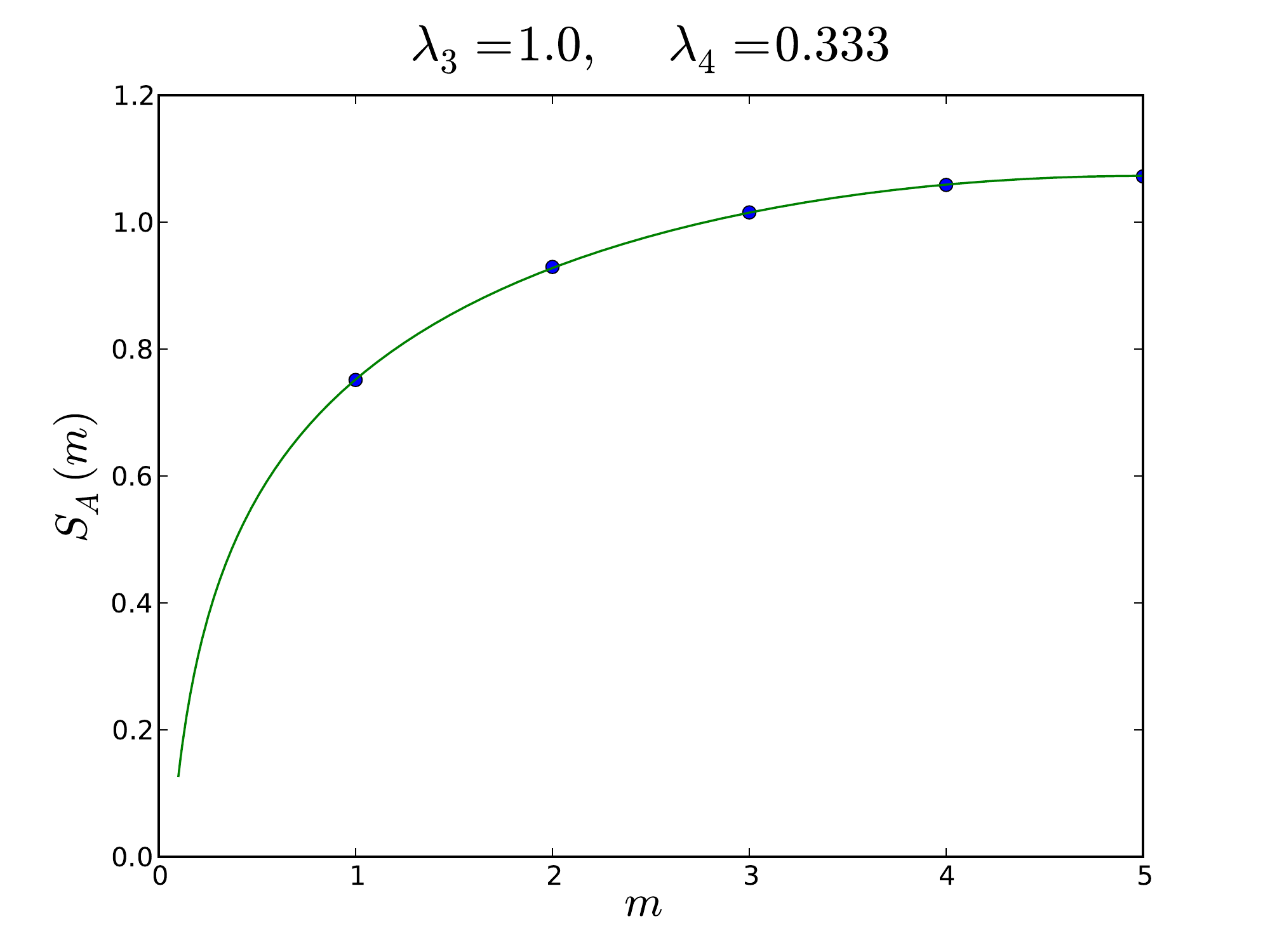}
\end{minipage}
\begin{minipage}{0.53\textwidth}
\includegraphics[scale=0.45]{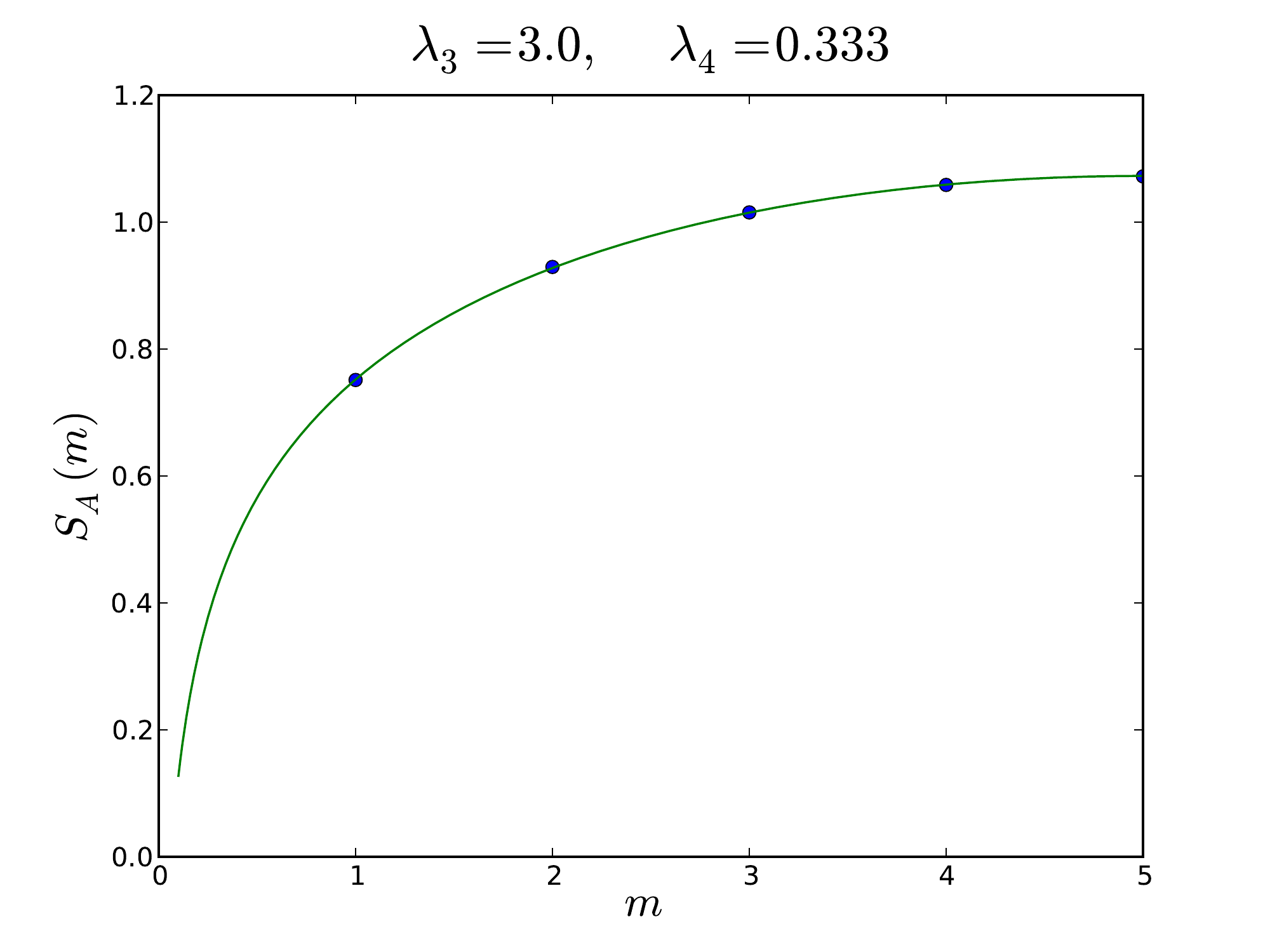}
\end{minipage}}
\caption{{\em Entanglement entropy for the $q=5$ coprime chain of $M=10$ sites for the two values $g_3=1.0$, $g_3=3.0$, varying the size $m$ of the subsystem $A$ at the transition point $g_4 = 1/3$. The fit with \eqref{entent} produces the value of the central charge $c=0.8194..$ both when $g_3=1.0$ and when $g_3=3.0$. }}
\label{ee32}
\end{figure}

\vspace{3mm}
\noindent 
{\bf No quantum phase transitions with an exponential number of ground states}. In the previous examples, making use of appropriate operators we have first 
reduced the exponentially large number of ground states of the classical coprime chain~\eqref{classical} to a {\em finite} value. The final degeneracy could be then completely lift by another operator, a phenomenon that leads eventually to a quantum phase transition. A natural question is now: what happens if we only {\em partially} reduce the original degeneracy of the coprime chain,
still remaining with an exponentially large number of ground states that can be further perturbed? Does the system reach criticality or not?
Let us examine the $q=5$ coprime chain once we add to its classical Hamiltonian the operators $ B^{(5)}_i$ containing the one-site  matrices
\be 
{\cal B}^{(5)} = 
\begin{pmatrix}
1 & 0 & 0 & 0 \\
0 & 0 & 0 & 0 \\
0 & 0 & 1 & 0 \\
0 & 0 & 0 & 0 
\end{pmatrix} \equiv {\mathcal D}^{(1)} + {\mathcal D}^{(2)}\,\,\,. 
\ee
The two magnetic operators ${\mathcal D}^{(1)}$ and ${\mathcal D}^{(2)}$ privilege the occupation numbers $\ket{2}$ and $\ket{4}$ and therefore they remove only the  states $|\tilde 3 \rangle = \ket{3\,3\,3\, \dots 3 }$ and $| \tilde 5\rangle =\ket{5\,5\,5\, \dots 5 }$ from the infinite set of the classical ground states. 

We can still mix the (exponentially degenerate) ground states left  by means of the operators $B^{(6)}_i$
expressed in terms of the one-site matrix 
\be
{\cal B}^{(6)} = 
\begin{pmatrix}
0 & 0 & 1 & 0 \\
0 & 0 & 0 & 0 \\
1 & 0 & 0 & 0 \\
0 & 0 & 0 & 0 
\end{pmatrix}\equiv {\mathcal S}^{(13)} \,\,\,. 
\ee
The final Hamiltonian is 
\be 
H \,=\,-\sum_{i=1}^M \left[\Phi(n_i,n_{i+1}) + 
g_{5} B^{(5)}_i  + g_6 B^{(6)} \right] \,\,\,.
\label{quantumhamiltonian56}
\ee
Will be possible varying the corresponding coupling constant $g_6$  to reach now a quantum phase transition? 
The answer is negative: contrary to what happened in the Ising and Potts chains this time
the ordered phase  characterised by the exponential ground state degeneracy is completely unstable under the mixing term $B^{(6)}_i$, namely it disappears for {\em arbitrarily small} values of $g_6$. This is shown in in Fig.~\ref{gap4} where we computed the mass gap of the theory.
\begin{figure}[H]
\center
\includegraphics[scale=0.5]{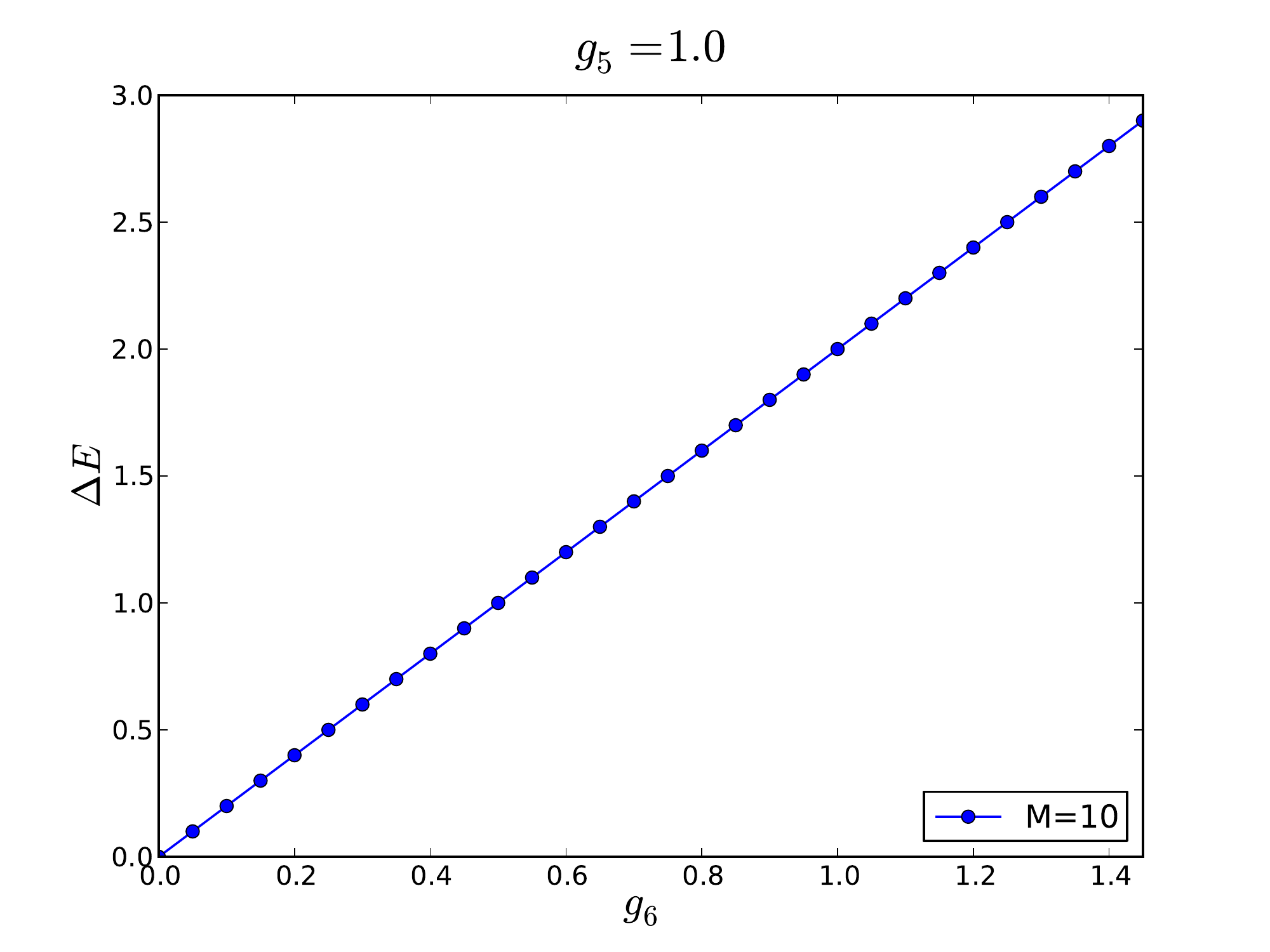}
\caption{{\em Gap of the $q=5$ quantum coprime Hamiltonian with the operators $B^{(5)}_i$  and $B^{(6)}$ switched on, for fixed $g_5=1.0$ and varying 
$g_6$, on a chain of $M=10$ sites. A gap opens immediately as soon as $g_6$ is non-zero. Moreover $\Delta E$ depends linearly on $g_6$ as $\Delta E = 2 g_6$, indicating that perturbation theory is exact at its first order. }}
\label{gap4}
\end{figure}
\noindent
The non-existence of a stable low-temperature phase under the switching of $g_6$ can be explained already at first order in perturbation theory, 
considering the term with $\sum_i B^{(6)}_i$ as a perturbation of the Hamiltonian $H = -\sum_i \left[\Phi(n_i,n_{i+1}) + g_5 B^{(5)}\right]$ 
\be \label{pert}
\delta H_p = -g_6 \sum_{i=1}^M B^{(6)}_i \,\,\,.
\ee
It is easy to compute the matrix associated to this perturbation in the $2^M$-degenerate ground state subspace,
composed of all the factorized states which are product of $\ket{2}$ and $\ket{4}$: apart
from the overall factor $(-g_6)$, such a matrix -- which is the one that determines the splitting of this energy level -- is nothing
but the adjacency matrix of a regular graph of degree $M$. 
Indeed, acting with \eqref{pert} on a state that contains  $2$'s and $4$'s, 
one obtains $M$ different states belonging to the same degenerate subspace\footnote{The states are obtained exchanging in only one of the $M$ possible site a $2$ with a
$4$ and vice-versa.}. Then each row of the perturbation in this subspace will contain $M$ non-zero entries, all equal to $(-g_6)$ and 
the remaining $2^M - M$ entries equal to zero. Since the regular graph associated to this matrix is also connected,
it follows, via the Perron-Frobenius theorem, that the lowest eigenvalue is unique and equal to $-M g_6$. This implies that the first order 
correction completely removes the ground state degeneracy, explaining the sudden opening of the gap as soon as $g_6 \ne 0$. It is worth noticing
that this behaviour is in contrast to what happens when the ground state subspace has only a finite degeneracy in the $M \to \infty$ limit, as in the
case of the Ising and the $3$-state Potts chains.
In these latter models, the perturbing operator has only zero entries in the two-fold and three-fold degenerate subspaces 
relative to the lowest eigenvalue of the unperturbed Hamiltonian: thus degeneracy is not lifted in first-order perturbation theory. 

Although the graph theory argument given above is pretty elegant and concise, it gives no information on the gap of $\delta H_p$ in the thermodynamic limit. Indeed one could think that the spectral gap of a regular graph might even close
when the number of vertices goes to infinity. However in this case it is easy to write down the whole spectrum of $\delta H_p$ in the degenerate subspace for every $M$. First observe that $H_p$ restricted to the ground state subspace is simply given by
\be 
\left. \delta H_p \right|_{GS} = -g_6 \sum_{i=1}^M \sigma^x_i \,\,\,, 
 \ee
where $\sigma_x$ is the usual Pauli matrix whose eigenvectors will be denoted
\be 
\ket{ + } = \frac{1}{\sqrt{2}} \begin{pmatrix}
1 \\ 1
\end{pmatrix}
\qquad \qquad \qquad 
\ket{ - } = \frac{1}{\sqrt{2}} \begin{pmatrix}
\,\,1 \\ -1
\end{pmatrix}\,\,\,.
\ee
Then we can construct the spectrum of $\delta H_p$ starting form the product state $\ket{ + \, + \, + \, \dots + }$, which is the unique eigenstate associated to the lowest eigenvalue $-g_2 M$, and flipping one spin at a time. In this way
it is easy to realise that all the eigenvalues are organised as
\begin{align*}
 & E_{ + \, + \, + \, \dots + } = -g_6 M  & \qquad    \qquad &  1-\mathrm{fold~degenerate} & \\[2.5mm]
 & E_{ + \, + \, \dots  + \, - \, + \dots + } = -g_6 (M - 2)  & \qquad  \qquad & M-\mathrm{fold~degenerate} & \\[1mm]
 & E_{ + \, + \, \dots  + \, - \, - \, + \dots + } = -g_6 (M - 4)  & \qquad  \qquad &  \binom{M}{2}-\mathrm{fold~degenerate} & \\
 &  \qquad \qquad \vdots &  &  \qquad  \qquad  \vdots  & \\
 & E_{ + \, + \, \dots  + \, -  \, \dots \, - \, + \dots + } = -g_6 (M - 2 p)&  \qquad   \qquad & \binom{M}{p}-\mathrm{fold~degenerate} & \\
 &  \qquad \qquad \vdots &  &  \qquad  \qquad  \vdots  & \\[1.5mm]
 & E_{ - \, - \, - \, \dots - } = g_6 M  & \qquad    \qquad &  1-\mathrm{fold~degenerate} & 
\end{align*}
Thus the gap is given by $\Delta E = 2 g_6$ for any $M$. Moreover from the data in Fig.~\ref{gap4} we can see that this does hold to all order.
In conclusion, in presence of the two set of operators $B^{(5)}$ and $B^{(6)}$ there is only one stable phase of chain, its high temperature phase, 
and therefore we cannot have phase transition.  

\section{Classical two-dimensional model and Hamiltonian limit}\label{classicalcoprimeprime}
In this section we describe how to identify the quantum coprime Hamiltonian which is associated to the homogeneous classical two-dimensional coprime model with Hamiltonian defined later in eq.~\eqref{classham}, compare also with eq.~\eqref{classical}. We will also examine how to use this mapping in order to infer some properties of the spectrum of the coprime quantum chain.

 For this purpose consider the operators $B^{(7)}_i$ and 
$B^{(8)}_i$ expressed in terms of the matrices 
\be
{\cal B}^{(7)} = 
\begin{pmatrix}
0 & 1 & 0 & 1 \\
1 & 0 & 1 & 1 \\
0 & 1 & 0 & 1 \\
1 & 1 & 1 & 0 
\end{pmatrix} \equiv {\mathcal S}^{(12)} + {\mathcal S}^{(14)} + {\mathcal S}^{(23)} + {\mathcal S}^{(24)} 
\qquad 
,
\qquad
{\cal B}^{(8)} = 
\begin{pmatrix}
0 & 0 & 1 & 0 \\
0 & 0 & 0 & 0 \\
1 & 0 & 0 & 0 \\
0 & 0 & 0 & 0 
\end{pmatrix} \equiv {\mathcal S}^{(13)}.
\ee
The close expressions of these two matrices for general values of $q$ can be written as follows 
\be \label{hamlimbs}
[{\cal B}^{(7)}]_{a,b}  = 1 - \Phi_{a,b}
\qquad \qquad \qquad
[{\cal B}^{(8)}]_{a,b} = \Phi_{a,b} - \delta_{a,b} \,\,\,.
\ee
Let us show that the coprime quantum chain with these operators included has an Hamiltonian related to the homogeneous
two-dimensional classical coprime model~\eqref{classham} on a square lattice. This correspondence is via the so called Hamiltonian limit \cite{kogut}. The $2$d classical coprime model is defined by the classical two-dimensional Hamiltonian
\be \label{classham}
\mathcal{ H } ( \{ \sigma \} ) \,=\, - J \sum_{ \langle i , j \rangle } \Phi ( \sigma_i , \sigma_j )
\qquad ,
\qquad
\sigma = 2,3,\dots q \,\,\,,
\ee
that is an obvious generalization of eq.~\eqref{classical}.
The form \eqref{hamlimbs} of the transverse operators comes out starting from the quantum Hamiltonian and \lq\lq Trotterizing'' 
the finite temperature quantum partition function
\begin{align}
\mathrm{Tr} \( e^{-\beta H } \) & \underset{ M \to \infty}{ \sim }
\sum_{ \sigma_1 \dots \sigma_N }
\braket{ \sigma_1 \dots \sigma_N }{ \prod_{i=1}^M e^{ \frac{\beta}{M} \sum_{j=1}^N \Phi(n_j,n_{j+1})} \, e^{  \frac{\beta}{M} \sum_{j=1}^N ( g_7 B^{(7)}_j + g_8 B^{(8)}_j ) } }{  \sigma_1 \dots \sigma_N } =  \nonumber 
\\[1mm]
 &  =  \sum_{ \{ \sigma_{i,j} \}_{i,j=1}^{M,N} }
 e^{ \frac{\beta}{M} \sum_{i,j=1}^{M,N} \Phi (\sigma_{i,j}\sigma_{i,j+1}) }
 \prod_{i=1}^M \( \otimes_{j=1}^{N}  \bra{  \sigma_{i,j} } \)
 { e^{  \frac{\beta}{M} \sum_{j=1}^N ( g_7 B^{(7)}_j + g_8 B^{(8)}_j ) } } 
 \( \otimes_{j=1}^{N}  \ket{  \sigma_{i,j} } \) = 
 \nonumber \\[1mm]
  & = \sum_{ \{ \sigma_{i,j} \}_{i,j=1}^{M,N} } e^{ \frac{\beta}{M} \sum_{i,j=1}^{M,N} \Phi (\sigma_{i,j}\sigma_{i,j+1}) }  \prod_{i,j=1}^{M,N} \braket{ \sigma_{i,j}  }{ e^{  \frac{\beta}{M} ( g_7 B^{(7)}_j + g_8 B^{(8)}_j ) } }{ \sigma_{{i+1},j}  } \label{quant}
\end{align}
The two-site diagonal matrix $\Phi_{n_i,n_{i+1}}$ then gives the
coupling in one of the two directions on the 2d lattice,
while the transverse part can be obtained matching the last line of this equation with the statistical partition function whose Hamiltonian is \eqref{classham} with different couplings in the two directions
\be \label{class}
Z = \sum_{ \{ \sigma_{i,j} \}_{i,j=1}^{M,N} } e^{ J_x \sum_{i,j=1}^{M,N} \Phi (\sigma_{i,j}\sigma_{i,j+1}) } \prod_{i,j=1}^{M,N} e^{ J_\tau \Phi (\sigma_{i,j}\sigma_{i+1,j}) }.
\ee
Comparing \eqref{quant} and \eqref{class} we obtain for the matrix elements of the $B_\alpha$
\be \label{match}
\braket{ \sigma   }{ e^{  \frac{\beta}{M} ( g_7 B^{(7)} + g_8 B^{(8)} ) } }{ \bar{\sigma}  } =  A e^{ J_\tau \phi (\sigma,\bar{\sigma}) }
\ee
where $A$ is a positive constant. To obtain now the exact expression of the $B_\alpha$ operators,
we consider the so-called Hamiltonian limit \cite{kogut},  $\beta/M \rightarrow 0$. In this way, taking
\be 
A e^{J_\tau} = 1
\qquad \qquad
\frac{\beta g_7}{M} = e^{-J_\tau }
\qquad \qquad
\frac{\beta g_8}{M} = 1
\ee
we reproduce exactly the operators in \eqref{hamlimbs}. Note that $g_7$ and $g_8$ are both positive in the quantum to classical correspondence.

The absence of a quantum phase transition in the  coprime quantum chain  which contains both the operators $B^{(7)}_i$ and $B^{(8)}_i$ 
with positive coupling constants can be argued on the basis of a Peierls argument for the classical two-dimensional coprime model.  
This argument, in its most concise form, goes as follows. Consider one of the configurations $\{ \sigma \}$ of lowest energy, in which the spin variable $\sigma$ is the same on all the lattice sites. They will be of course the only relevant ones at zero temperature. As soon as we move from zero temperature, we have to determine which are the typical configurations which affect the thermodynamics and whether the original ``magnetization'' still persists or not, for certain non-zero range of values of the temperature. The first guess is to excite the ``ground state'' configuration by placing an island of different numbers in the middle. Let $L$ be the length of the domain wall, the free energy difference is given by
\be 
\Delta F = 2 L J - T \Delta S
\ee
Here is the key point: in the case of the Ising model the entropy difference is given underestimating the number of this simple excited configurations with $2^L$, so that
\be \label{df}
\Delta F \le 2 L J - T \log 2^L = L ( 2 J - T \log 2 ) \le 0
\qquad \Rightarrow \qquad
T \ge \frac{2 J}{ \log 2 } \simeq T_c
\ee 
\begin{figure}[t]
\hspace{-0.7cm}
\includegraphics[scale=0.95]{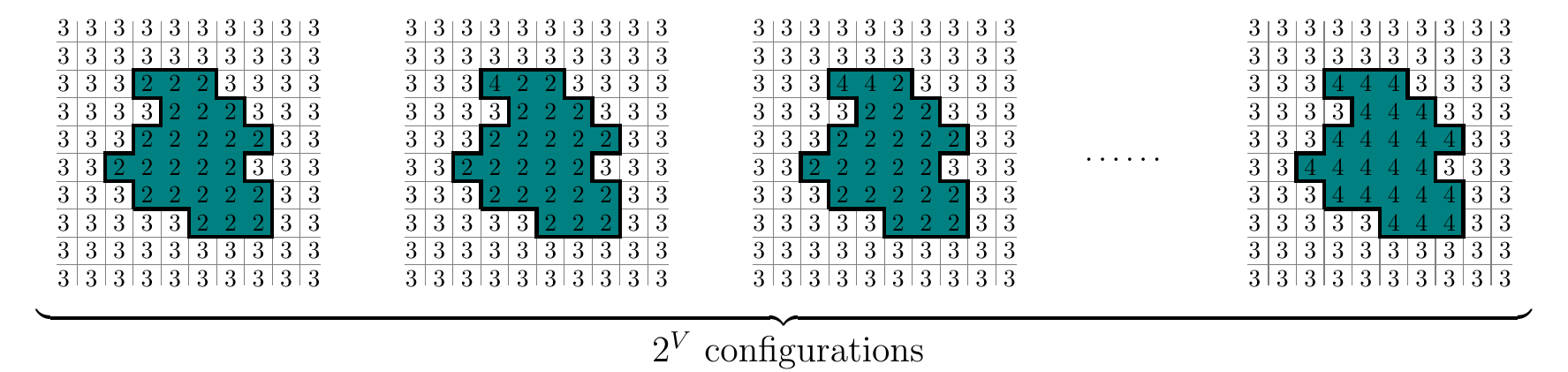}
\caption{{\em For fixed length $L$ of the domain wall, there are $2^V$ possible configurations having the same energy difference with the ground state energy, being $V$ the number of cells inside the contour. Thus the number of states made of one Peierls island of length $L$ is always greater than $2^V 2^L$, where $V \sim L^2$ in the limit $L \to \infty$. }}
\label{peierls}
\end{figure}
Thus the excited states becomes relevant only at a finite value of $T$. However in the coprime model \eqref{classham} the excited states are exponentially many more and their entropy difference with the ground state do not scale as the length $L$ of the domain wall, but as the area of the island \emph{inside} the wall. Indeed let us consider the coprime statistical model with $q=5$. Pick up the ground state made up e.g. of $3$s. The latter can be excited with an island composed of all equal integers sharing no common divisors with $3$. However, since the coprime interaction makes no distinction between $2$ and $4$, there are $2^V$ possible excited states composed of all possible combinations of $2$ and $4$ (see Fig.~\ref{peierls}), where $V$ is the number of sites inside the domain wall. Then for fixed length $L$ the entropy difference can be underestimated as
\be 
\Delta S \ge \log ( 2^L 2^V ) \underset{L \to \infty}{\sim} L^2 \log 2
\ee
Going back to \eqref{df} it is thus clear that the entropy contribution becomes dominant in the thermodynamic limit as soon as the temperature is switched on. Then the transition temperature $T_c$ shrinks to $0$ when the length of domain wall is sent to infinity and the model is always disordered.

\begin{figure}[b]
\center
\makebox[0pt][c]{
\hspace{-10mm}
\begin{minipage}{0.52\textwidth}
\includegraphics[scale=0.45]{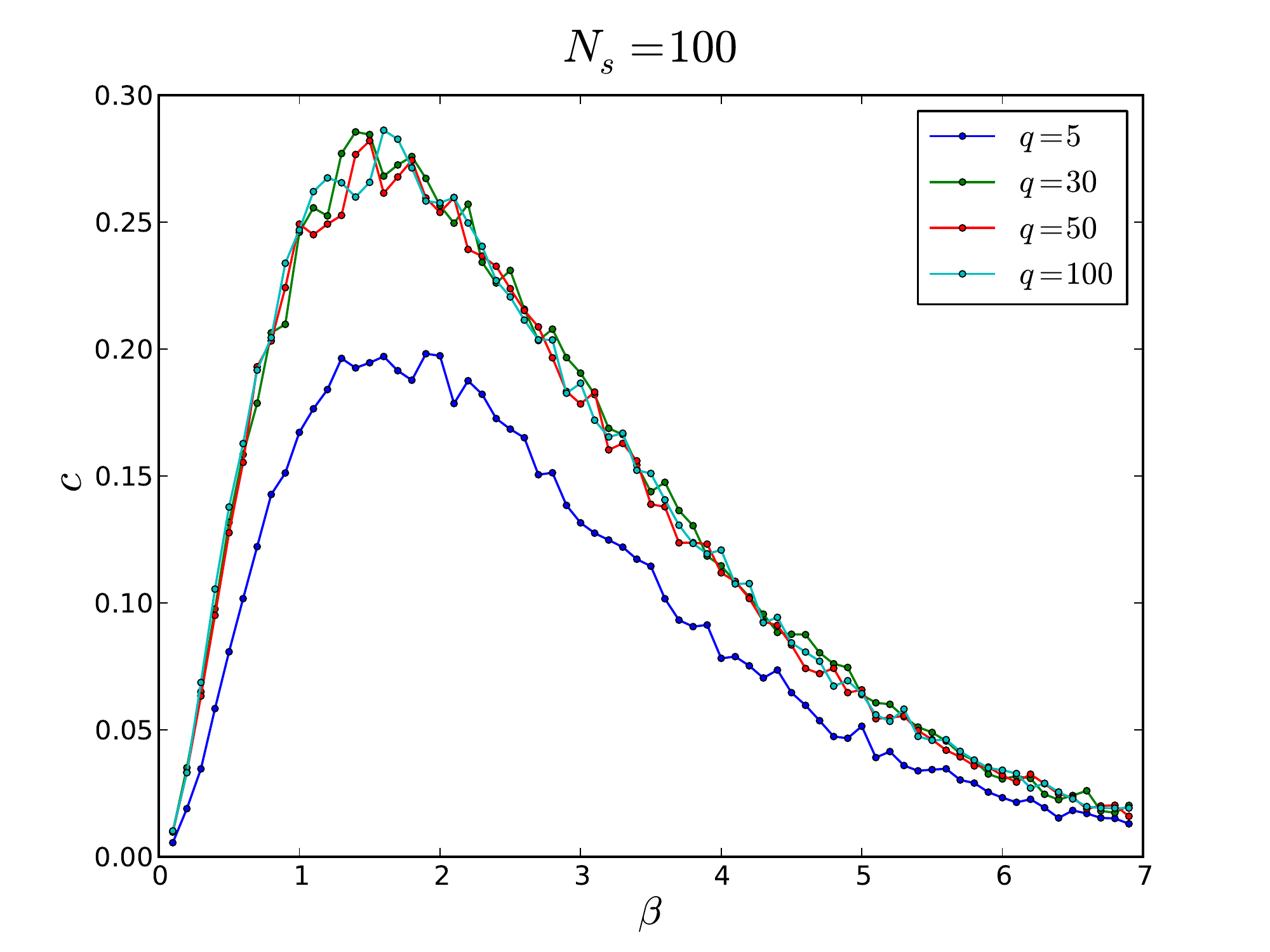}
\end{minipage}
\begin{minipage}{0.53\textwidth}
\includegraphics[scale=0.45]{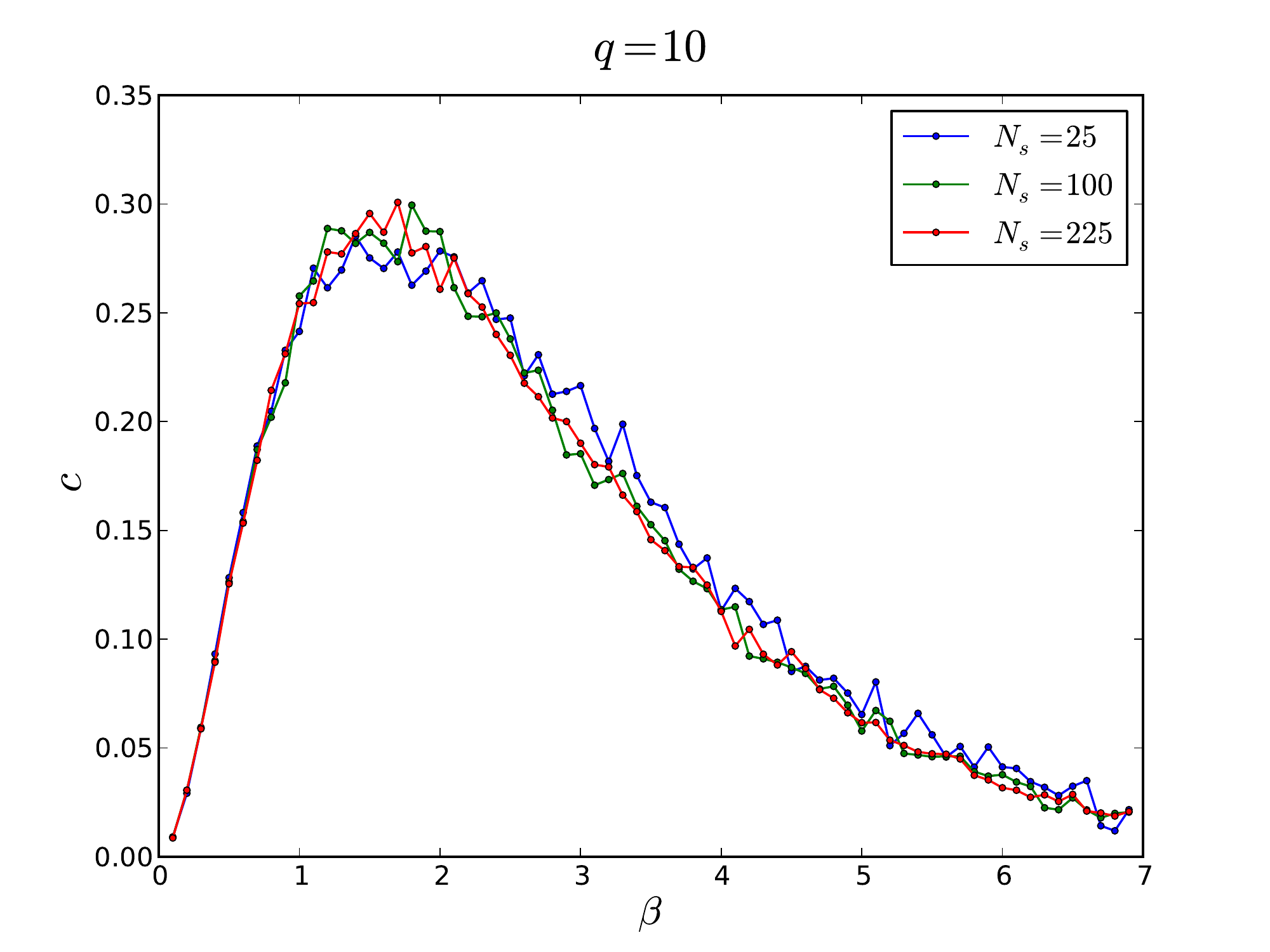}
\end{minipage}}
\caption{{\em In both plots the samples are taken from a single stochastic chain starting from a random configuration. The samples are taken after a Monte Carlo time $t_{eq} = 10^4 \cdot N_s$, one each time interval of length $\Delta t = 600$, to ensure their independence. Here $N_s$ is the number of lattice sites. Each average necessary to compute the specific heat is performed over $3000$ realizations of the Boltzmann distribution. The plot on the left shows the specific heat per lattice site as a function of the inverse temperature $\beta$ for fixed $N_s$ and varying $q$, while the finite size scaling for fixed $q$ is shown on the right. The absence of a continuous PT is made evident by the analyticity of the numerical curve.     }}
\label{specheat}
\end{figure}

The absence of phase transition on the 2d statistical model can be numerically checked by computing the specific heat as a function of the temperature $T$ 
\be 
c(T) = \frac{\partial \med{ \mathcal{H} } }{ \partial T } = \frac{ \med{ \mathcal{H}^2 } - \med{ \mathcal{H} }^2 }{ T^2 }\,\,\,.
\ee
The latter is easily obtained by sampling the Boltzmann distribution using a simple Metropolis algorithm \cite{metro} in which the move consists of changing randomly the integer on a random site. The presence of a critical point would be signalled by a spike in the graph of $c(T)$ for some value of $T$, representing a divergence of this function with some critical exponent $\alpha$. The results are shown in Fig.~\ref{specheat}. The specific heat per lattice site exhibits a maximum for a value of $\beta$ between $1$ and $2$, but no sign of divergences, therefore the two-dimensional statistical coprime model is always 
in its disordered phase. 
 
Coming now to the  coprime quantum chain, the numerical study of the mass gap of the quantum  Hamiltonian 
\begin{equation}
H = - \sum_{i=1}^M \left[\Phi(n_i,n_{i+1}) + g_7 B^{(7)}_i + g_8 B^{(8)}_i \right] \,\,\,,
\end{equation}
for arbitrary sign of the two coupling constants is shown in Fig.~\ref{hamgap}. This figure is interesting because it shows that in first quadrant (when $g_7 > 0$ and $g_8 >0$) the is system is gapped, as predicted by the Peierls argument for the classical model. However, there are two lines, respectively in the 
third and four quadrants where, instead, there could be a vanishing mass gap. This results seems to originate in these cases from a  competition between two interactions with coupling constants of opposite sign.     
\begin{figure}[t]
\center
\includegraphics[scale=0.5]{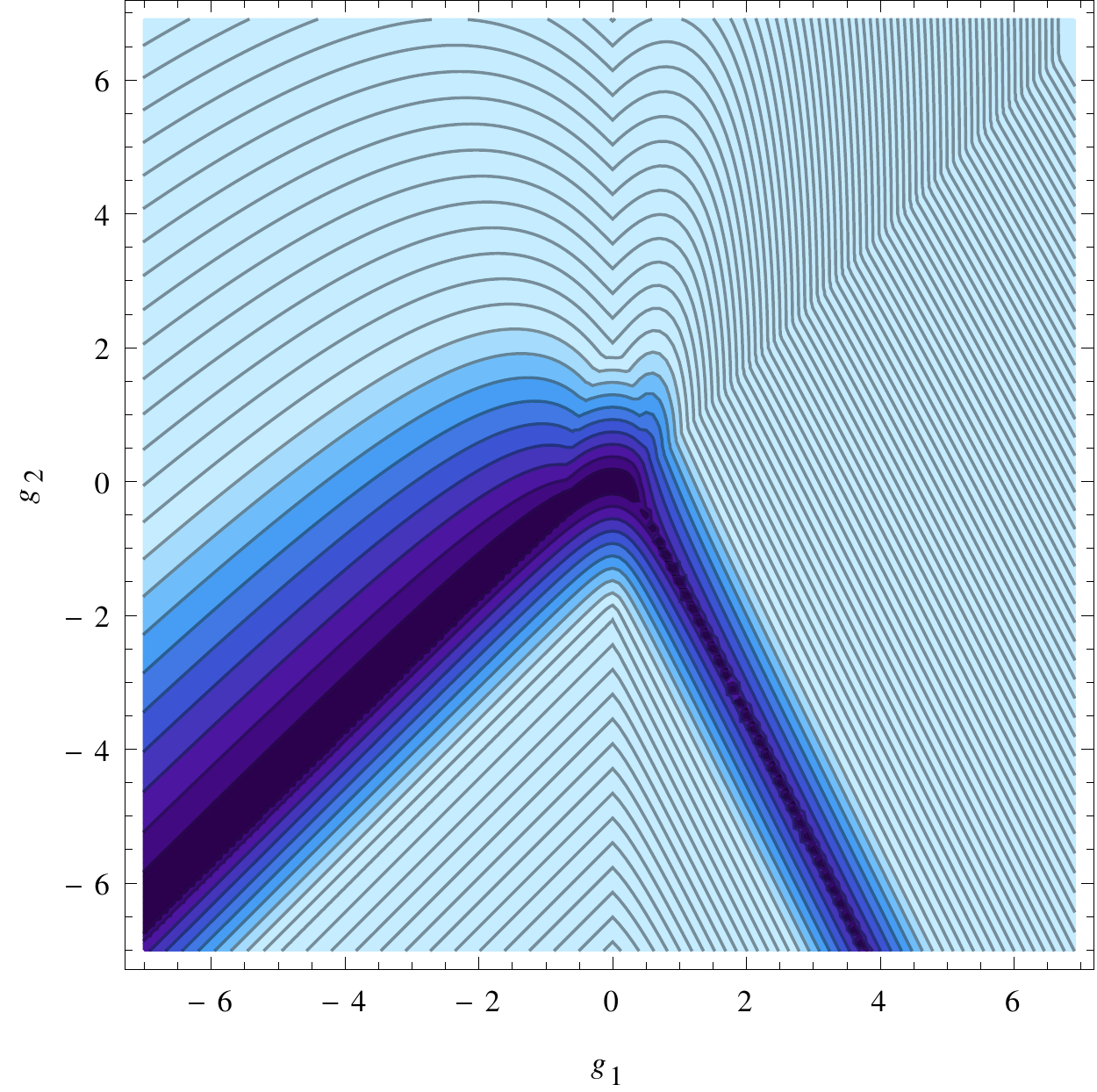}
\caption{{\em Magnitude and contour lines of the gap of the quantum Hamiltonian with the operators \eqref{hamlimbs} as a function of the two coupling constants. Dark colors stand for lower value of the gap. The data are taken from a chain of $N=8$ sites, varying $g_1$ and $g_2$ on a square of size $7 \times 7$ with a step $\Delta g = 0.1$. When $g_1,g_2 > 0$ it is evident that the gap is non-zero, as discussed in the text. It is interesting to notice the presence of two lines over which the gap seems to close. However these lines lie into regions of the plane where the correspondence with the classical model ceases to apply.}}
\label{hamgap}
\end{figure}
For the classical two-dimensional statistical system the absence of phase transitions can be explained in terms of the failure of the Peierls argument. 
 
\section{Conclusions}\label{conclusions}
In this paper we have introduced the coprime quantum chain: a strongly correlated bosonic system characterised by local interactions sensible to  the prime number backbone of the single-site occupation numbers. We have initially showed that at the classical level the model  has a number of ground states exponentially large. Their actual value depends on the boundary conditions and can be computed exactly using a blend of spectral graph theory and number theory. This 
is particularly evident in the limit $q \rightarrow \infty$,  a situation analysed by Don Zagier in Appendix C. 
We have also shown that in the ferromagnetic case there could also be frustration phenomena. At the quantum level, the most important
property that emerges from our analysis is the possibility to come across several quantum critical points in the phase space of the coprime chain. 
These critical points are characterised by their corresponding class of universality and can be clearly identified by calculating the ground state entanglement entropy.
We have discussed, in particular, the emergence of the classes of universality of the Ising and 3-state Potts chains and we expect that, 
with a proper tuning of the coupling constants, one should be able to reach the critical point of the $\mathbb{Z}_n$ quantum chains.

There are many unexplored features inherent in the  coprime quantum chain that would be interesting to investigate in the future.
First of all, it is important to establish the existence of lines of integrability in the parameter space of the coupling constants $\beta_{\alpha}$ of the Hamiltonian (\ref{quantumhamiltonian}). If those lines exist, one can expect to solve exactly the model through Bethe Ansatz techniques. This could lead in particular to an exact expression for its free energy and possibly also for its correlation functions. 

Secondly, it would be useful to study (even numerically) the  coprime quantum model defined on higher dimensional lattices and check whether also in higher dimensions there is the possibility to drive the system toward criticality by switching on proper operators. 

Thirdly, it is intriguing to determine the surfaces (in coupling constant space) where the mass gap of the theory vanishes, as it happens for instance along some lines shown in Fig.~\ref{hamgap}. 

Finally it would be relevant to look deeper into the quantum properties of the coprime chain.
The diagonal coprime interaction term $\Phi(n_i,n_{i+1})$ has the peculiarity of being sensible to the prime content of the occupation numbers, a circumstance that, as we have at length discussed in this paper, leads to an exponential degeneracy of the ground states, contrary to more familiar quantum one-dimensional chains. 
Out-of-equilibrium protocols such as global or local quantum quenches could be realised without too many differences respect to what already considered in a well developed literature, see for instance the papers published in the special issue of JSTAT {\em Quantum Integrability in Out of Equilibrium Systems} \cite{JSTAT}
and references therein. The chain could be prepared in one of its ground state and its time-evolution studied switching on some of
the  parameters $\beta_{\alpha}$. It will be then possible to analyse problematic connected with relaxation toward equilibrium, entanglement spreading and energy transport, mimic previous studies in the Ising spin chain, \cite{RSMS, CEF, FC, EP, SD, DVBD}. 
Unfortunately it is not clear whether answers to these questions can be formulated with techniques based on integrability or one has to necessarily resort to DMRG simulations.  In this latter case however truncation of the one-site Hilbert space might actually be implemented quite easily as routinely done when dealing with dynamics in the Bose-Hubbard model~\cite{KLA}. Let us also mention that the function  $\Phi$ could be generalized further to include $k$-site interactions based on pairwise coprimality
conditions. 
 
\vspace{1cm}
\begin{flushleft}\large
\textbf{Acknowledgements}
\end{flushleft}

One of us (GM) would like to thank the International
Institute of Physics of Natal, where this work started, for the 
warm hospitality.  

\newpage
\appendix
\section{Basic elements of graph theory}\label{Appgraphtheory}

\noindent 
In this appendix we recall some basic ingredients of graph theory used in the main text.\\

\vspace{1mm}
\noindent
{\bf Main Definitions}. 

\begin{itemize}
\item An (undirected) graph $G$ is a set of of vertices $V=\{1,\dots,n\}$ and edges $E$ that connect them. An edge is a pair of
vertices $(i,j)$, chosen among the total number of possible pairing $\binom{n}{2}$.

\item The degree of a vertex $i$ is denoted by  $d_i$ and it is the total number of edges  touching it. Two vertices can be connected by multiple edges, if the
number of edges connecting vertex $i$ to vertex $j$ is $d_{ij}$ then clearly $d_i=\sum_{j}d_{ij}$. 

\item A graph $G$ is fully specified by its adjacency matrix $A$ with elements $A_{ij}=d_{ij}$, $1\leq i,j\leq n$ and the convention that $d_{ij}=0$ if two vertices are disconnected. 

\item A graph is called regular (or otherwise irregular) when every vertex has the same degree $d_i=k$, $\forall i=1,\dots,n$.  

\item The complement graph $\bar{G}$ of $G$ is the graph with the same vertices $V$ and that contains as edges the pairs $(i,j)$ missing in the graph $G$. Obviously the complement of a regular graph is regular with degree $n-1-k$. Finally, if we denote by $J$ the matrix with all entries equal to $1$ and $I$ the identity matrix, the adjacency matrix of the complement graph $\bar{G}$ is $\bar{A} = J - I - A$, being $A$ the adjacency matrix of $G$. 
\end{itemize}

As an example we consider the graph $G$ showed in Fig.~\ref{fig:graphs}.
\begin{center}
 \begin{figure}[b]
 \includegraphics[scale=1]{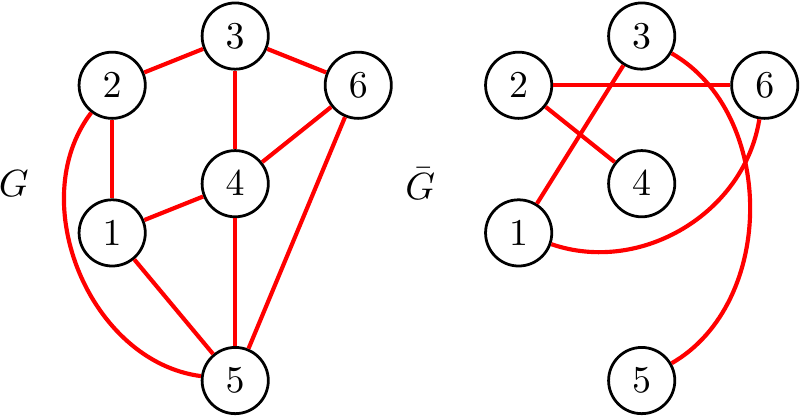}
\caption{An example of an undirected graph $G$ with 6 vertices and 10 edges and its complement $\bar{G}$.}  
 \label{fig:graphs}
 \end{figure}
\end{center}
Such a graph  contains 6 vertices and 10 edges, its adjacency matrix is given by
\begin{equation}
 A=\begin{pmatrix}
    0& 1 & 0 & 1 & 1 & 0 \\
      1&  0 &  1& 0 & 1 & 0 \\
       0&1 & 0&1 & 0 & 1 \\
        1& 0 & 1 &  0& 1& 1 \\
        1 &1 &0 &1 & 0& 1 \\
         0& 0 & 1 & 1 & 1 & 0\\
   \end{pmatrix}.
\end{equation}
The graph $G$ is irregular, since in general different nodes have different degrees:
 $d_4=d_5=4$ and $d_1=d_2=d_3=d_6=3$.\\ 

\vspace{3mm}
\noindent
{\bf Spectral problem for the adjacency matrix.}
The adjacency matrix of an undirected graph $G$ is a symmetric non-negative $n\times n$ matrix and its spectral problem is a classical problem in graph theory 
(see for instance \cite{Cvit}). Let us start by considering the case of a regular graph $G$ of degree $k$. For simplicity we will also assume the graph to be connected, meaning that there is 
always a path connecting two arbitrary vertices: in this case the adjacency matrix $A$ is irreducible and Perron-Frobenius theorem can be applied. 

It easy to show that the $n$-dimensional vector $|f\rangle$, with all entries equal 1
\begin{equation}
| f \rangle \,=\, \left(
\begin{array}{c} 
1 \\
1 \\
..\\ 
\\
.. \\
1\\
1
\end{array}
\right)
\end{equation}
is an eigenvector of the adjacency matrix $A$ with eigenvalue $k$. Since all components of this vector are positive, $|f\rangle$ is actually the (unique) Perron Frobenius eigenvector of $A$ and therefore $k$ is also the largest eigenvalue of the matrix $A$. Moreover the one-dimensional eigenspace generated by $|f\rangle$ is orthogonal to all other eigenspaces of $A$ (being symmetric). If $|v\rangle$ is an eigenvector of $A$ with eigenvalue $\theta<k$ then  
we also have
\begin{equation}
\label{duality}
\bar{A}|v\rangle=(J-I-A)|v\rangle=(-1-\theta)|v\rangle.
\end{equation}
Hence note that we have 
\begin{equation}
\bar{A}|f\rangle=(n-1-k)|f\rangle \,\,\,.
\end{equation}
If the $n$ real numbers 
\begin{equation}
k > \theta_1 \geq \theta_2 \geq \dots \geq \theta_{n-1}
\end{equation}
are the spectrum of $A$, then the eigenvalues of $\bar{A}$ are 
\begin{equation}
n-k-1> -1-\theta_{n-1} \geq -1-\theta_{n-2} \geq \dots \geq -1-\theta_1 \,\,\,.
\end{equation}
This observation furnishes an equation satisfied by the characteristic 
polynomials $P_{G}(x)$ and $P_{\bar{G}}(x)$  of the adjacency matrices of $G$ and $\bar{G}$. The characteristic polynomial of $G$ is indeed
\begin{equation}
P_{G}(x)\, =\, (x-k)\prod_{i=1}^{n-1}(x-\theta_i)\,\,\,,
\end{equation}
and one obtains 
\begin{equation}
\label{car_pol_reg}
 P_{\bar{G}}(x)=(-1)^{n}\frac{x-n+k+1}{x+k+1}P_{G}(-1-x)\,\,\,.
\end{equation}

\vspace{3mm}
Unfortunately the simple analysis outlined above for the spectral problem of a regular graph does not apply to the graph whose adjacency matrix $A$ is the coprime matrix since this graph is irregular. Instead of presenting the complete analysis for the spectral theory of the irregular graphs, here we simply quote the results relative to the largest eigenvalue $\eta_{max}$ of the adjacency matrix as well as the result which concerns the characteristic polynomial of the  complementary graph. 

The largest eigenvalue $\eta_{max}$ of an irregular connected graph $A$ (that is unique and positive by the Perron Frobenius theorem) is bounded by the average degree $\bar{d}$ of the graph and its maximum degree $d_{max}$
\begin{equation}
 \bar{d}\leq\eta_{max}\leq d_{max}\,\,\,.
\end{equation}
Notice that for a regular graph $G$ of degree $k$, $\bar{d}=d_{max}=k$ and therefore we recover the previous result. 

Let us now state that it is possible to generalize eq.\,\eqref{car_pol_reg} also to irregular graphs. Firstly we observe that $|f\rangle$ is no longer an eigenstate of $A$ but we can nevertheless define the so-called \textit{main spectrum} of the graph $G$ as the vector space $M$ generated by all the eigenvectors of $A$ that are not orthogonal to $|f\rangle$. The main spectrum of a regular graph would be one-dimensional and would coincide with the eigenspace of $|f\rangle$. 

If $|v_1\rangle,\dots,|v_{m}\rangle$ are the vectors of an orthonormalized basis for the main spectrum $M$, namely $A|v_i\rangle=\mu_i|v_i\rangle$ and $\langle v_i|f\rangle\not=0$, it is customary to introduce the angles $\beta_i$ of an irregular graph as 
\begin{equation}
\beta_i = \frac{1}{\sqrt{n}}\langle f|v_i\rangle 
\,\,\,\,\,\,
,
\,\,\,\,\,\, i=1,\dots,m \,\,\,.
\end{equation}
Starting from the definition of the characteristic polynomial of the complementary graph $\bar{G}$, $P_{\bar{G}}(x)=\det(\bar{A}-xI)$ and using the spectral decomposition of $A$, together with definition of $\bar{A}$, it is then possible to show that
\begin{equation}
\label{car_pol_irreg}
 P_{\bar{G}}(x)=(-1)^{n}P_{G}(-1-x)\left(1-n\sum_{i=1}^{m}\frac{\beta_i^2}{x+1+\mu_i}\right)\,\,\,,
\end{equation}
an equation that generalizes \eqref{car_pol_reg}.

\section{Maximum degree of the coprime graph in the limit $q \rightarrow \infty$}\label{Appmaxdegree}

\vspace{1mm}
\noindent
It is possible to estimate the rate of growth of the maximum degree of the coprime graph with the number $q$  based using as hypothesis the statistical independence of the primes. The argument goes as follows. Given a number $q$, let us firstly determine which is the maximum index $s$ such that the 
number made of the product of the first $s$ consecutive primes is less than $q$ 
\be 
\hat n \,=\, p_1 \times p_2 \ldots \times p_s \,< \, q \,\,\,.
\label{goodnumber}
\ee
Such number $\hat n$ is associated to one of the vertices with the maximum degree in the range $[2,q]$. In fact, this number divides all numbers multiples of $p_1 =2$, all those which are multiples of $p_2 = 3$, all those multiples of $p_3 = 5$, etc. We now estimate how many numbers have common factors with 
$\hat n$ using a probability argument. 

The total number of numbers that are divisible by $2$ is given by $q$ times the probability that a number is divisible by $2$, which is $\frac{1}{2}$, so $q \times \frac{1}{2}$. The total number of those which are divisible by $3$ are given, naively, by $q \times \frac{1}{3}$. However this is an over-counting since among these multiples of $3$ there are those we have already counted as divisible by $2$ (as, for instance, the number $6$) and therefore we have to subtract them. So the genuine numbers which are divisible by $3$ but not also by $2$ are given by $q$ multiplied for the probability $\hat p_3$ that a number is divisible by $3$ but not by $2$ 
\be 
\hat p_3 \,=\,\frac{1}{3} \, \left(1 - \frac{1}{2}\right) \,\,\,. 
\ee
Analogously, we can count the genuine numbers divisible by $5$ but not divisible for $2$ and $3$, in term of the corresponding probability 
\be 
\hat p_5 \,=\,\frac{1}{5} \,\left(1 - \frac{1}{2}\right) \, \left(1 - \frac{1}{3}\right) 
\ee 
and, more generally, 
\be 
\hat p_k \,=\,\frac{1}{p_k} \,\prod_{m=1}^{k-1} \left(1 - \frac{1}{p_m}\right) \,\,\,.
\ee
In this way, we predict that the maximum degree grows as 
\be 
d_{max} \,\sim\, q \, \sum_{k=1}^s \hat p_k \,\,\,.
\label{estmd}
\ee 
Notice that the number of terms included in the sum depends on the number $q$ itself: therefore there will be a sequence of discontinuities of the 
corresponding slope each time that $q$ overpass the values 
\be
2\, ,6 \,, 30\,, 210 \,, 2310 \,, 30030 \, \ldots 
\ee 
associated to the sequence of numbers $\hat n$. This explains why the slope of the maximum degree changes by increasing $q$,
giving rise to a pronounced finite-size dependence as discussed in the text.  

However, in the asymptotic limit $q\rightarrow \infty$, we have that also the maximum index $s$ goes to infinity, $s \rightarrow \infty$, and therefore the 
ratio $d_\mathrm{max}/q$ is given by the infinite series 
\be \label{pseries}
\frac{d_\mathrm{max}}{q} \underset{ q \to \infty }{ \sim } \sum_{n=1}^\infty \,\frac{1}{p_n} \prod_{m=1}^{n-1} \left(1 - \frac{1}{p_m}\right)
\ee
The series is convergent since
\be \label{hseries}
\sum_{n=1}^\infty \,\frac{1}{p_n}  \exp \( \sum_{m=1}^{n-1} \left(1 - \frac{1}{p_m} \right) \) < \sum_{n=1}^\infty \,\frac{1}{p_n}  \exp \( - \sum_{m=1}^{n-1}  \frac{1}{p_m}  \) < \sum_{n=1}^\infty \,\frac{1}{p_n}  \frac{ e^{ \frac{\pi^2}{6} }}{ \log {p_n}  }\,\,\,,
\ee 
where the last inequality follows from the fact that the truncated sum of the reciprocals of the primes is bounded by
\be 
\sum_{p < n } \frac{1}{p} > \log \log n - \log \frac{\pi^2}{ 6 } \,\,\,.
\ee
The series in the last member of \eqref{hseries} converges as a consequence of the prime number theorem, which states that the magnitude of the $n$-th prime number goes as
\be 
p_n \underset{ n \to \infty }{ \sim }  n \log n\,\,\,,
\ee
so that the general term behaves asymptotically as
\be 
\frac{1}{ p_n \log p_n } \underset{ n \to \infty }{ \sim } \frac{1}{ n (\log n)^2 }
\ee
which is enough to make the series convergent.

Since the maximum degree of a graph cannot exceed the number of vertices of the graph, the series \eqref{pseries} must converge to a value smaller or equal to $1$. As a matter of fact the result is exactly $1$, as the truncation of the series \eqref{pseries} can be put in the telescopic form
\begin{align*}
\sum_{n=1}^N \, \(1 - 1 + \frac{1}{p_n} \) \prod_{m=1}^{n-1} \left(1 - \frac{1}{p_m}\right)  & = 
\sum_{n=1}^N {} \prod_{m=1}^{n-1} \left(1 - \frac{1}{p_m}\right) - \sum_{n=2}^{N+1} {} \prod_{m=1}^{n-1} \left(1 - \frac{1}{p_m}\right)  =  \\[1mm]
& =  1 - \prod_{m=1}^{N} \left(1 - \frac{1}{p_m}\right)  \quad \underset{ N \to \infty }{  \rightarrow } \quad  1  - \frac{1}{ \zeta(1) } = 1 
\end{align*}
where
\be 
\zeta(z) = \prod_{k=0}^{\infty} \(1 - p_k^{-z} \)^{-1}
\ee
is the product representation of the Riemann Zeta function, whose only pole is $z=1$.

In summary, for asymptotically large values of $q$ the maximum degree of the coprime graph scales exactly as $q$ 
\begin{equation}
d_m \simeq q 
\,\,\,\,\,\,\,\,\,
,
\,\,\,\,\,\,\,\,\,
q \rightarrow \infty
\end{equation}

\newpage

\def\={\;=\;}  \def\+{\,+\,}  \def\i{^{-1}}  \def\la{\langle} \def\ra{\rangle} \def\wt{\widetilde}
\def\Z{\Bbb Z}  \def\N{\Bbb N}  \def\R{\Bbb R}  \def\C{\Bbb C} \def\H{\mathcal H} \def\bH{\widetilde{\H}}
\def\l{\lambda} \def\ve{\varepsilon} \def\a{\alpha} \def\b{\beta} \def\z{\zeta} \def\k{\kappa}  \def\L{\xi} \def\D{\Delta}
\def\PF{_{\text{PF}}}  \def\lo{\l_1} \def\lD{\l_D} \def\vo{v^{(1)}} \def\vD{v^{(D)}} \def\vDN{v^{(D,N)}}
\def\CN{C^{(N)}}  \def\bC{\overline C} \def\bCN{\bC^{(N)}}  \def\wB{\wt B}  \def\bB{\overline B} \def\wv{\wt v}
\def\NMc{N_M^{\text{cyc}}}  \def\Av{\text{Av}} \def\Avm{\Av^{(m)}}  \def\NN{\mathcal N}

\section{Eigenvalues of the coprimality matrix}\label{AppZagier}  \centerline{\bf by Don Zagier} \bigskip

Let $N$ be a natural number and define an $N\times N$ matrix $C=\CN$ by
$$ C_{mn}\= \begin{cases} 1 &\text{ if $(m,n)=1$,} \\  0 &\text{ if $(m,n)>1$,}\end{cases}\qquad(1\le m,\,n\le N), $$
i.e., as the top $N\times N$ part of the infinite matrix $C^{(\infty)}$ whose $(m,n)$-entry is~1 if
$m$ and~$n$ are coprime and 0 otherwise.  The matrix $\CN$ is real and symmetric, so has~$N$ real
eigenvalues, and since the sum of the entries of each of its rows is of the order of~$N$, it is
natural to suppose that these eigenvalues also grow roughly linearly in~$N$. In fact numerical
experiments suggested that the eigenvalues of $N\i\CN$ converge to a well-defined spectrum
as $N\to\infty$. In this appendix we will first discuss these computations and then use an Ansatz
suggested by them to derive a formula or the limiting eigenvalues and the corresponding
eigenvectors.  We then show that after a suitable choice of basis the matrices $N\i\CN$ themselves
converge pointwise to a well-defined infinite matrix having the predicted eigenvalues and eigenvectors,
and finally describe the corresponding results for the complementary matrix $\bCN$ whose $(m,n)$-entry 
is~0 if $m$~and~$n$ are coprime and~1 if they are not.  

Since the results of this appendix may be of independent interest to number theorists or others not familiar
with statistical models, we have made the text self-contained, giving all needed notations or definitions 
from scratch rather than quoting them from the main text or Appendix~A of the paper, and giving references
where needed to related results or discussion in the main text.  We have also changed some notations and
terminology to be more conformal with standard number-theoretical practise. For instance, 
the number we call~``$N$" is denoted by~``$q$" in the main paper, and what we call the coprimality matrix
and the complementary coprimality matrix are referred to in the paper as the coprimality matrices for the 
antiferromagnetic and ferromagnetic cases, respectively.

\subsection{Numerical results}\label{C1} We first discuss the Perron-Frobenius eigenvalue, since
this is the easiest one to calculate numerically.  Recall (cf.~Appendix~A) that the Perron-Frobenius 
theorem says that a square matrix with strictly positive entries has a unique eigenvector with positive
entries and that the corresponding eigenvalue is real and is larger than the absolute value of any other
eigenvalue. The same theorem holds for a matrix that has non-negative entries and is ``primitive,"
meaning that some power of it has strictly positive entries.  The matrix $C=\CN$ is primitive
because $(C^2)_{mn}=\sum_{k=1}^NC_{mk}C_{kn}\ge C_{m1}C_{1n}=1$, so it has a Perron-Frobenius
eigenvalue $\l\PF$ and eigenvector $v\PF$, which can be calculated easily by starting with
the vector~$f_1=(1,\dots,1)$ and applying $C$ to it repeatedly, renormalizing each time to make
the leading coefficient equal to~1.  This process converges very quickly, with 20 iterations
sufficing even for $N=10000$ to give 9-digit accuracy for the eigenvalue and the first few components 
of the eigenvector.  The Perron-Frobenius eigenvalues for a few values of $N$ are given by the table
  \smallskip \begin{center} \begin{tabular}{|c||c|c|c|c|l|c|c|} \hline 
  \;$N$\qquad &  100 & 200 & 500 & 1000 & 2000 & 5000 & 10000 \\ \hline $\;\l\PF(N^{-1}\CN)\;$
   & \;0.67643\; &\; 0.68139\; & \;0.67904\; & \;0.67869\; & \;0.67872\; & \;0.67853\; & \;0.67846\; \\
  \hline \end{tabular} \end{center} \smallskip
and the beginning of the (normalized) Perron-Frobenius eigenvectors for $N=10,\,10^2,\,10^3,\,10^4$ by
 \smallskip  \begin{center} \begin{tabular}{|c||c|c|c|c|c|c|c|c|c|} \hline 
  $N$ & $v_1$ & $v_2$ & $v_3$ & $v_4$ & $v_5$ & $v_6$ & $v_7$ & $v_8$ & $v_9$\\ \hline 
 10  & \;1.0000\; & \,0.6057\, & \,0.7293\, & \,0.6057\, & \,0.8103\, & \,0.3923\, & \,0.8724\, & \,0.6057\,& \,0.7293\, \\
 100  & \;1.0000\; & \,0.6162\, & \,0.7310\, & \,0.6162\, & \,0.8250\, & \,0.4463\, & \,0.8715\, & \,0.6162\,& \,0.7310\, \\
 1000 & 1.0000 & 0.6179 & 0.7320 & 0.6179 & 0.8281 & 0.4520 & 0.8734 & 0.6179 &  0.7320 \\
\;10000\; & 1.0000 & 0.6180 & 0.7320 & 0.6180 & 0.8284 & 0.4524 & 0.8730 & 0.6180 & 0.7320 \\
\hline  \end{tabular} \end{center} \smallskip \noindent
These tables strongly suggest that the limiting value of $\l\PF$ exists and equals roughly 0.6785 and that 
the corresponding normalized eigenvectors also converge to a vector close to the one for~$N=10000$.   
 
For the other eigenvectors and eigenvalues the calculation is slower and we only went up
to~$N=1000$, calculating the characteristic polynomial of $\CN$ and its roots in each case. 
The eigenvalue~0 occurs with a high multiplicity (this is obvious, because if $e_n$ denotes the 
standard $n$th basis element of $\Z^N$ then $e_n-e_{n'}$ is in the kernel of~$\CN$ whenever
$n$ and~$n'$ have the same prime factors), and there are many more eigenvalues whose quotient
by~$N$ tend to~0 as~$N$ grows, but for fixed~$i$ the $i$th largest and $i$th smallest eigenvalues 
of $N\i\CN$ seem to converge numerically to well-defined non-zero limits: 
\smallskip  \begin{center} \begin{tabular}{|c||ccc|} \hline  
$N$ &\qquad(smallest three eigenvalues,\;& \,\dots\,, & \;largest three eigenvalues)\;\quad \\ \hline 
100  & $\;(-0.26167,\;-0.18533,\;-0.12076,$ & \,\dots\,, & \; 0.04784,\;0.07035,\;0.67643)\; \\
200  & $\;(-0.26037,\;-0.18097,\;-0.11708,$ & \,\dots\,, & \; 0.04597,\;0.06976,\;0.68139)\; \\
500  & $\;(-0.25953,\;-0.18146,\;-0.11645,$ & \,\dots\,, & \; 0.04495,\;0.06963,\;0.67904)\; \\
1000 & $\;(-0.25937,\;-0.18206,\;-0.11685,$ & \,\dots\,, & \; 0.04468,\;0.06949,\;0.67869)\;  \\
\hline  \end{tabular} \end{center} \smallskip \noindent
In the rest of this appendix we will identify all of these numbers and discuss some related results.

\subsection{Exact results} \label{C2} 
The numerical data given above is confirmed by the following more precise statement, in which we have used 
the standard number-theoretical notation $a|b$ to mean that the integer~$a$ divides the integer~$b$.

\medskip 
\noindent {\bf Theorem 1.}  {\bf a.\,}
{\it The Perron-Frobenius eigenvalue of $N\i\CN$ converges for $N\to\infty$ to the number}
\be\label{PF}\lo\=\prod_{\text{\rm $p$ prime}} \frac{p-1+\sqrt{(p-1)(p+3)}}{2p} \= 0.67846225243465570728\cdots\;.\ee
 {\bf b.\,} {\it More generally, the entire spectrum of $N\i\CN$ converges as $N\to\infty$ to the countable
subset \hbox{$\{0\}\,\cup\,\{\lD\mid \text{\rm$D$ squarefree}\}$} of~$\R$,
where $\lD$ is an algebraic multiple of~$\lo$ given by}
\be\label{lD} \lD \= \lo\,\prod_{\text{\rm $p|D$, $p$ prime}} \frac{-p-1+\sqrt{(p-1)(p+3)}}2 \;. \ee
 {\bf c.\,} {\it The normalized eigenvector of $\CN$ corresponding to $\lD$ converges for $N\to\infty$ 
to a well-defined vector $\vD\in\R^\infty$ with components given by}
\be\label{vD} \vD_n \= \prod_{\text{\rm $p$ prime}\atop p|n,\,p\nmid D} \frac{-p+1+\sqrt{(p-1)(p+3)}}2\;\cdot
\prod_{\text{\rm $p$ prime}\atop p|n,\,p|D}\frac{-p+1-\sqrt{(p-1)(p+3)}}2  \qquad(n\in\N). \ee

\medskip
Before continuing we discuss the numerical values.  The convergence of the Euler product defining~$\lo$ is 
rather slow, but can be accelerated by a standard trick: if we define $f(x)\in\Z[[x]]$ to be the power series
$$ f(x) \= \frac{(1-x^3)(1-x^5)^5}{(1-x^2)(1-x^4)^2}\cdot\frac{1-x+\sqrt{(1-x)(1+3x)}}2
   \=  1 \,-\, 12\,x^6 \+ 28\,x^7 \+ \cdots \;,$$
then we can express $\lo$ using the rapidly convergent infinite product
$$ \lo \= \frac{\zeta(2)\zeta(4)^2}{\zeta(3)\zeta(5)^5}\,\prod_{\text{$p$ prime}} f(p) \,,$$
where now only a few thousand terms are needed to give the above 20-digit numerical value.  The values of
the three smallest and the two second largest eigenvalues~$\lD$ are then given by
$$ \begin{aligned} \l_2 &\= \lo\cdot\tfrac{-2}{3+\sqrt5} &=\; -0.2591495203462274\cdots\,, \\
\l_3 &\= \lo\cdot\tfrac{-1}{2+\sqrt3} &=\; -0.1817934126348665\cdots\,, \\
\l_5 &\= \lo\cdot\tfrac{-1}{3+2\sqrt2} &=\; -0.1164057194013901\cdots\,, \\
\l_6 &\= \lo\cdot\tfrac{-2}{3+\sqrt5}\cdot\tfrac{-1}{2+\sqrt3} &=\;\phantom-0.0694389046956844\cdots\,, \\
\l_{10} &\= \lo\cdot\tfrac{-2}{3+\sqrt5}\cdot\tfrac{-1}{3+2\sqrt2} &=\;\phantom-0.0444630283264479\cdots\,, 
\end{aligned} $$
in excellent agreement with the values for~$N=1000$ tabulated above.  Finally, the first few components 
of the first eigenvector $v^{(1)}$ as given in part~{\bf c.} of the theorem have the exact and numerical values
\begin{align*} v^{(1)} &\= \bigl(1,\,\tfrac{\sqrt5-1}2,\,\sqrt3-1,\,\tfrac{\sqrt5-1}2,\,2\sqrt2-2,\,
 \tfrac{(\sqrt5-1)(\sqrt3-1)}2,\,\sqrt{15}-3,\,\tfrac{\sqrt5-1}2,\;\sqrt3-1,\;\dots\,\bigr) \\
 &\;\simeq\;\bigl(1.00000,\, 0.61803,\, 0.73205,\, 0.61803,\, 0.82843,\, 0.45243,\, 0.87298,\, 0.61803,\, 0.73205,\;\dots\,\bigr)\,,
\end{align*}
in perfect agreement with the values for~$N=10000$ given above, while the second eigenvector begins
$$ v^{(2)} \= \bigl(1,\,\tfrac{-\sqrt5-1}2,\,\sqrt3-1,\,\tfrac{-\sqrt5-1}2,\,2\sqrt2-2,\,
 \tfrac{(-\sqrt5-1)(\sqrt3-1)}2,\,\sqrt{15}-3,\,\tfrac{-\sqrt5-1}2,\;\sqrt3-1,\;\dots\,\bigr) \;.$$

\subsection{Ansatz via Dirichlet series}\label{C3}  We begin by giving a heuristic argument leading to
the results stated in the theorem above.  Based on the numerical data, we assume as an Ansatz that the
Perron-Frobenius eigenvalue of $N\i\CN$ has a limiting value $\l=\lo$ as $N\to\infty$ and that the corresponding
eigenvector $v\PF$, normalized to have first component~1, has a limiting value $v=(v_1,v_2,\dots)\in\R_+^\infty$.
Then for each fixed integer~$m\ge1$ we have
\be\label{AA} \l\,v_m \= \lim_{N\to\infty}\biggl(\frac1N\sum_{1\le n\le N\atop(m,n)=1}v_n\biggr) \;=:\;\Avm(v)\,, \ee
the ``prime-to-$m$ average" of the infinite vector~$v$.  For $m=1$ this says in particular that the partial sum
$S_N=v_1+\cdots+v_N$ equals $\l N+\text o(N)$ as~$N\to\infty$, and therefore that the Dirichlet series
$$ V(s) \= \sum_{n=1}^\infty\frac{v_n}{n^s}\=\sum_{N=1}^\infty\frac{S_N-S_{N-1}}{N^s}
   \=\sum_{N=1}^\infty S_N\Bigl(\frac1{N^s}-\frac1{(N+1)^s}\Bigr)
   \=s\,\sum_{N=1}^\infty \frac{\l\+\text o(1)}{N^s}$$
converges for $s>1$ (or for $s\in\C$ with $\Re(s)>1$) and satisfies 
 \be\label{BB} \text{Res}_{s=1}\bigl(V(s)\bigr) \= \lim_{\ve\searrow0}\bigl(\ve V(1+\ve)\bigr)\= \l\;. \ee
From~\eqref{AA} it follows that $v_m$ depends only on the prime factors of~$m$, i.e. $v_m=v_D$ where $D$ is
the {\it radical} of~$m$ (defined as the largest squarefree integer dividing~$m$, or as the product of the
prime divisors of~$m$). As a second Ansatz, again based on the numerical data
(e.g.~$v_2v_3\approx0.6290\times0.7320\approx0.4524\approx v_6$), we assume that these numbers are
multiplicative, i.e., that
$$ v_m \= \prod_{p|m} \a_p $$
for some positive real numbers $\a_p$. (Here and from now on we make the convention that the letter~$p$ always 
denotes a prime, and no longer write this in the products.) Then for $\Re(s)>1$ we have 
$$ V(s) \= \prod_p\biggl(1\+\frac{\a_p}{p^s}\+\frac{\a_p}{p^{2s}}\+\frac{\a_p}{p^{3s}}\cdots\biggr) 
 \= \prod_p\biggl(1\+\frac{\a_p}{p^s-1}\biggr) \= \z(s) \prod_p\biggl(1\+\frac{\a_p-1}{p^s}\biggr)\,,$$
and together with~\eqref{BB} this implies that the infinite product $\prod_p\Bigl(1+\frac{\a_p-1}p\Bigr)$
converges and equals~$\l$.  The same argument applied to~\eqref{AA} with~$m\ge1$ arbitrary gives
$$ \l\,\prod_{p|m}\a_p \= \lim_{s\searrow1}\biggl((s-1)\prod_{p\nmid m}\Bigl(1\+\frac{\a_p}{p^s-1}\Bigr)\biggr)
 \= \l\,\Bigl/\,\prod_{p|m}\Bigl(1\+\frac{\a_p}{p-1}\Bigr)\,.$$
Since $\l\ne0$, this equation is consistent if and only if $\a_p$ for every prime~$p$ is the positive root
of the quadratic equation $\a(1+\a/(p-1))=1$, i.e., if and only if we choose
\be\label{CC}   \a_p \= \frac{-p+1+\sqrt{(p-1)(p+3)}}2 \;. \ee
This gives the result stated in the theorem for both $\lo=\l$ and the corresponding eigenvector~$\vo$.
The general case follows the same way, since if~$\l$ is {\it any} limiting eigenvalue of $N\i\CN$ and
$v=(v_1,v_2,\dots)$ the corresponding eigenvector, then the entire argument goes through unchanged except
that it is no longer required that the numbers $v_p$ are positive, so that we can choose $v_p=\b_p$ rather
than $v_p=\a_p$ for finitely many primes~$p$, where 
\be\label{CD} \b_p= \frac{-p+1-\sqrt{(p-1)(p+3)}}2 \ee
is the Galois conjugate of~$\a_p$.  If we denote by~$D$ the product of the primes for which we have made this
alternative choice, then we get the further eigenvalues
$$ \lD\= \prod_{p\nmid D}\Bigl(1+\frac{\a_p-1}p\Bigr)\cdot\prod_{p|D}\Bigl(1+\frac{\b_p-1}p\Bigr)
     \= \lo\prod_{p|D}(\a_p-1) $$
(in accordance with~\eqref{lD}) and corresponding eigenvectors~$\vD$ as given in the theorem. 

Although the above argument was based on several heuristic assumptions, it is not hard to prove that it in fact gives
the correct answer.  First of all, each of the vectors $\vD$ that we produced satisfies the two Ans\"atze that
we made, namely, that the $n$th component of the vector is the product over all primes $p|n$ of its $p$th
component, and that the averages $\Avm(\vD)$ exist and equal $\lD\vD_m$ for all~$m\ge1$, the latter statement
being true because $\a_p=1+\text O(1/p)$. Secondly, this averaging property implies that the truncated 
vector $\vDN=(\vD_n)_{1\le n\le N}$ is a near eigenvector of~$\CN$ with near eigenvalue~$\lD$, in 
the sense that $\|(N\i\CN-\lD)\vDN\|\le\ve\|\vDN\|$ for $D$ and $\ve>0$ fixed and $N$ 
sufficiently large, and this in turn implies that the matrix $N\i\CN$ has an eigenvalue near~$\lD$,
because if $V_1,\dots,V_N$ denote an orthonormal system of eigenvectors 
of $N\i\CN$ with corresponding eigenvalues $\L_1,\dots,\L_N$, then from 
$$\|(N\i\CN-\lD)\vDN\|^2\=\sum_{i=1}^N(\L_i-\lD)^2\,(\vD,V_i)^2\;\ge\;\min_{1\le i\le N}(\L_i-\lD)^2\cdot\|\vDN\|^2$$
we obtain $\min_{1\le i\le N}|\L_i-\lD|\le\ve$.  This shows that the eigenvalues~$\lD$ indeed belong to
the limiting spectrum of~$N\i\CN$. To see that there are no others, we observe that
$$ \sum_{i=1}^N\L_i^2 \= \tr\bigl((N\i\CN)^2\bigr) 
 \= \frac1{N^2}\sum_{1\le m,n\le N\atop (m,n)=1} \= \frac1{\z(2)}\+\text o(1) $$
(where for the last equality we have used the easy and well-known fact that the probability of two large
random integers being coprime is equal to $\prod_p(1-p^{-2})=\z(2)\i$), while
$$\begin{aligned} \sum_{D\ge1\atop\text{$D$ squarefree}} \lD^2 &\=
  \prod_p\biggl[\biggr(\frac{p-1+\sqrt{(p-1)(p+3)}}{2p} \biggr)^2 \+
\biggr(\frac{p-1-\sqrt{(p-1)(p+3)}}{2p} \biggr)^2\biggr] \\
  &\= \prod_p\Bigl(1\,-\,\frac1{p^2}\Bigr) \= \frac1{\z(2)} \end{aligned}$$
(where we have used the fact that $\sum_{\text{$D$ squarefree}}f(D)=\prod_p(1+f(p))$ for any multiplicative 
function~$f(n)$).  The inequality of these two numbers show that no non-zero eigenvalues have been missed, 
since all of the~$\L_i$ are real and therefore have non-negative squares.

\subsection{Second approach via moments}\label{C4}  The argument just given for the trace of
$(\CN)^2$ can be extended to other powers.  For $M\ge1$ we set
\be\label{tM}  t_M \= \lim_{N\to\infty} \frac{\text{tr}\bigl((\CN)^M\bigr)}{N^M}\,. \ee
For $M=1$ this is~0 since $\CN$ has trace~1, and for $M=2$ it equals $\z(2)\i$ ($=6/\pi^2$), as we just saw.
In general, it is not hard to see that the limit defining $t_M$ exists for any~$M$ and gives the probability
that a cycle of $M$ random large integers has every pair of neighbors coprime.  Since the coprimality of two
integers is equivalent to the non-existence of a prime dividing both of them, we have $t_M=\prod_p t_M(p)$,
where the product is over all primes~$p$ and $t_M(p)$ denotes the probability that a random cycle of $M$
integers (mod~$p$) has no pair of adjacent 0's, i.e.~$t_M(p)=\NMc(p)/p^M$ where $\NMc(p)$ is the number
of $M$-tuples $(n_1,\dots,n_M)$ with $0\le n_i\le p-1$ and none of the $M$ pairs 
$(n_1,n_2),\dots,(n_{M-1},n_M),(n_M,n_1)$ equal to~$(0,0)$. Similarly, the limiting value
\be\label{tsM}  t_M^* \= \lim_{N\to\infty}\frac{f_1^t\,(\CN)^{M-1}f_1}{N^M}\, \ee
where $f_1=(1,\dots,1)$ as before, exists and equals the probability that
a random $M$-tuple (rather than $M$-cycle) of large integers has only coprime neighbors, which again
factors as $\prod_p(N_m(p)/p^M)$ with $N_M(p)$ being the number of $M$-tuples of integers in~$\{0,\dots,p-1\}$
with no two adjacent~0's. Both $\NMc(p)$ and $N_m(p)$ are polynomials in~$p$, with the first values
being given by
 \smallskip \begin{center} \begin{tabular}{|c||c|c|c|c|c|} \hline 
  \;$M$\qquad &  1 & 2 & 3 & 4 & 5 \\ \hline 
  \hline $\;\NMc(p)\;$ &\;$p-1$\;&\;$p^2-1$\;&\;$p^3-3p+2$\;&\;$p^4-4p^2+4p-1$\;&\;$p^5-5p^3+5p^2-1$ \\ 
  \hline $\;N_M(p)\;$ & \;$p$\;&\;$p^2-1$\;&\;$p^3-2p+1$\; & \;$p^4-3p^2+2p$\; & \;$p^5-4p^3+3p^2+p-1$\; \\
  \hline \end{tabular} \end{center} \smallskip
and the further ones by the recursive formula
 $$ \NMc(p) \= (p-1)\,\bigl(N_{M-1}^{\text{cyc}}(p)\+N_{M-2}^{\text{cyc}}(p)\bigr)\,, 
   \;\quad N_M(p) \= (p-1)\,\bigl(N_{M-1}(p)\,+N_{M-2}(p)\bigr) $$ 
for all~$M\ge3$. To see this, we denote by $N_{M,a,b}(p)$ the number of $M$-tuples of integers
in $\{(0,\dots,p-1\}$ beginning with~$a$ and ending with~$b$.  Since this number only depends on
whether $a$~and~$b$ are equal to~0 or not, and is symmetric in $a$~and~$b$, we have
$$\begin{aligned}
 \NMc(p)&\;=\sum_{0\le a,b\le p-1\atop(a,b)\ne(0,0)}N_{M,a,b}(p)\=(p-1)^2N_{M,1,1}(p)\+2(p-1)N_{M,1,0}(p)\,,\\ 
  N_M(p)&\;=\sum_{0\le a,b\le p-1\atop\hphantom{(a,b)\ne(0,0)}}N_{M,a,b}(p
     \=(p-1)^2N_{M,1,1}(p)\+2(p-1)N_{M,1,0}(p)\+N_{M,0,0}(p)  \end{aligned}$$
together with the recursive formula
$$  N_{M,a,b}(p) \= (p-1)N_{M-1,a,1}(p) \+ N_{M-1,a,0}(p)\cdot 
  \begin{cases}1&\text{if $b\ne0$,} \\ 0&\text{if $b=0$,} \end{cases} $$
from which we obtain by induction on~$M$ the closed formula
$$ \begin{pmatrix}N_{M,1,1}(p) & N_{M,0,1}(p) \\ N_{M,1,0}(p) & N_{M,0,0}(p)\end{pmatrix}
\= {\begin{pmatrix}p-1&1\\p-1&0\end{pmatrix}}^{M-2}\begin{pmatrix}1&1\\1&0\end{pmatrix}\,. $$
Combining these formulas, we obtain the special values and recursions for $N_M(p)$ and $\NMc(p)$ given above.
As a further consequence, we also find
$$\begin{aligned}t_M &\= \prod_p\frac{\NMc(p)}{p^M}
 \=\prod_p\text{tr}\Bigl(\Bigl(\begin{matrix}1-1/p&1/p\\1-1/p&0\end{matrix}\Bigr)^M\Bigr) \\
 &\= \prod_p\biggl(\biggl(\frac{p-1+\sqrt{(p-1)(p+3)}}{2p} \biggr)^M 
    \+ \biggr(\frac{p-1-\sqrt{(p-1)(p+3)}}{2p} \biggr)^M\biggr) \= \sum_{D\ge1\atop\text{$D$ squarefree}}\lD^M
\end{aligned}$$
for all $M\ge2$. This equality of traces gives another way to see that the limiting non-zero spectrum
of $N\i\CN$ is the set of real numbers~$\lD$. Finally, we can use the above formulas for $\NMc(p)$
and~$N_M(p)$ to calculate the values of $t_M=\prod_p(\NMc(p)/p^M)$ and $t^*_M=\prod_p(N_M(p)/p^M)$ numerically
by the same method as was used in~\S2 for~$\lo$, obtaining the approximate values
\smallskip \begin{center} \begin{tabular}{|c||c|c|c|c|c|c|c|} \hline 
  \;$M$\qquad & \;1 \quad  & 2 & 3 & 4 & 5 & 6 \\ \hline 
  \hline $\;t_M\;$ & 0 &\;0.60792710185\;&\;0.28674742843\;&\;0.21777871662\;&\;0.14236414403\;&\;0.09787564575\;\\ 
  \hline $\;t^*_M\;$ & 1 &\;0.60792710185\;&\;0.42824950568\; & \;0.28674742843\; & \;0.19548347937\;&\;0.13239358404\; \\
  \hline \end{tabular} \end{center} \smallskip
for small~$M$, with $t^*_2=t_2=1/\z(2)=6/\pi^2$ and $t^*_4=t_3$. We can also use them to give a closed formula for 
$N_M(p)$ for any~$M$ by diagonalizing the matrix $\bigl(\begin{smallmatrix}p-1&1\\p-1&0\end{smallmatrix}\bigr)$,
and from this deduce a formula for the ``universal ratio" $\lim_{M\to\infty}(t_M^*/t_M)$
as an Euler product, but this will be done in an easier way in the next section.

\subsection{Third approach: change of base}\label{C5}  Since the matrix $\CN$ has all components
equal to 0~or~1, the rescaled matrix $N\i\CN$ tends pointwise to~0, even though its spectrum converges.
However, we can make a change of basis over~$\Z$ in such a way that new matrix of converges pointwise to
a well-defined operator of Hilbert-Schmidt type whose spectrum is the limiting spectrum of~$N\i\CN$.
Specifically, we denote by $e_n$ ($1\le n\le N$) the $n$th standard basis element of~$\Z^N$ and by 
$f_d$ ($1\le d\le N$) the vector of length $N$ with $n$th component 1 if $d|n$ and 0 if $d\!\nmid\!n$, 
the first few values being
   $$ \begin{aligned} e_1\=(1,0,0,0,0,0,\dots)\,, & \qquad f_1\=(1,1,1,1,1,1,\dots)\,, \\
     e_2\=(0,1,0,0,0,0,\dots)\,, & \qquad f_2\=(0,1,0,1,0,1,\dots)\,, \\
     e_3\=(0,0,1,0,0,0,\dots)\,, & \qquad f_3\=(0,0,1,0,0,1,\dots)\,. \end{aligned}$$
The relationship between the bases can be expressed algebraically as
\be\label{fe}  f_d\=\sum_{1\le n\le N\atop d|n}e_n\,, \qquad e_\ell\=\sum_{1\le m\le N\atop \ell|m}\mu(m/\ell)\,f_m \ee
where in the second equation $\mu(k)$ denotes the M\"obius function (equal to $(-1)^r$ if $k$ is the product 
of $r$ distinct primes and to 0 if $k$ is not squarefree) and we have used the standard M\"obius inversion formula.
Hence the $(d,m)$-component of the matrix $B^{(N)}$ representing $\CN$ with respect to the basis~$\{f_d\}$ is 
$$ (B^{(N)})_{dm} \= \sum_{1\le n\le N\atop d|n}\sum_{\ell|m\atop(\ell,n)=1}\mu(m/\ell) 
  \= \mu(m)\sum_{1\le n\le N\atop d|n,\;m|n}1 \= \mu(m)\,\Bigl(\frac N{[m,d]}\+\text O(1)\Bigr)\,, $$
where $[d,m]=\frac{dm}{(d,m)}$ denotes the least common multiple of~$d$ and~$m$.  Thus $N\i B^{(N)}$ 
converges pointwise to the infinite matrix $B=\bigl(\frac{\mu(m)}{[m,d]}\bigr)_{d,m\in\N}$.
The $m$th column of this matrix vanishes identically if $m$ is not squarefree and its $d$th and $d'$th
rows are proportional if $d$ and $d'$ have the same prime factors, so the non-zero eigenvalues of~$B$ are the 
same as those of the $\wB$ consisting of the rows and columns of~$B$ with square-free indices (i.e.,
$\wB=(B_{dm})_{\text{$d,m$ squarefree}})$. This reduced matrix $\wB$ is simply the tensor product over all 
primes~$p$ of the $2\times2$ matrix 
$\bigl(\begin{smallmatrix}1&-1/p\\1/p&-1/p\end{smallmatrix}\bigr)$ corresponding to $m,\,d\in\{1,p\}$,
and since the eigenvalues of this matrix are equal to $\frac{p-1\pm\sqrt{(p-1)(p+3)}}{2p}$ we see that the
eigenvalues of $\wB$ are indeed precisely the numbers $\l_D$ ($D$~squarefree) defined in Theorem~1.

The following theorem gives a more precise version of this. Recall that an operator
on a Hilbert space is called {\it Hilbert-Schmidt} if the trace of its product with its adjoint 
(\,=\,the sum of the squares of the absolute values of its matrix coefficients with
respect to any orthonormal basis of the space) converges.  

\medskip 
\noindent {\bf Theorem 2.} {\bf a.} {\it Let $\H$ be the Hilbert space defined as the completion of
the vector space of finite linear combinations of vectors $f_n$ $(n\ge1)$ with respect to the norm defined
by the positive definite bilinear form $\la f_m,f_n\ra=1/[m,n]$. Then the linear operator $B:\H\to\H$ defined by
\be\label{defB}  B(f_m) \= \sum_{n=1}^\infty\frac{\mu(n)}{[m,n]}\,f_n \qquad(m\ge1) \ee
is a Hilbert-Schmidt operator and the spectrum of $N\i\CN$ converges as~$N\to\infty$ to the spectrum of~$B$.}
\newline\noindent {\bf b.\,} {\it The eigenvector $\vD$ of $B$ with eigenvalue~$\l_D$ ($D$ squarefree) is given by} 
\be\label{vtD} v^{(D)} \= \sum_{n\ge1 \atop\text{$n$ squarefree}} \wv_n^{(D)}\,f_n\,, \qquad
\wv_n^{(D)} \= \prod_{p\mid n,\;p\nmid D}(\a_p-1)\cdot\prod_{p\mid n,\;p\mid D}(\b_p-1)\,. \ee
\newline\noindent {\bf c.\,} {\it The vectors $\vD$ are orthogonal, with the scalar product $H_D:=\la\vD,\vD\ra$
given for~$D=1$ by}
\be\label{H1} H_1 \= \prod_p\Bigl(1\+\frac{\a_p^2-1}p\Bigr) \= 0.49957899226467847066\cdots  \ee
{\it and for general $D$ by}
\be\label{HD} H_D \= \prod_{p\nmid D}\Bigl(1\+\frac{\a_p^2-1}p\Bigr)\,\prod_{p\mid D}\Bigl(1\+\frac{\b_p^2-1}p\Bigr)
   \= H_1\,\prod_{p|D}(p+\a_p)\,. \ee

{\it Proof.} The bilinear form in the theorem is positive definite because the scalar product of the vectors 
$f_m$ and $f_n$ in~$\R^N$ as defined in~\eqref{fe} equals $N/[m,n]+\text O(N)$ for $N$ large and the 
number $\la f_m,f_n\ra$ is just the limiting value of $1/N$ times this as~$N\to\infty$.  
(In other words, if we rescale the scalar product in~$\R^N$ by~$1/N$ then the vectors~$f_n$ as defined in~\eqref{fe}
have well-defined limits in~$\H$, which we have denoted by the same letter, whereas the original basis
vectors~$e_n$ all tend to~0 in the limit.) The formula~\eqref{defB} for the matrix~$B$  with respect to the basis
$\{f_n\}$ of~$\H$ follows from the discussion preceding the theorem.  The matrix~$B$ is self-adjoint because 
$\la B(f_m),f_\ell\ra$ is given by the sum $\sum_{n=1}^\infty\mu(n)/[m,n][\ell,n]$, which converges
and is symmetric in $m$ and~$\ell$ (its value, which will not be need, is $\z(2)\i/\prod_{p|mn}(1+1/p)$ if 
$m$~and~$n$ are coprime and~0 if they are not), and is then of Hilbert-Schmidt type because $\tr(B^2)$ is given 
by the sum $\sum_{m,n\ge1}\mu(m)\mu(n)/[m,n]^2$, which is convergent (with value $\prod_p(1-2/p^2+1/p^2)=1/\z(2)$).  
It is perhaps worth noting that a direct proof of the positive-definiteness property, without using the
interpretation of the~$f_n$ as limits of rescaled vectors in~$\R^N$, can be given by making the change of basis 
$$  f_m^* \= \sum_{d|n} \mu(n/d)\,d\,f_d\,, \qquad f_n \= \frac1n\,\sum_{d|n}f^*_d\,, $$
since a simple calculation shows that this new basis is orthogonal, with $\la f_m^*,f_n^*\ra=\delta_{m,n}\varphi(n)$.
(Here $\varphi(n)$ is the ``Euler totient function", defined as the number of residue classes modulo~$n$ prime to~$n$
or by the formulas $\varphi(n)=\sum_{d|n}\mu(n/d)d=n\prod_{p|n}(1-1/p)$.) This gives an alternative description 
of~$\H$ as the space of vectors $\sum a_nf_n^*$ with $\sum|a_n|^2\varphi(n)=1$. Another remark is that
the operator $B$ defined by~\eqref{defB} maps~$\H$ to the subspace~$\wt\H$ spanned by the $f_n$
with square-free~$n$, so that the non-zero spectrum of~$B$ coincides with that of $\wB=B|_{\wt H}$.
(Compare the discussion of~$\wB$ preceding the theorem, or the discussion of ``equivalence classes" in
Section~III of the main paper.)

This completes the proof of part~{\bf a.} of the theorem. The formula in~{\bf b.} for the limiting value of the
coefficients of the eigenvector~$\vD$ with respect to the basis $\{f_n\}$ is an easy consequence of 
equations~\eqref{vD} and~\eqref{fe}, and is left to the reader. Equation~\eqref{H1} follows from~\eqref{vtD} because
$$ \la\vo,\vo\ra \= \sum_{m,n\ge1\atop\text{$m,\,n$ squarefree}}\frac{\wv_m^{(1)}\wv_n^{(1)}}{[m,n]}
 \= \prod_p\Bigl(1\+2\,\frac{\a_p-1}p\+\frac{(\a_p-1)^2}p\Bigr) \= \prod_p\Bigl(1\+\frac{\a_p^2-1}p\Bigr) $$
(where the numerical value is computed by the same method as already used for~$\l_1$ in~\S2 and for $t_M$ 
and~$t_M^*$ in~\S4), and~\eqref{HD} is proved in the same way with $\a_p$ replaced by~$\b_p$
whenever~$p|D$. Finally, the orthogonality of $\vD$ and $v^{(D')}$ for $D\ne D'$ follows because
$\la \vD,v^{(D')}\ra$ is given by an Euler product whose $p$th Euler factor for a prime~$p$ dividing
exactly one of $D$ and~$D'$ equals $1+\frac{\a_p-1}p+\frac{\b_p-1}p+\frac{(\a_p-1)(\b_p-1)}p=0$.

\medskip
We can use the results of Theorem~2 to give alternative formulas for the traces $t_M$ and~$t_M^*$
considered in the last section. Indeed, by a calculation similar to the one for $\la \vD,\vD\ra$ we find
\be\label{vDf1} \la \vD,f_1\ra \= \sum_{n\ge1\atop\text{$n$ squarefree}}\frac{\wv_n^{(D)}}n
  \= \prod_{p\nmid D}\Bigl(1\+\frac{\a_p-1}p\Bigr)\,\prod_{p|D}\Bigl(1\+\frac{\b_p-1}p\Bigr) \= \lD \ee
for all squarefree $D$, and hence
\be\label{tMtsM}  t_M\,=\,\tr(B^M)\,=\!\sum_{D\ge1\atop\text{$D$ squarefree}}\lD^M\,, \qquad
t_M^*\,=\,\la B^{M-1}f_1,f_1\ra\;=\!\sum_{D\ge1\atop\text{$D$ squarefree}}\frac{\lD^{M+1}}{H_D}\,, \ee
(By~\eqref{lD} and~\eqref{HD} both of these numbers have Euler products that can be checked to agree with those
given in~\S4.) Since $\lo>|\lD|$ for all~$D\ne1$, this also gives the value
\be\label{univrat} \lim_{M\to\infty}\frac{t_M^*}{t_M} \=  \frac{\lo}{H_1} \=  1.35806801915162265452\cdots \ee
for the ``universal ratio" mentioned at the end of~\S4.

\subsection{Degeneracies}\label{C6} In the previous section we gave an algebraic proof of the
orthogonality of the eigenvectors~$\vD$ for distinct squarefree integers~$D$. This orthogonality would 
of course be automatic from the eigenvalue property of the vectors $\vD$ if we knew that the eigenvalues~$\lD$
as defined in Theorem~1 are all distinct.  Surprisingly enough, however, this ``multiplicity one" statement 
is {\it false}, and in fact there are infinitely many degenerate eigenvalues, and almost certainly also
eigenvalues having an arbitrarily large multiplicity \hbox{(degeneracy)}.  Although it is something of a digression,
we include a brief discussion of this phenomenon here, especially as the multiple eigenvalues of~$B$ play a 
role in the study of the spectrum of the complementary coprimality matrix considered in the next section.

The degeneracy of eigenvalues is a very rare phenomenon, the only example under 250000 being given
by $\l_{207949}=\l_{238141}\approx-2.85\times10^{-6}$, and the only other examples
with coprime indices under 10000000 being given by $\l_{479695}=\l_{492331}\approx-1.78\times10^{-6}$
and $\l_{420595}=\l_{561511}\approx1.78\times10^{-6}$. (There are 35 other pairs of indices
under~$10^7$ with equal eigenvalues, but they are all multiples of one of these three pairs and
are therefore not interesting, since if $\l_{D_1}=\l_{D_2}$ for some squarefree numbers $D_1$ and~$D_2$
with $(D_1,D_2)=D_3>1$, then $\l_{D_1/D_3}$ and $\l_{D_2/D_3}$ are already equal and $\l_{D_1}$
and $\l_{D_2}$ are just obtained by multiplying them by~$\l_{D_3}/\lo$.)  To see why these examples
hold, we note that $\lD$ for general squarefree~$D$ equals $\lo\mu(D)\prod_{p|D}\ve_p$ where $\ve_p$ 
is the quadratic unit~$\frac{p+1-\sqrt{(p-1)(p+3)}}2$ of norm~1 and trace~$p+1$.  In the first 
example we have $207949=7\times61\times487$, while $238141$ is prime, and also
$\ve_7=\a$, $\ve_{61}=\a^2$, $\ve_{487}=\a^3$ and $\ve_{238141}=\a^6$, where $\a=4-\sqrt{15}$, 
so that $\l_{207949}=-\lo\a\,\a^2\a^3=-\lo\a^6=\l_{238141}$. In the second and third examples
we have the prime factorizations $479695=5\times197\times487$, $492331=7\times61\times1153$,
$420595=5\times7\times61\times197$ and $561511=487\times1153$ and also $\ve_5=\b$, $\ve_{197}=\b^3$
and $\ve_{1153}=\b^4$, where $\b=3-2\sqrt2$, so $\l_{479695}=\l_{492331}=-\l_{420595}=-\l_{561511}=-\lo\a^3\b^4$.

The general picture is as follows.  For any integer $c\ge1$ we denote by $\NN(c)$ the set of integers
$n\ge1$ for which $2T_n(c)-1$ is prime, where $T_n(c)$ denotes the $n$th Chebyshev polynomial (defined
by~$T_n(c)=\cos n\theta$ if $c=\cos\theta$). Standard conjectures of number theory imply that
\smallskip\newline \phantom{X} (a) the set $\NN(c)$ is infinite for every~$c$, and
\smallskip\newline \phantom{X} (b) any finite subset $I\subset\N$ is contained in~$\NN(c)$ 
  for infinitely many values of~$c$.
\smallskip\newline Indeed, for~(a) the standard heuristics of number theory imply that the probability 
of each number $2T_n(c)-1$ being prime should be inversely proportional to its logarithm, which grows linearly 
with~$n$, so that (assuming that these probabilities are independent) the number of integers $n\le n_0$ 
belonging to~$\NN(c)$ should grow roughly like $\log n_0$ as~$n_0\to\infty$, while for~(b) we use that
any fixed finite collection of irreducible polynomials in one variable are expected to have all of their 
values prime for infinitely many integer values of their argument (Hardy-Littlewood conjecture). Now
if $c$ is an integer for which the set $\NN(c)$ contains (say) the set $\{1,2,3,4\}$, which by~(b)
should occur infinitely often, then we have $\l_{p_1p_4}=\l_{p_2p_3}$, where $p_n$ for $n\in\NN(c)$
denotes the prime~$2T_n(c)-1$, because from the definition of the Chebyshev polynomials we have
$\ve_{p_n}=\ve_{p_1}^n$. This already gives us a conjectural set of approximately $X^{1/5}/(\log X)^4$ 
pairs of squarefree integers~$\le X$ with equal eigenvalues, and a larger set of expected cardinality
about $X^{1/3}/(\log X)^3$ can be obtained by taking $c$ with $\NN(c)\supset\{1,2,3\}$ and then
using the equality $\l_{487p_3}=\l_{427p_1p_2}$ (or $\l_{427p_3}=\l_{487p_1p_2}$) 
if $\{p_1,p_2,p_3\}\cap\{7,61,487\}=\emptyset$. As a more elaborate example, we have that
$\NN(c)\supseteq\{4,6,8,10,16,18,24,36\}$ for~$c=2870$, so that we have $\l_D=\ve^{54}\lo$ for
each $D\in\{p_{18}p_{36},\,p_6p_8p_{16}p_{24},\,p_4p_8p_{18}p_{24},\,p_4p_6p_8p_{36}\}$ and $\l_D=-\ve^{58}$ 
for each $D\in\{p_{16}p_{18}p_{24},\, p_6p_{16}p_{36},\,p_4p_{18}p_{36},\,p_4p_6p_8p_{16}p_{24}\}$,
where $\ve=2870-\sqrt{2870^2-1}$. However, the indices of these multiplicity~4 examples are fairly
huge, since the primes factors involved are large, ranging from $p_4=1085544213969601$ 
to $p_{36}\approx 2\times10^{135}$, and the eigenvalues with these high multiplicities
are corresponding tiny: $\ve^{54}\approx10^{-203}$ and $-\ve^{58}\approx-10^{-218}$.
Finally, we note that (assuming the conjectural statement~(b) above) we can get
eigenvalues of large multiplicities either by taking $c$ with $\NN(c)\supset\{1,\dots,k\}$ with $k$
large or by taking many values of~$c$ with $\NN(c)\supset\{1,2,3\}$ and multiplying the corresponding indices.
But for the statement that there are infinitely many degenerate eigenvalues no conjecture
is needed, since for this we can simply take any pair of the form $(207949p,238141p)$ with $p$ a large prime.

\subsection{The complementary coprimality matrix}\label{C7}  So far we have been studying the spectrum 
of the matrix $\CN$ for large~$N$, corresponding in the terminology of the main body of the paper to the
``antiferromagnetic" case.  But one is also interested in the ``ferromagnetic" case, corresponding to the 
{\it complementary coprimality matrix} $\bC=\bCN$ defined by
$$ \bC_{mn}\= 1\,-\,C_{mn}
 \= \begin{cases} 0 &\text{ if $(m,n)=1$} \\  1 &\text{ if $(m,n)>1$}\end{cases}\qquad(1\le m,\,n\le N). $$
In this final section of the appendix we discuss the asymptotic spectrum of $N\i\bCN$ for large~$N$.

Since the first row and column of this symmetric $N\times N$ matrix vanish identically, we could omit
them, in which case $\bCN$ would coincide with the $(N-1)\times(N-1)$ matrix $\Phi$ studied in the main 
body of the paper (where $N$ was denoted by~$q$).  The $p$th row and column of $\bC$ also vanish identically
if $p$ is a prime with $N/2<p\le N$, so this matrix is not primitive (we have $(\bC^M)_{pp}=0$ for all~$m$),
but if we remove these offending rows and columns, which only removes some 0's from the spectrum of~$\bC$,
then the resulting matrix is primitive (its 4th power has strictly positive entries, as one checks easily)
and hence again has a Perron-Frobenius eigenvalue $\k\PF>0$ and eigenvector $w\PF$.  We can again find
these rapidly even for quite large~$N$, and again we find that the numbers $\k\PF/N$ and the components
of~$w=w\PF$ (now normalized by $w_2=1$, since $w_1$ vanishes identically) converge numerically to the
values $0.54636$ and $w=(0,\,1,\,0.5626,\,1,\,0.3003,\,1.19394,\,0.2043,\,1,\,0.5626,\,\dots)$ (again
corresponding to $N$=10000, and where again the value of $w_n$ depends only on the radical of~$n$). Similarly,
by computing the characteristic polynomial of the whole matrix $N\i\bCN$ (now only up to $N=1000$), we find 
that its eigenvalues and eigenvectors again converge, with the three smallest and three largest limiting 
eigenvalues having the approximate values $(-0.07348,\,-0.04605,\,-0.03358)$ and $(0.1264,\,0.2041,\,0.5464)$. 

We were not able to find any ``closed form" expression for any of these numbers.  However, the following
theorem and its proof give an explicit (and, as we will see afterwards, numerically effective) description
of the limiting spectrum of $N\i\bCN$ and of the corresponding eigenvectors.  To state it, we first
observe that, by the same arguments as in the case of the original coprimality matrix, the limiting
spectrum of~$N\i\bCN$ is the same as the spectrum of the map $\bB:\H\to\H$ defined by
\be \label{defbB}  \bB(w) \= \la w,f_1\ra\,f_1 \,-\, B(w) \qquad(w\in\H)\,, \ee
which is again a Hilbert-Schmidt operator since it differs from~$-B$ by an operator with a finite-
(actually, one-) dimensional image. To describe the spectrum of this operator, we introduce the function
\be \label{defF} F(x) \= \sum_{D} \frac{\l_D^2/H_D}{x+\lD}\;, \ee
with $\l_D$ and $H_D$ as in Theorems~1 and~2.  (Here and from now on we write simply~$\sum_D$ to denote 
a sum over positive squarefree integers~$D$.)  Since $\l_D^2/H_D=\text O(1/D^3)$, the sum 
in~\eqref{defF} is absolutely convergent for $x\in\C^*\smallsetminus\{-\lD\}$ and defines a meromorphic
function in~$\C^*$ with simple poles at the negatives of the eigenvalues~$\lD$ and no other singularities.
The spectrum of~$\bB$ is then given as follows.

\medskip 
\noindent {\bf Theorem 3.}  {\it The spectrum of $\bB$ consists of the roots~$\k$ of the equation $F(\k)=1$ 
together with the negatives of the of the multiple eigenvalues of $B$. Moreover, the eigenvalues of~$\bB$
are intertwined with the eigenvalues of~$-B$ in the sense that in each connected component 
of $(-\lo,\infty)\smallsetminus\{-\l_2,-\l_3,\dots\}$  there is precisely one eigenvalue of~$\bB$, with 
multiplicity one, while if $\l\ne0$ is an eigenvalue of~$B$ with multiplicity~$m\ge2$ then $-\l$ is an
eigenvalue of~$\bB$ with multiplicity exactly~$m-1$.}

\smallskip{\it Proof.} The Hilbert-Schmidt property of~$\bB$ guarantees that $\H$ has a basis
consisting of orthogonal eigenvectors for this operator.  Suppose that $w$ is an eigenvector with
eigenvalue~$\k\ne0$. We have to distinguish two cases, according as $\la w,f_1\ra$ vanishes or not.
In the first case it follows from~\eqref{defbB} that $w$ is also an eigenvector of~$B$, with eigenvalue
$-\k$, so $\k=-\lD$ for some positive squarefree integer~$D$. This integer need not be unique, as we saw
in the last section, but (since $|\lD|\to0$ as~$D\to\infty$) there are at most finitely many indices
$D_1,\dots,D_m$ with $\l_{D_i}=-\k$.  By~\eqref{vDf1} we know that all of the $v^{(D_i)}$ have the same
scalar product with~$f_1$, so~$w$, which is orthogonal to~$f_1$ by assumption, belongs to the 
$(m-1)$-dimensional subspace of~$\H$ spanned by the vectors $v^{(D_i)}-v^{(D_m)}$ with $1\le i\le m-1$, 
and conversely any vector $w$ in this subspace satisfies $\bB(w)=\k w$. This takes care of the cases
in the theorem where the eigenvalue of~$\bB$ coincides with some eigenvalue of~$-B$.  We can therefore
now assume that the scalar product $\la w,f_1\ra$ is non-zero, and then by rescaling~$w$ that it is 
equal to~1. Write $w$ in terms of the eigenbasis $\{\vD\}$ of~$\H$ as $\sum_Dc_D\,\vD$ with coefficients
$c_D\in\R$ satisfying $\sum_Dc_D^2H_D<\infty$. Then equations~\eqref{defbB} and~\eqref{vDf1} give
$$ \sum_D(\k+\lD)\,c_D\,\vD \= B(w)\+\bB(w) \= f_1 \= \sum_D\frac{\lD}{H_D}\,\vD\,, $$
from which we deduce $\k+\lD\ne0$ and $c_D=\frac{\lD/H_D}{\k+\l_D}$ for all~$D$. Now substituting
this formula into the equation $\la w,f_1\ra=1$, and using~\eqref{vDf1} once again, we find that $F(\k)=1$
as claimed. The fact that there is precisely one value of~$\k$ between any two consecutive distinct
eigenvalues of~$-B$ follows from the fact that $F'(x)$ is negative whenever it is finite, so that
$F(x)$ is strictly monotone decreasing between any two consecutive poles and hence assumes the value~1
precisely once in any such interval. 

\medskip
We can use Theorem~3 to compute the first few eigenvalues~$\k$.  The sum~\eqref{defF} already
converges like $\sum_DD^{-3}$, as mentioned above, but if we want to obtain the roots of $F(x)=1$
to high precision then this is not good enough.  (For instance, with 10000 terms we would only get
about~8 digits of precision, and to get 30 digits we would need an unrealistic $10^{15}$ terms.)
To get faster convergence, we write
$$ F(x) \= \sum_D\frac{\lD^2}{H_D}\,\biggl(\sum_{M=1}^A\frac{(-\lD)^{M-1}}{x^M}
 \+\frac{(-\lD)^A}{x^A(x+\lD)} \biggr)
 \= \sum_{M=1}^A\frac{(-1)^{M-1}t_M^*}{x^M} \+ \frac{(-1)^A}{x^A}\sum_D\frac{\lD^{A+1}/H_D}{x+\lD}$$
(here we have used~\eqref{tMtsM} for the second inequality) for any integer $A>0$, where the infinite 
sum now converges like $\sum D^{-A-3}$. We can now calculate $F(x)$ to very high precision and 
compute the points where it takes on the value~1, finding in particular that the largest one 
(Perron-Frobenius eigenvalue of~$\bB$) has the numerical value $0.54637892502940275779044781\dots$,
in very good agreement with the experimental values quoted at the beginning of the section, and
that the next five values are equal roughly to 0.203916198, 0.126014340, 0.090504259, $-0.073640081$,
and $-0.045951975$, again in good agreement with the numerical data. 

There is in fact yet another way to calculate $F(x)$, in which the convergence becomes exponential 
rather than merely (inverse) polynomial, and we give this as well since it leads naturally to our
last topic, the study of the moments of~$\bB$.  Denote by $T^*(x)=\sum_{M\ge1}t_M^*x^M$ the generating
series of the numbers~$t_M^*$ discussed in~\S4.  Since $t_M^*=\text O(\lo^M)$, this series
converges for $|x|<\lo\i$, and from equation~\eqref{tMtsM} we find
\be\label{TsF} T^*(x) \= \sum_{M=1}^\infty\sum_D\frac{\lD^{M+1}}{H_D}\,x^M
   \= \sum_D\frac{\lD^2}{H_D}\,\frac x{1-\lD x} \= -F(-1/x) \ee
for such~$x$.  We can then combine the two expansions by writing
$$ T^*(x) \= \sum_{|\lD|>c}\frac{\lD^2}{H_D}\,\frac x{1-\lD x} 
\+ \sum_{M=1}^\infty\biggl(t_M^*\,-\,\sum_{|\lD|>c}\frac{\lD^{M+1}}{H_D}\biggr)\,x^M$$
for any constant~$c>0$, where the first sum is finite and the second converges with exponential
rapidity for~$|x|<c\i$, making it evident that $T^*(x)$ has a meromorphic continuation to all of
$\C^*$ with simple poles at $x=\lD\i$ as its only singularities.  We end by showing how to use
the function $T^*(x)$ to compute the cyclic and free-boundary-condition moments
\be\label{uMusM} s_M \,=\, \tr({\bB}^M) \= \sum_j\k_j^M\,, \qquad
   s_M^*\= \la\bB^M\!f_1,f_1\ra \= \sum_j\frac{\la w_j,f_1\ra^2}{\la w_j,w_j\ra}\,\k_j^{M-1} \ee
(the analogues for~$\bB$ of the numbers~$t_M$ and $t_M^*$ studied in~\S4), where in the second
expression in each case $\{w_j\}$ denotes an orthogonal system of eigenvectors of~$\bB$ with 
eigenvalues~$\k_j$.  A simple combinatorial argument (which we omit) starting with~\eqref{defbB} 
shows that these numbers are computed in terms of the original moments $t_M$ and $t_M^*$ by
\be\label{sM} s_M \=(-1)^Mt_M
 \+ M\sum_{k=1}^M\frac{(-1)^{M-k}}k\sum_{M_1,\dots,M_k\ge1\atop M_1+\cdots+M_k=M}t_{M_1}^*\cdots t_{M_k}^*\ee
and
\be\label{ssM} s^*_M\=\sum_{k=1}^M(-1)^{M-k}\sum_{M_1,\dots,M_k\ge1\atop M_1+\cdots+M_k=M}t_{M_1}^*\cdots t_{M_k}^*\,. \ee
Introduce the characteristic power series $\D_B(x)=\prod_D(1-\lD x)$ and $\D_{\bB}(x)=\prod_j(1-\k_jx)$.
From the expansion $\log(1-x)=-\sum_M\frac{x^M}M$ we deduce that $\D_B(x)=\exp\bigl(-\sum_Mt_M\frac{x^M}M\bigr)$
and similarly for~$\D_{\bB}$ with $t_M$ replaced by~$s_M$, so multiplying~\eqref{sM} by $(-1)^{M-1}x^M/M$
and summing over $M\ge1$ we get
$$ \log\D_{\bB}(-x)\=\log\D_B(x)\+\sum_{k=1}^\infty\frac{(-1)^{k-1}T^*(x)^k}k \= \log\D_B(x)\+\log(1+T^*(x))\,. $$
or $\D_{\bB}(-x)=\D_B(x)(1+T^*(x))$.
This gives another proof of the theorem, since the zeros of \hbox{$\D_B(x)(1+T^*(x))$} are the zeros of $\D_B(x)$
with multiplicity reduced by~1 (an eigenvalue of multiplicity~$m$ of~$B$ corresponds to an $m$-fold zero of
$\D_B$ and a simple pole of~$T^*$) together with the zeros of $1+T^*(x)$, which by~\eqref{TsF} are precisely
the roots of $F(-1/x)=1$. It also gives another way to obtain the eigenvalues~$\k$ numerically, since the
function $\D_B(x)(1+T^*(x))$ is entire and hence has a power series expansion with coefficients tending to
zero more than exponentially quickly (as we can check numerically using the values computed in~\S4; for
instance, the numbers $t_{100}$ and $t^*_{100}$ are of the order of $\lo^{100}\approx10^{-17}$, while the
100th coefficient of the product of the power series $\exp(-\sum t_Mx^M/M)$ and $1+T^*(x)$ is of the order
of~$10^{-186}$) and hence can be used to compute~$\D_{\bB}(x)$ accurately for any~$x$. In the same way,
multiplying~\eqref{ssM} by $(-x)^M$ and summing over~$M$ we obtain the formula
$$  1\+S^*(-x) \= \sum_{k=0}^\infty(-1)^kT^*(x)^k\=\frac1{1+T^*(x)} $$
for the generating series $S^*(x)=\sum_{M\ge1}s_M^*x^M$ of the moments~$s_M^*$ in terms of the generating
series~$T^*(x)$ of the~$t_M^*$.  This in turn allows us to calculate the ``universal ratio" 
$\lim\limits_{M\to\infty}\frac{s_M^*}{s_M}$, since $-\sum_Ms_Mx^{M-1}$, the logarithmic derivative of $\D_{\bB}(x)$,
has a simple pole with residue~1 at its smallest singular point $x=1/\k\PF$ (inverse Perron-Frobenius 
eigenvalue) while $\sum s^*_Mx^M$ has a simple pole of residue $-1/{T^*}'(-1/\k\PF)$ at the same point, giving
\be\label{UR}  \lim_{M\to\infty}\frac{s_M^*}{s_M} \= \frac{-1}{\k\PF\,F'(\k\PF)} \= 1.294408632903071274\cdots\;, \ee
in good numerical agreement with the value obtained experimentally in the main body of the paper. More
generally, from the expression $w=\sum_D\frac{\lD/H_D}{\k+\lD}\vD$ proved above for the $\vD$-expansion of an
eigenvector~$w$ of~$\bB$ with eigenvalue~$\k$, normalized by~$\la w,f_1\ra=1$, we obtain the formula
$\la w,w\ra=\sum\frac{\lD^2/H_D}{(\k+\lD)^2}=-F'(\k)$ for any~$\k$, so that~\eqref{UR} also follows from~\eqref{uMusM}.


\end{document}